\documentclass[11pt,a4]{article}
\pdfoutput=1
\usepackage{jheppub}
\usepackage{graphicx,amssymb,amsmath,amsfonts}

\title{Rotating strings confronting PDG mesons}

\author{Jacob Sonnenschein and Dorin Weissman}
\affiliation{The Raymond and Beverly Sackler School of Physics and Astronomy,\\
	Tel Aviv University, Ramat Aviv 69978, Israel}
\emailAdd{cobi@post.tau.ac.il}
\emailAdd{dorinw@mail.tau.ac.il}

\abstract{We revisit the model of mesons as rotating strings with massive endpoints and confront it with meson spectra. We look at Regge trajectories both in the \((J,M^2)\) and \((n,M^2)\) planes, where $J$ and $n$ are the angular momentum  and  radial excitation number respectively. We start from states comprised of \(u\) and \(d\) quarks alone, move on to trajectories involving \(s\) and \(c\) quarks, and finally analyze the trajectories of the heaviest observed \(\bbb\) mesons. The endpoint masses provide the needed transition between the linear Regge trajectories of the light mesons to the deviations from linear behavior encountered for the heavier mesons, all in the confines of the same simple model. From our fits we extract the values of the quark  endpoint masses, the Regge slope (string tension) and quantum intercept. The model also allows for a universal fit where with a single value of the Regge slope we fit all the \((J,M^2)\) trajectories involving \(u\), \(d\), \(s\), and \(c\) quarks. We include a list of predictions for higher mesons in both $J$ and $n$.}
\def\be{\begin{equation}}
\def\ee{\end{equation}}
\newcommand{\del}{\partial}
\newcommand{\alp}{\ensuremath{\alpha'\:}}
\newcommand{\ssb}{s\bar{s}}
\newcommand{\ccb}{c\bar{c}}
\newcommand{\bbb}{b\bar{b}}
\newcommand{\rchi}[1]{\ensuremath{\chi^2_m/\chi^2_l = #1}}
\newcommand{\ten}[1]{\times10^{#1}}
\newcommand{\mud}{m_{u/d}}
\begin{document}
\maketitle
\flushbottom
\section{Introduction}
The stringy description of mesons, which was one of the founding motivations of string theory, has been thoroughly investigated since the seventies of the last century\cite{Collins:book}. In this note we reinvestigate this issue.
What is the reason then to go back to ``square one" and revisit this question? There are at least three reasons for reinvestigating the stringy nature of mesons: (i) Holography, or gauge/string duality, provides a bridge between the underlying theory of QCD (in certain limits) and a bosonic string model of mesons. (ii) There is a wide range of  heavy mesonic resonances that have been discovered in recent years, and (iii) up to date we lack a full exact procedure of quantizing a rotating string with massive endpoints.

In this note we will not add anything new about (iii) but rather combine points (i) and (ii). Namely, we describe a model of spinning bosonic strings with massive endpoints that follows from a model of spinning strings in holographic confining backgrounds. Leaving aside the regime where holography applies, we then confront this model with experimental data of meson spectra. We use $\chi^2$ fits to check the validity of the model and to extract its defining parameters.

The passage from the original AdS/CFT duality to the holographic description of hadrons in the top down approach  includes several steps. First one has to deform the background, namely the geometry and the bulk fields, so that the corresponding dual gauge field theory is non-conformal and non-supersymmetric. Prototype backgrounds of such a nature are that of a $D_4$ brane compactified on $S^1$\cite{Witten:1998zw} (and its non-critical analogous model\cite{Kuperstein:2004yf}). The fundamental quark degrees of freedom are then injected to the gravity models via ``flavor probe branes". For instance for the compactified D4 brane model D8 anti D8 branes are incorporated\cite{SakSug}.

The spectra of hadrons has been determined in these models by computing the spectra of the fluctuations of bulk fields corresponding to glueballs and scalar and vector fluctuations of the probe-branes which associate with scalar and vector mesons respectively. (See for instance \cite{SakSug},\cite{Casero:2005se},\cite{oai:arXiv.org:0806.0152}).

Both for the  glueballs and for the mesons the spectra deduced from the gravitational backgrounds and the probe branes do not admit Regge behavior, neither the linear relation between $M^2$ and the angular momentum $J$ nor the linearity between $M^2$ and the radial excitation number $n$. In fact in terms of the bulk fields one can get also scalar and vector mesons. To get higher spin mesons one has to revert to a stringy configuration. There is an unavoidable big gap between the low spin mesons described by the gravity and probe modes and the high spin one described by holographic strings\cite{Peeters:2006iu}.\footnote{The bottom-up approach of the soft-wall model has been proposed in order to admit $(n,M^2)$ linearity \cite{oai:arXiv.org:hep-ph/0602229}.This model suffers from certain other drawbacks and does not admit $(J,M^2)$ linearity.  It seems fair to say that generically the spectra of the bulk and probe modes associated with confining backgrounds do not admit the Regge behavior.}

An alternative approach to extract the spectra of mesons and glueballs, both low and high spin ones, is to study rotating open strings connected to the probe branes\footnote{This approach was used also in \cite{Imoto:2010ef}.} or folded closed strings for mesons. Regge trajectories of the latter in various confining backgrounds were analyzed in \cite{oai:arXiv.org:hep-th/0311190}. It is a very well known feature of rotating stringy configurations in flat space-time.

		\begin{figure}[t!] \centering
					\includegraphics[width=.90\textwidth, trim = 10 80 20 200]{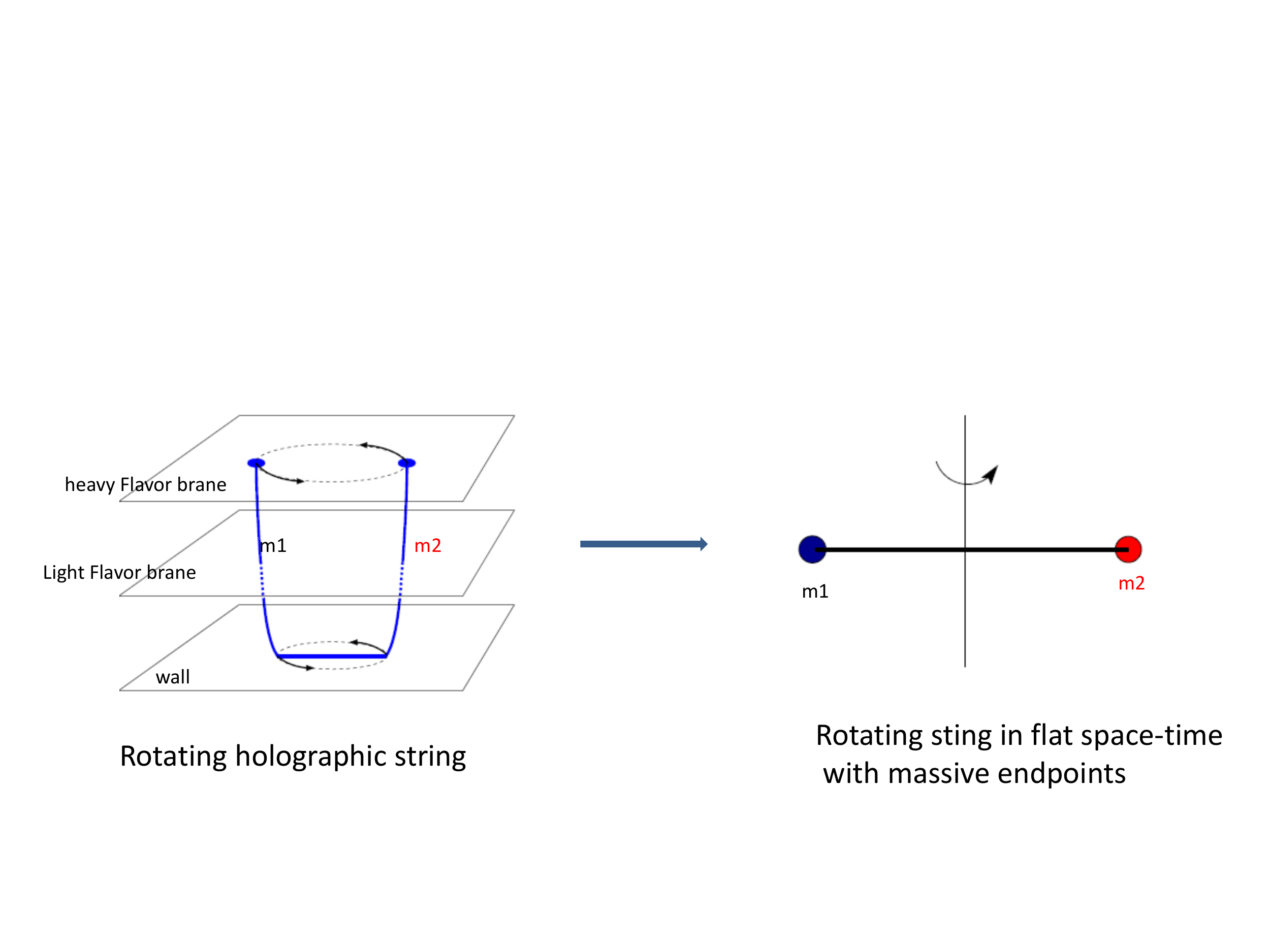}
					\caption{\label{fig:mapholflat} On the left: rotating holographic open string and on the right the corresponding open string with massive endpoints in flat space-time with $m_1=m_2$.}
		\end{figure}

The major difference between rotating open strings in holographic backgrounds and those in flat space-time is that the former do not connect the two endpoints along the probe brane, but rather stretch along the ``wall"\footnote{In top down models the ``wall" refers to the minimal value of the holographic radial direction.} and then connect vertically to flavor branes (See figure (\ref{fig:mapholflat})). The figure depicts the special case of $m_1=m_2$. In a similar manner we can have $m_1\neq m_2$ by attaching the ``vertical" strings segments to different flavor branes.

In \cite{Kruczenski:2004me} it was shown that classically the holographic rotating string can be mapped into one in flat space-time with massive endpoints.\footnote{In \cite{Kruczenski:2004me} the map was shown for a particular class of models. One can generalize this map to rotating open strings in any confining background\cite{cobitalks}.} Basically it was shown that the equations of motion of the two systems are equivalent.

The string endpoint mass is given approximately by the string tension times the ``length" of the string along one of the two ``vertical" segments. This reduces to\cite{Kruczenski:2004me}
 \be
 m_{sep}= T\int_{u_\Lambda}^{u_f}du\sqrt{g_{00}g_{uu}}
 \ee
   where $T$ is the string tension, $u$ is the holographic radial coordinate, $u_\Lambda$ is its minimal value (the ``wall"), $u_f$ is the location of the flavor branes and $g_{00}$ and $g_{uu}$ are the metric components along the time and holographic radial directions respectively.

   Obviously this mass is neither the QCD physical mass nor the constituent quark mass. We would like to argue that both for the spectra as well as for decays\cite{Peeters:2005fq} of mesons this is the relevant physical mass parameter.

In this note we assume this map, consider a bosonic string rotating in flat four dimensional space-time with massive endpoints as a model for mesons and leave aside holography altogether.\footnote{Approximating the ``vertical segment'' with the massive endpoints is reminiscent of a similar approximation done with holographic Wilson lines \cite{oai:arXiv.org:hep-th/9911123}. A comparison between mesons and holographic rotating strings, rather than massive strings in flat space-time, is deferred to future work.}

The theoretical models we use are rather simple. We start from an action that includes a Nambu-Goto term for the string and two terms that describe relativistic massive chargeless particles. We write down the corresponding classical equations of motion and the Noether charges associated with the energy $E$ and angular momentum  $J$ of the system. Unlike the massless case, for massive endpoints there is no explicit relation between for instance $E$ and $(J,m,T)$, but rather $E$ and $J$ can be written in terms of $T$, $m$ and $\omega l$, where $\omega $ is the angular velocity and $l$ is the string length. For two limits of light massive endpoints where $\frac{2m}{E} \ll 1$ and heavy ones where $\frac{E-2m}{2m}\ll 1$ one can eliminate $\omega l$ (the two limits involve taking \(\omega l \rightarrow 1\) and \(\omega l \rightarrow 0\) respectively) and get approximated direct relations between $E$ and $J$.

Going beyond the classical limit for rotating strings is a non-trivial task. The common lore for strings with massless endpoints, namely the linear trajectories, is that the passage from classical to quantum  trajectories is via the replacement
\be
J = \alp E^2 \qquad \rightarrow \qquad J + n - a = \alp E^2
\ee
where the slope  $\alp = (2\pi T)^{-1}$, $n$ is the radial excitation number and $a$ is the intercept.

In a recent paper\cite{Hellerman:2013kba} a precise analysis of the quantum massless string has been performed. It was shown there that for a case of a single plane of angular momentum, in particular in $D=4$ dimensions, an open string with no radial excitation ($n=0$) indeed admits $J - a = \alp E^2$ with $a=1$. This is a non-trivial result since the calculation of the intercept (to order $J^0$) yields in D dimensions the result $ a = \frac{D-2}{24} + \frac{26-D}{24} =1$, where the first term is the usual ``Casimir" term and the second is the Polchinski-Strominger term. For the rotating string with massive endpoints a similar determination of the intercept has not yet been written down even though certain aspects of the quantization of such a system have been addressed\cite{Chodos:1973gt}\cite{Baker:2002km}\cite{Zahn:2013yma}.

Falling short of the full quantum expression for the Regge trajectories one can use a WKB approximated determination of the trajectories\cite{Schreiber:2004ie}. The latter depends on the choice of the corresponding potential.

The models used in this paper to fit that experimental data are the following:
\begin{itemize}
\item The linear trajectory   $J + n = \alp E^2 +a $
\item The ``massive trajectory" which is based on the classical expressions for $E$ and $J$ where the latter includes assumed quantum correction, again in the form of $J\rightarrow J+n-a$. The trajectories then read
	\be E = 2m\left(\frac{q\arcsin(q)+\sqrt{1-q^2}}{1-q^2}\right) \ee
	\be J + n = a + 2\pi\alp m^2\frac{q^2}{(1-q^2)^2}\left(\arcsin(q)+q\sqrt{1-q^2}\right)   \ee
	These expressions reduce to the linear trajectory equation in the limit \(m \rightarrow 0\).
\item The WKB approximation for the linear potential $V=Tl$ which takes the form
	\be n = a + \alp E^2 \left(\sqrt{1-b^2}+b^2\log\left(\frac{1-\sqrt{1-b^2}}{b}\right)\right) \ee
where  $b \equiv (2m/E)$.
\end{itemize}

The parameters that we extract from the fits are the string tension (or the slope $\alp$), the string endpoint masses, and the intercept.

%%%%%%%%%%%%%%%%%%%%%%%%%%%

The main idea of this paper is to investigate  the possibility of constructing a unified description of mesons that covers mesons of light quarks as well as those built from heavy quarks. It is a common practice to view mesons of light quarks with the linear Regge trajectories (which correspond to rotating open strings with massless endpoints) and non-relativistic potential models for heavy quark mesons. Here we suggest and test a stringy model that interpolates between these two descriptions.

In a sequel paper we propose and confront with data in a similar manner a stringy rotating model for baryons.

The paper is organized as follows. In the next section we describe the basic theoretical model. We start with the action, equations of motion and Noether charges of the rotating bosonic string with massive endpoints. We then present a WKB approximation. Next we describe the fitting procedure. Section 4 is devoted to the results of the various fits. We separate the latter to fits of the $M^2$ as a function of the angular momentum $(J,M^2)$ and of the radial excitation $(n,M^2)$. In both categories we discuss light quark mesons, strange  mesons, charmed mesons and mesons containing $b$ quarks. We present also a universal fit. We then present our WKB fits. We discuss the issue of fits with respect to the orbital angular momentum $L$ and the total angular momentum $J$, and calculate the string lengths to verify the validity of a long string approximation for the fitted mesons. Section 5 is devoted to a summary, conclusions and open questions.
\section{Basic theoretical model}
	\subsection{Classical rotating string with massive endpoints}
		We describe the string with massive endpoints (in flat space-time) by adding to the Nambu-Goto action,
		\be S_{NG} = -T\int\!\!{d\tau d\sigma \sqrt{-h}} \ee
		\[ h_{\alpha\beta} \equiv \eta_{\mu\nu} \del_\alpha X^\mu \del_\beta X^\nu \]
		a boundary term - the action of a massive chargeless point particle
		\be S_{pp} = -m\int\!\! d\tau \sqrt{-\dot{X}^2} \ee
		\[ \dot{X}^\mu \equiv \del_\tau X^\mu \]
		at both ends. There can be different masses at the ends, but here we assume, for simplicity's sake, that they are equal. We also define \(\sigma = \pm l\) to be the boundaries, with \(l\) an arbitrary constant with dimensions of length.

		The variation of the action gives the bulk equations of motion
		\be \del_\alpha\left(\sqrt{-h}h^{\alpha\beta}\del_\beta X^\mu\right) = 0 \label{eq:bulk} \ee
		and at the two boundaries the condition
		\be T\sqrt{-h}\del^\sigma X^\mu \pm m\del_\tau\left(\frac{\dot{X}^\mu}{\sqrt{-\dot{X}^2}}\right) = 0 \label{eq:boundary} \ee
		
		It can be shown that the rotating configuration
		\be X^0 = \tau, X^1 = R(\sigma)\cos(\omega\tau), X^2 = R(\sigma)\sin(\omega\tau) \ee
		solves the bulk equations \eqref{eq:bulk} for any choice of \(R(\sigma)\). We will use the simplest choice, \(R(\sigma) = \sigma\), from here on.\footnote{Another common choice is $X^0=\tau, x^1=\sin(\sigma) \cos(\omega\tau), X^2= \sin(\sigma)\sin(\omega\tau)$.} Eq. \eqref{eq:boundary} reduces then to the condition that at the boundary,
		\be \frac{T}{\gamma} = \gamma m \omega^2 l \label{eq:boundaryRot}\ee
		with \(\gamma^{-1} \equiv \sqrt{1-\omega^2 l^2}\).\footnote{Notice that in addition to the usual term $\gamma m$ for the mass, the tension that balances the ``centrifugal force" is $\frac{T}{\gamma}$.}
		We then derive the Noether charges associated with the Poincar\'e invariance of the action, which include contributions both from the string and from the point particles at the boundaries. Calculating them for the rotating solution, we arrive at the expressions for the energy and angular momentum associated with this configuration:
	\be	E = -p_0 = 2\gamma m + T \int_{-l}^l\!\frac{d\sigma}{\sqrt{1-\omega^2\sigma^2}} \ee
	\be J = J^{12} = 2\gamma m \omega l^2 + T \omega \int_{-l}^l\! \frac{\sigma^2 d\sigma}{\sqrt{1-\omega^2\sigma^2}} \ee
		Solving the integrals, and defining \(q \equiv \omega l\) - physically, the endpoint velocity - we write the expressions in the form
		\be E = \frac{2m}{\sqrt{1-q^2}} + 2Tl\frac{\arcsin(q)}{q} \ee
		\be J = 2m l \frac{q}{\sqrt{1-q^2}} + Tl^2\left(\frac{\arcsin(q)-q\sqrt{1-q^2}}{q^2}\right) \ee
		The terms proportional to \(m\) are the contributions from the endpoint masses and the term proportional to \(T\) is the string's contribution. These expressions are supplemented by condition \eqref{eq:boundaryRot}, which we rewrite as
		\be Tl = \frac{mq^2}{1-q^2} \label{eq:Tm}\ee
		This last equation can be used to eliminate one of the parameters \(l, m , T,\) and \(q\) from \(J\) and \(E\). Eliminating the string length from the equations we arrive at the final form
		\be E = 2m \left(\frac{q\arcsin(q)+\sqrt{1-q^2}}{1-q^2}\right) \label{eq:massiveE} \ee
		\be J = \frac{m^2}{T}\frac{q^2}{(1-q^2)^2}\left(\arcsin(q)+q\sqrt{1-q^2}\right) \label{eq:massiveJ} \ee
		These two equations are what define the Regge trajectories of the string with massive endpoints. They determine the functional dependence of \(J\) on \(E\), where they are related through the parameter \(0 \leq q < 1\) (\(q = 1\) when \(m = 0\)). Since the expressions are hard to make sense of in their current form, we turn to two opposing limits - the low mass and the high mass approximations.
			In the low mass limit where the endpoints move at a speed close to the speed of light, so \(q \rightarrow 1\), we have an expansion in \((m/E)\):
		\be J = \alp E^2\left(1-\frac{8\sqrt{\pi}}{3}\left(\frac{m}{E}\right)^{3/2} + \frac{2\sqrt{\pi^3}}{5}\left(\frac{m}{E}\right)^{5/2} + \cdots\right) \label{eq:lowMass}\ee
		from which we can easily see that the linear Regge behavior is restored in the limit \(m\rightarrow 0\), and that the first correction is proportional to \(\sqrt{E}\). The Regge slope \(\alp\) is related to the string tension by \(\alp = (2\pi T)^{-1}\).
		The high mass limit, \(q \rightarrow 0\), holds when \((E-2m)/2m \ll 1\). Then the expansion is
		\be J = \frac{4\pi}{3\sqrt{3}}\alp m^{1/2} (E-2m)^{3/2} + \frac{7\pi}{54\sqrt{3}} \alp m^{-1/2} (E-2m)^{5/2}
			+ \cdots \label{eq:highMass} \ee
			
\subsection{The WKB approximation}
	We follow here the approach of  E. Schreiber\cite{Schreiber:2004ie}, where the string is treated as a ``fast'' degree of freedom that can be replaced by an effective potential between the ``slow'' degrees of freedom - the string endpoints. Then, we treat the endpoints as spinless point particles in a potential well. As such, the relativistic energy carried by the quarks is:
		\be (E-V(x))^2 - p^2 = m^2 \ee
	If the particle is at the end of a rotating rod of length \(x\), then \(p^2 = p_x^2 + (J_q/x)^2\). With the usual replacement of \(p_x \rightarrow -i\partial_x\), we arrive at the one dimensional differential equation to be solved
		\be -\partial_x^2\psi(x) = \left[\left(E-V(x)\right)^2-m^2-(J_q/x)^2\right]\psi(x) \ee
		If we define
		\be p(x) = \sqrt{(E-V(x))^2-m^2-(J_q/x)^2} \ee
	then the spectrum of the system is obtained, in the WKB approximation, by the quantization condition
		\be \pi n = \int_{x_-}^{x_+} p(x) dx \ee
	The limits of the integral \(x_-\) and \(x^+\) are the classical ``turning points'' - those points where the integrand, \(p(x)\), is zero. The condition that the integral be an integer multiple of \(\pi\) implies the relation between \(n\) and the energy eigenvalues \(E_n\).
	How we continue from here depends on our choice of the potential \(V(x)\). Also, we have to decide how to relate the total angular momentum \(J\) with the momentum carried by the point particles, \(J_q\), and the angular momentum carried by the string itself, which we'll call \(J_s\).

	If we treat the string as a classical rotating rod, then the (non-relativistic) expression for its energy is
		\be V(x) = Tx + \frac{3}{2}\frac{J_s^2}{Tx^3} \ee	
	Another option is the quantum mechanical expression for a string fixed at both ends\cite{Arvis:1983fp}
		\be V(x) = \sqrt{(Tx)^2 - T\frac{\pi(D-2)}{12}} \ee
	More general potentials can also include a spin-orbit interaction term, or an added Coulomb potential.
	The simplest option, and the one for which we can solve the integral analytically, is to set contributions from the string's angular momentum and the quantum corrections to the potential \(V(x)\) to zero. Namely, to take the linear potential \(V(x) = |Tx|\). To solve the integral, we also have to assume \(J_q = 0\), so the state has no angular momentum at all.
		\be n\pi = \frac{E^2}{T}\left[\sqrt{1-b^2}+b^2\log\left(\frac{1-\sqrt{1-b^2}}{b}\right)\right] \ee
		with \(b = m/E\). Now this is a result for only one particle - half our system. We modify the result to apply it to the two particle system (assuming the two particles are identical in mass) by the simple replacement \(n \rightarrow n/2\), and \(E \rightarrow E/2\). The equation is now of the form
		\be n = \alp E^2 \left[\sqrt{1-b^2}+b^2\log\left(\frac{1-\sqrt{1-b^2}}{b}\right)\right] \ee
		with \(\alp = (2\pi T)^{-1}\) as always, and \(b\) now redefined to be
		\be b \equiv \frac{2m}{E} \ee
		The high mass expansion (\(1-b \ll 1\)) of the above expression is similar to the one obtained for the classical rotating string:
		\be n = \frac{8}{3}\alp m^{1/2}(E-2m)^{3/2} + \frac{1}{5}\alp m^{-1/2}(E-2m)^{7/2} + \ldots \label{eq:highMassW} \ee
		Comparing this with the high mass limit for the classical rotating string, in eq. \eqref{eq:highMass}, we see that the only difference is in the expansion coefficients, a difference of about 10\% in the coefficients for the leading term, and 16\% in the next to leading order.
		The low mass expansion, on the other hand, results in a different kind of behavior from the classical rotating string:
		\be n = \alp E^2 \left(1 + 4\left(\frac{m}{E}\right)^2\log\left(\frac{m}{E}\right)
						- 2\left(\frac{m}{E}\right)^2 + 2\left(\frac{m}{E}\right)^4+\ldots\right) \label{eq:lowMassW} \ee
		The leading order term now being proportional to \(\alp m^2 \log(m/E)\), as opposed to the \(\alp m^{3/2} E^{1/2}\) of the expansion in eq. \eqref{eq:lowMass}.
\section{Fitting models}
	\subsection{Rotating string model}
		We define the \emph{linear} fit by
			\be J + n = \alp E^2 + a \label{eq:linear} \ee
		where the fitting parameters are the \emph{slope} \alp and the \emph{intercept}, \(a\).

		For the \emph{massive} fit, we use the expressions for the mass and angular momentum of the rotating string, eqs. \eqref{eq:massiveE} and \eqref{eq:massiveJ}, generalized to the case of two different masses, and we add to them, by hand, an intercept and an extrapolated \(n\) dependence, assuming the same replacement of \(J \rightarrow J + n - a \).
			\be E = \sum_{i = 1,2}m_i\left(\frac{q_i\arcsin(q_i)+\sqrt{1-q_i^2}}{1-q_i^2}\right) \label{eq:massFitE} \ee
			\be J + n = a + \sum_{i=1,2}\pi\alp m_i^2\frac{q_i^2}{(1-q_i^2)^2}\left(\arcsin(q_i)+q_i\sqrt{1-q_i^2}\right) \label{eq:massFitJ} \ee
		We relate the velocities \(q_1\) and \(q_2\) can be related using the boundary condition \eqref{eq:boundaryRot}, from which we have
		\be \frac{T}{\omega} = m_1\frac{q_1}{1-q_1^2} = m_2\frac{q_2}{1-q_2^2} \label{eq:boundaryTwo}\ee
		so the functional dependence between \(E\) and \(J\) is still through only one parameter \(0 \leq q_i < 1\).
		With the two additions of \(n\) and \(a\), the two equations reduce to that of the linear fit in \eqref{eq:linear} in the limit where both masses are zero.
		Now the fitting parameters are \(a\) and \(\alp\) as before, as well as the the two endpoint masses \(m_1\) and \(m_2\). For a lot of the cases we assume \(m_1 = m_2\) and retain only one free mass parameter, \(m\).
\subsection{WKB model}
		The third fitting model is the WKB. It is defined by
		\be n = a + \frac{1}{\pi}\int_{x_-}^{x^+}dx\sqrt{(E-V(x))^2-m^2-(J_q/x)^2} \label{eq:fitWKB} \ee
		where \(x_{\pm}\) are the points where the integrand is zero and again we have added an intercept as an independent parameter by hand.	The potential we chose was simply the linear potential \(V(x) = Tx\) with \(T = (2\pi\alp)^{-1}\). The angular momentum is then carried only by the quarks. We chose to identify \(J_q\) with the orbital angular momentum \(L\). For those states with \(J_q = 0\) we solve the integral and use the resulting formula,
		\be n = a + \alp E^2 \left(\sqrt{1-b^2}+b^2\log\left(\frac{1-\sqrt{1-b^2}}{b}\right)\right) \label{eq:fitWKB0}\ee
		where \(b \equiv (2m/E)\). If we can't make that assumption we solve eq. \eqref{eq:fitWKB} numerically. The fitting parameters are again \(m\), \alp, and \(a\).
\subsection{Fitting procedure}
		We measure the quality of a fit by the dimensionless quantity \(\chi^2\), which we define by
				\be \chi^2 = \frac{1}{N-1}\sum_i\left(\frac{M_i^2-E_i^2}{M_i^2}\right)^2 \label{eq:chi_def} \ee
			\(M_i\) and \(E_i\) are, respectively, the measured and calculated value of the mass of the \(i\)-th particle, and \(N\) the number of points in the trajectory. We will also use the subscripts \(l\), \(m\), or \(w\) to denote which fitting model a given value of \(\chi^2\) pertains to. So, for instance, \(\chi^2_l/\chi^2_m\) is the ratio of the value of \(\chi^2\) obtained by a linear fit to that of a massive fit of the same trajectory.
			A more common definition of \(\chi^2\) would have the standard deviation \(\sigma_i = \Delta M_i^2\) in the denominator, but we have used \(M_i^2\). We do this mostly for reasons of practicality. The high accuracy to which some of the meson's masses are known makes \(\chi^2\) (when defined using \(\sigma_i\) as the denominator) vary greatly with very small changes in the fitting parameters. \footnote{For example, the mass of the \(\ssb\) \(\phi\) is \(1019.455\pm0.020\) MeV. Fixing the mass and slope at values near the minimum for \(\chi^2\) as defined in \eqref{eq:chi_def}, a change to the intercept from \(a = 0.8210\) (the minimum using our definition) to \(0.8211\) takes \(\chi^2\) (using the standard definition) from \(7.21\) to \(0.02\), and going to \(a = 0.8212\) takes us to \(\chi^2 = 8.66\). This type of behavior may also result in our fitting algorithms missing the optimum entirely.} We feel the kind of precision required then in the fits is unnecessary for the purposes of our work. By using definition \eqref{eq:chi_def} for \(\chi^2\) we can still extract reasonably accurate values for the fitting parameters from the different trajectories, and identify those deviations from the linear Regge behavior which we will attribute to the presence of massive endpoints.
\section{Fit results}
This section discusses the results of our fits. The fits to the trajectories in the \((J,M^2)\) plane and the trajectories in the \((n,M^2)\) plane are presented separately. For the radial trajectories, where we have used both the massive model and the WKB model, the results are further separated between the two different types of fits. In each subsection, we describe the lightest quark trajectories first and move on gradually to the heaviest.
The details of the fits to each of the individual trajectories, including the specification of all the states used and the plots of each of the trajectories in the \((J,M^2)\) or \((n,M^2)\) planes, can be found in appendix \ref{app:individual}.
\paragraph{A note on units and notation:} When units are not explicitly stated, they are GeV\(^{-2}\) for \(\alp\) and MeV for masses. The intercept \(a\) is dimensionless. If the letters \(l\), \(m\), or \(w\), are used as subscripts, they will always refer to the linear, massive, and WKB fits respectively.
\subsection{Trajectories in the \texorpdfstring{$(J,M^2)$}{(J,M2)} plane}
		\subsubsection{Light quark mesons}	
		\begin{figure}[t!] \centering % chi_3d_light
						\includegraphics[natwidth=1200bp, natheight=900bp, width=.40\textwidth]{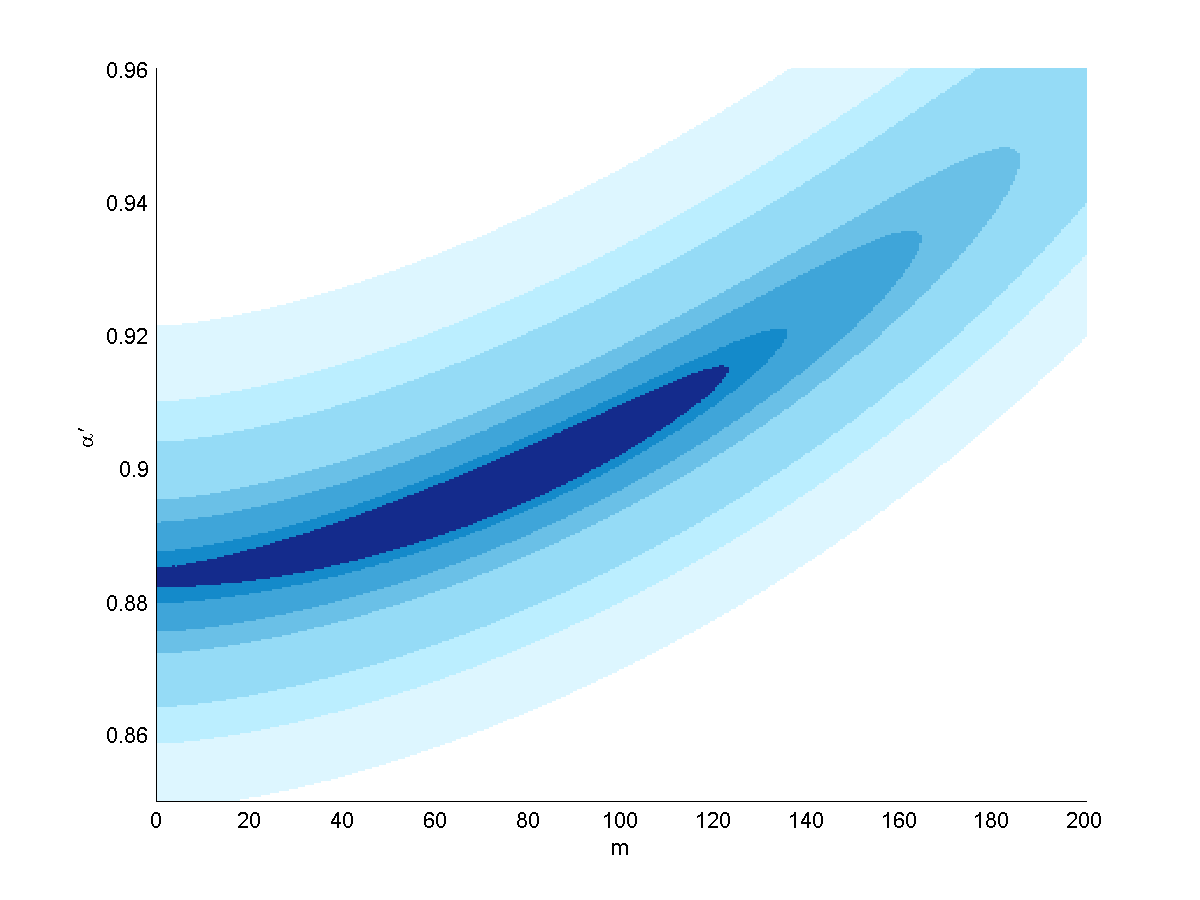}
						\includegraphics[natwidth=392bp, natheight=900bp, width=.13\textwidth]{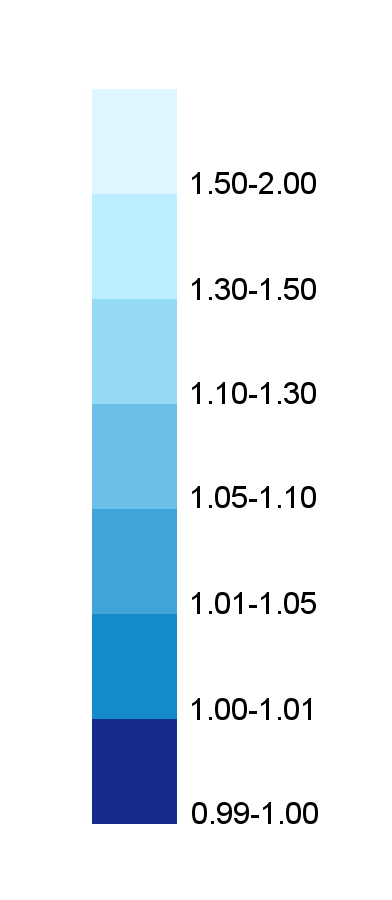}
						\includegraphics[natwidth=1200bp, natheight=900bp, width=.40\textwidth]{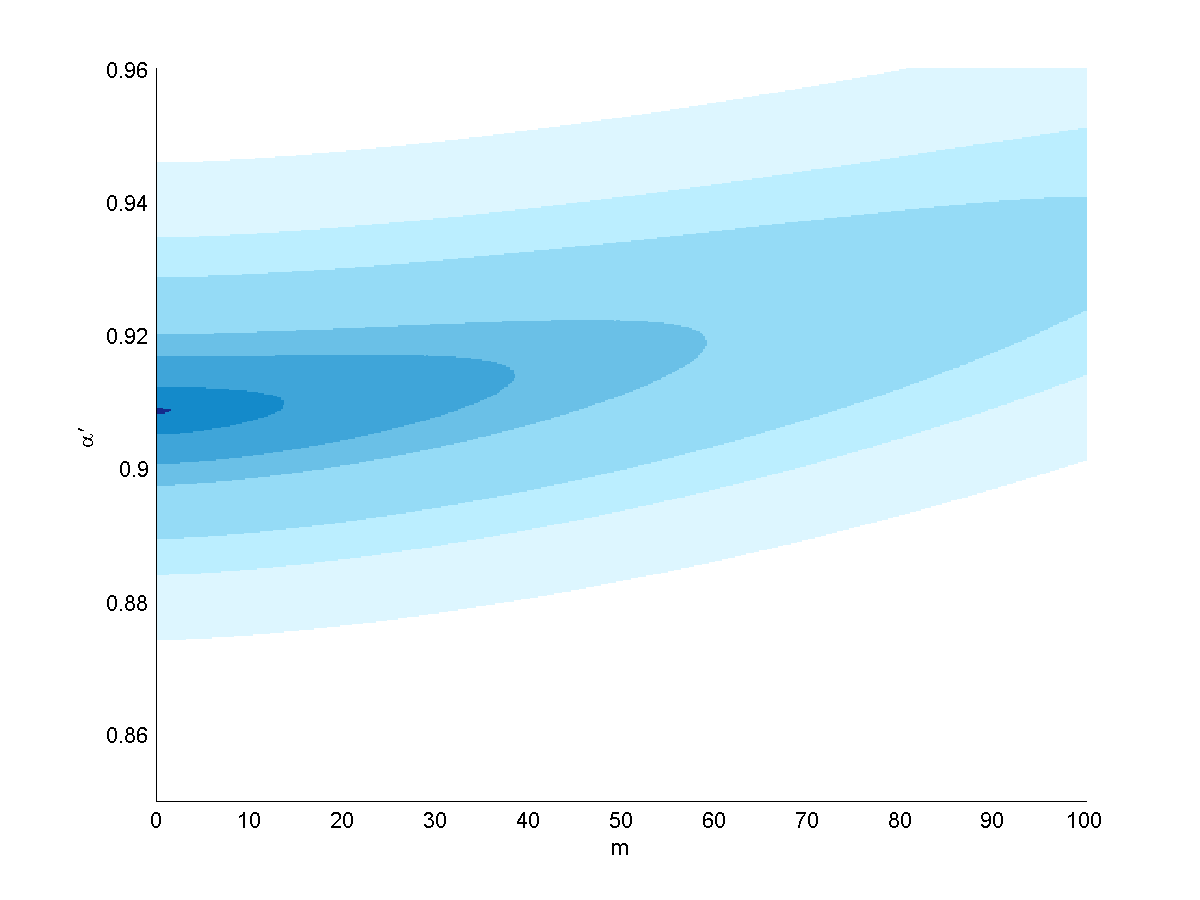}
						\caption{\label{fig:chi_3d_light} \(\chi^2\) as a function of \(\alp\) and \(m\) for the \((J,M^2)\) trajectory of the \(\rho\)  (left) and \(\omega\) (right) mesons. The intercept \(a\) is optimized to get a best fit for each point in the \((\alp,m)\) plane. \(\chi^2\) in these plots is normalized so that the value of the optimal linear fit \((m=0)\) is \(\chi^2 = 1\).}
\end{figure}
			We begin by looking at mesons consisting only of light quarks - \(u\) and \(d\). We assume for our analysis that the \(u\) and \(d\) quarks are equal in mass, as any difference between them would be too small to reveal itself in our fits.
			This sector is where we have the most data, but it is also where our fits are the least conclusive. The trajectories we have analyzed are those of the \(\pi/b\), \(\rho/a\), \(\eta/h\), and \(\omega/f\).
			
			Of the four \((J,M^2)\) trajectories, the two \(I = 1\) trajectories, of the \(\rho\) and the \(\pi\), show a weak dependence of \(\chi^2\) on \(m\). Endpoint masses anywhere between \(0\) and \(160\) MeV are nearly equal in terms of \(\chi^2\), and no clear optimum can be observed. For the two \(I = 0\) trajectories, of the \(\eta\) and \(\omega\), the linear fit is optimal. If we allow an increase of up to \(10\%\) in \(\chi^2\), we can add masses of only \(60\) MeV or less. Figure (\ref{fig:chi_3d_light}) presents the plots of \(\chi^2\) vs. \(\alp\) and \(m\) for the trajectories of the \(\omega\) and \(\rho\) and shows the difference in the allowed masses between them.
			
			The slope for these trajectories is between \(\alp = 0.81-0.86\) for the two trajectories starting with a pseudo-scalar (\(\eta\) and \(\pi\)), and \(\alp = 0.88-0.93\) for the trajectories beginning with a vector meson (\(\rho\) and \(\omega\)). The higher values for the slopes are obtained when we add masses, as increasing the mass generally requires an increase in \(\alp\) to retain a good fit to a given trajectory. This can also be seen in figure (\ref{fig:chi_3d_light}), in the plot for the \(\rho\) trajectory fit.
			
		\subsubsection{Strange and \texorpdfstring{$\ssb$}{s-sbar} mesons}
			We analyze three trajectories in the \((J,M^2)\) involving the strange quark. One is for mesons composed of one \(s\) quark and one light quark - the \(K^*\), the second is for \(\ssb\) mesons - the trajectory of the \(\phi\), and the last is for the charmed and strange \(D^*_s\), which is presented in the next subsection with the other charmed mesons.
					\begin{figure}[t!] \centering % chi_j_strange
						\includegraphics[natwidth=1200bp, natheight=900bp, width=.48\textwidth]{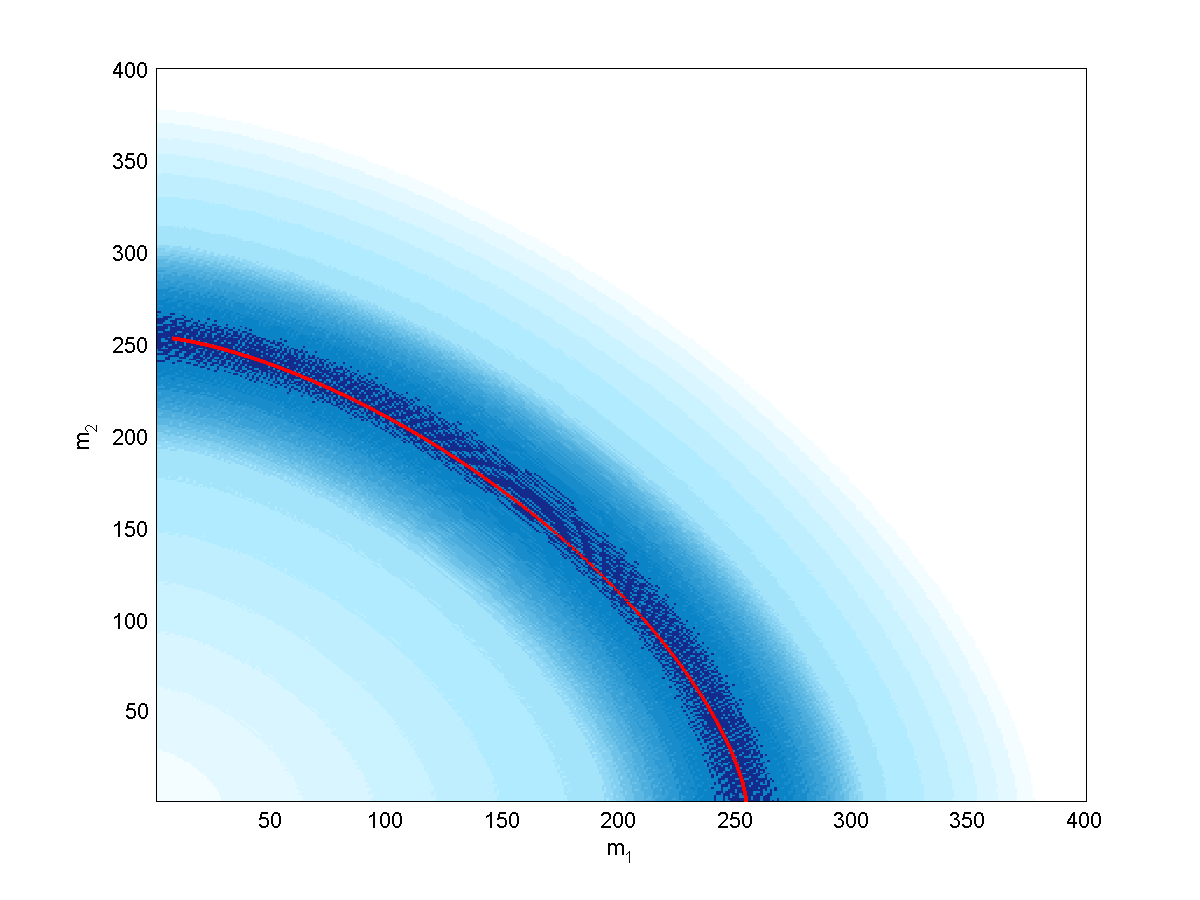}	 \hfill
						\includegraphics[natwidth=1200bp, natheight=900bp, width=.48\textwidth]{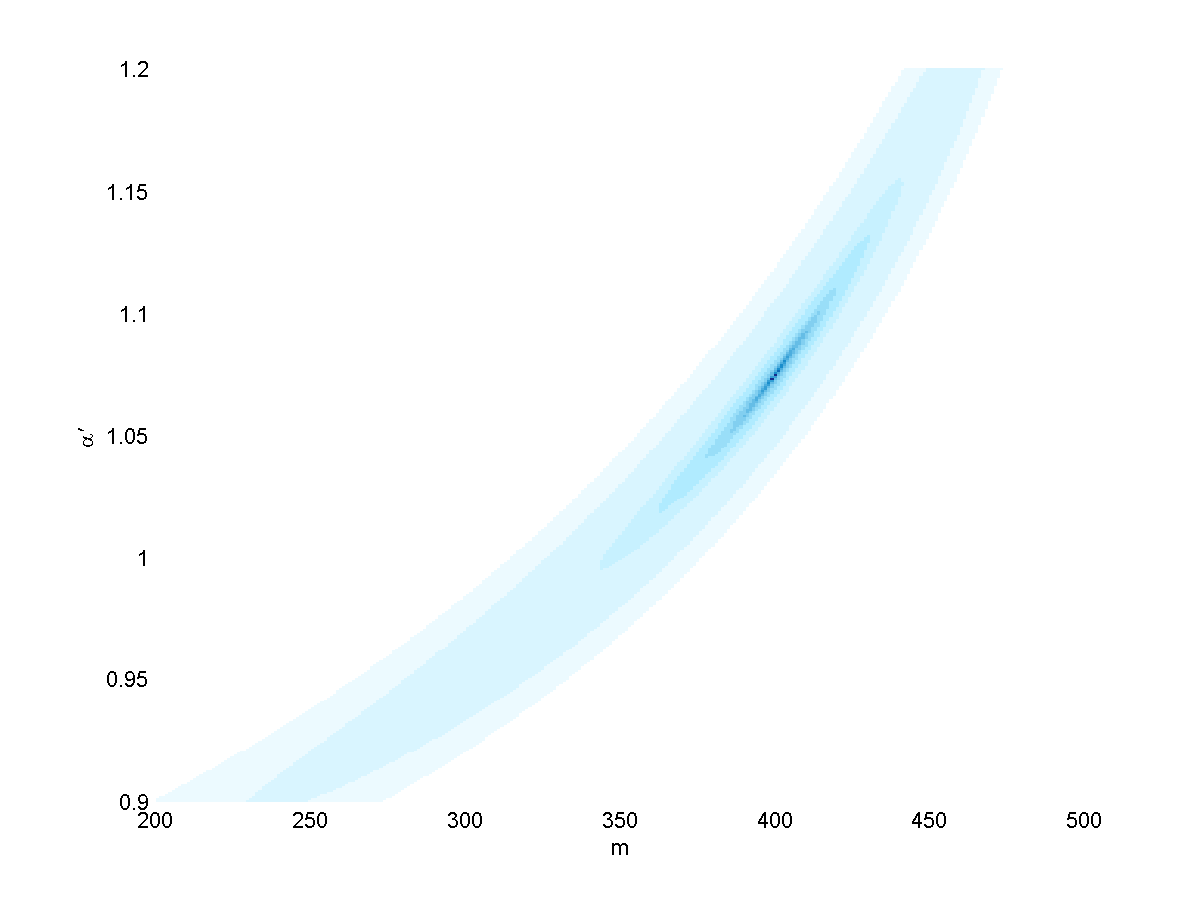}
						\caption{\label{fig:chi_j_strange} Left: \(\chi^2\) as a function of two masses for the \(K^*\) trajectory. \(a\) and \(\alp\) are optimized for each point. The red line is the curve \(m_1^{3/2}+m_2^{3/2} = 2\times(160)^{3/2}\) along which the minimum (approximately) resides. The minimum is \(\rchi{0.925}\) and the entire colored area has \(\chi^2_m/\chi^2_l < 1\). On the right is \(\chi^2\) as a function of \(\alp\) and \(m\) for the \((J,M^2)\) trajectory of the \(\phi\). The intercept \(a\) is optimized. The minimum is at \(\alp = 1.07, m = 400\) with \(\chi^2_m/\chi^2_l < 10^{-4}\) at the darkest spot. The lightest colored zone still has \(\chi^2_m/\chi^2_l < 1\), and the coloring is based on a logarithmic scale.}
				\end{figure}
			
			The \(K^*\) trajectory alone cannot be used to determine both the mass of the \(u/d\) quark and the mass of the \(s\). The first correction to the linear Regge trajectory in the low mass range is proportional to \(\alp\left(m_1^{3/2}+m_2^{3/2}\right)\sqrt{E}\). This is the result when eq. \eqref{eq:lowMass} is generalized to the case where there are two different (and small) masses. The plot on the left side of figure (\ref{fig:chi_j_strange}) shows \(\chi^2\) as a function of the two masses.
			
			The minimum for the \(K^*\) trajectory resides along the curve \(\mud^{3/2} + m_s^{3/2} = 2\times(160)^{3/2}\). If we take a value of around \(60\) MeV for the \(u/d\) quark, that means the preferred value for the \(m_s\) is around 220 MeV. The higher mass fits which are still better than the linear fit point to values for the \(s\) quark mass as high as 350 MeV, again when \(\mud\) is taken to be 60 MeV. The slope for the \(K^*\) fits goes from \(\alp = 0.85\) in the linear fit to \(0.89\) near the optimum to \(0.93\) for the higher mass fits.
			
			The trajectory of the \(\ssb\) mesons includes only three states, and as a result the optimum is much more pronounced than it was in previous trajectories. It is found at the point \(m_s = 400\), \(\alp = 1.07\). The value of \(\chi^2\) near that point approaches zero. The range in which the massive fits offer an improvement over the linear fit is much larger than that, as can be seen in the right side plot of figure (\ref{fig:chi_j_strange}). Masses starting from around \(m_s = 250\) MeV still have \rchi{0.50} or less, and the slope then has a value close to that of the other fits, around \(0.9\) GeV\(^{-2}\).
		
		\subsubsection{Charmed and \texorpdfstring{$\ccb$}{c-cbar} mesons}
		
		\begin{figure}[t!] \centering
						\includegraphics[natwidth=1200bp, natheight=900bp, width=.48\textwidth]{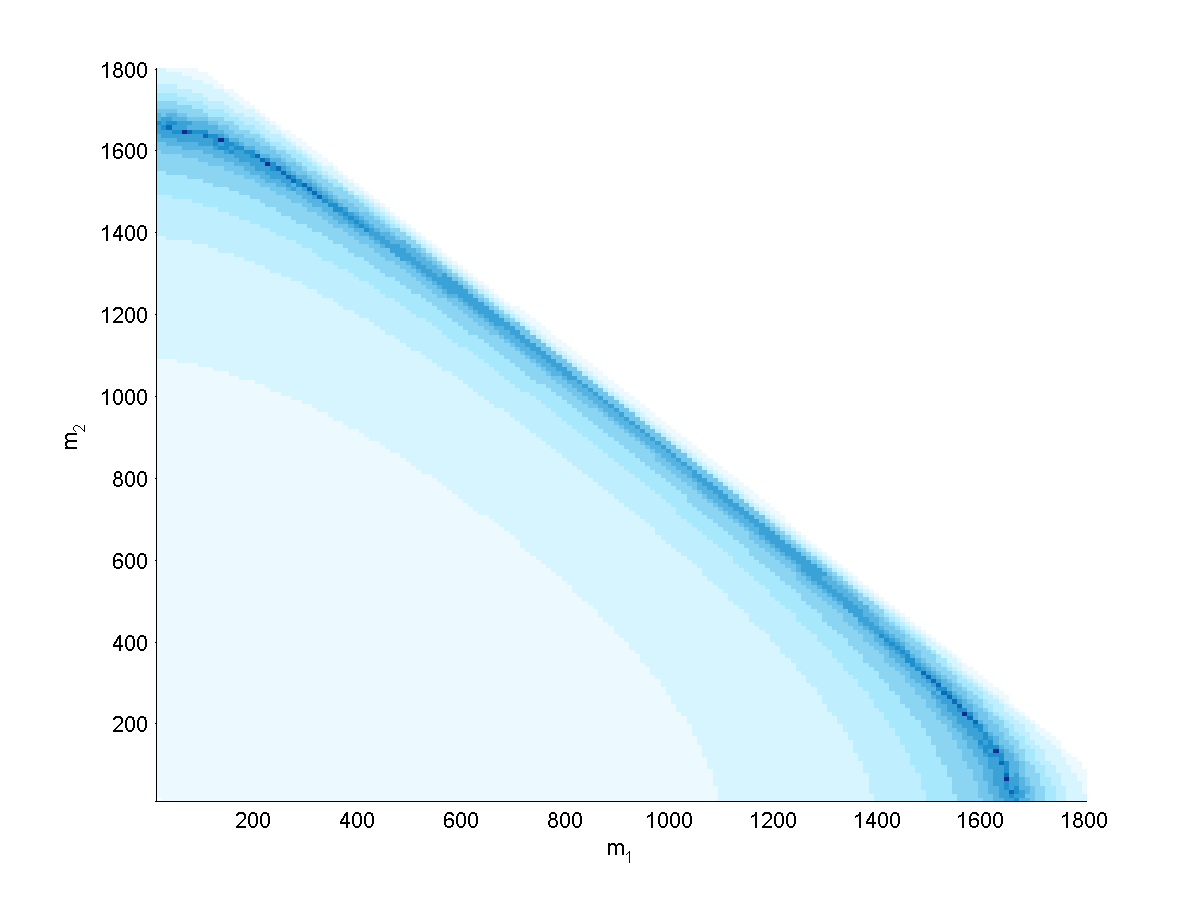}	 \hfill
						\includegraphics[natwidth=1200bp, natheight=900bp, width=.48\textwidth]{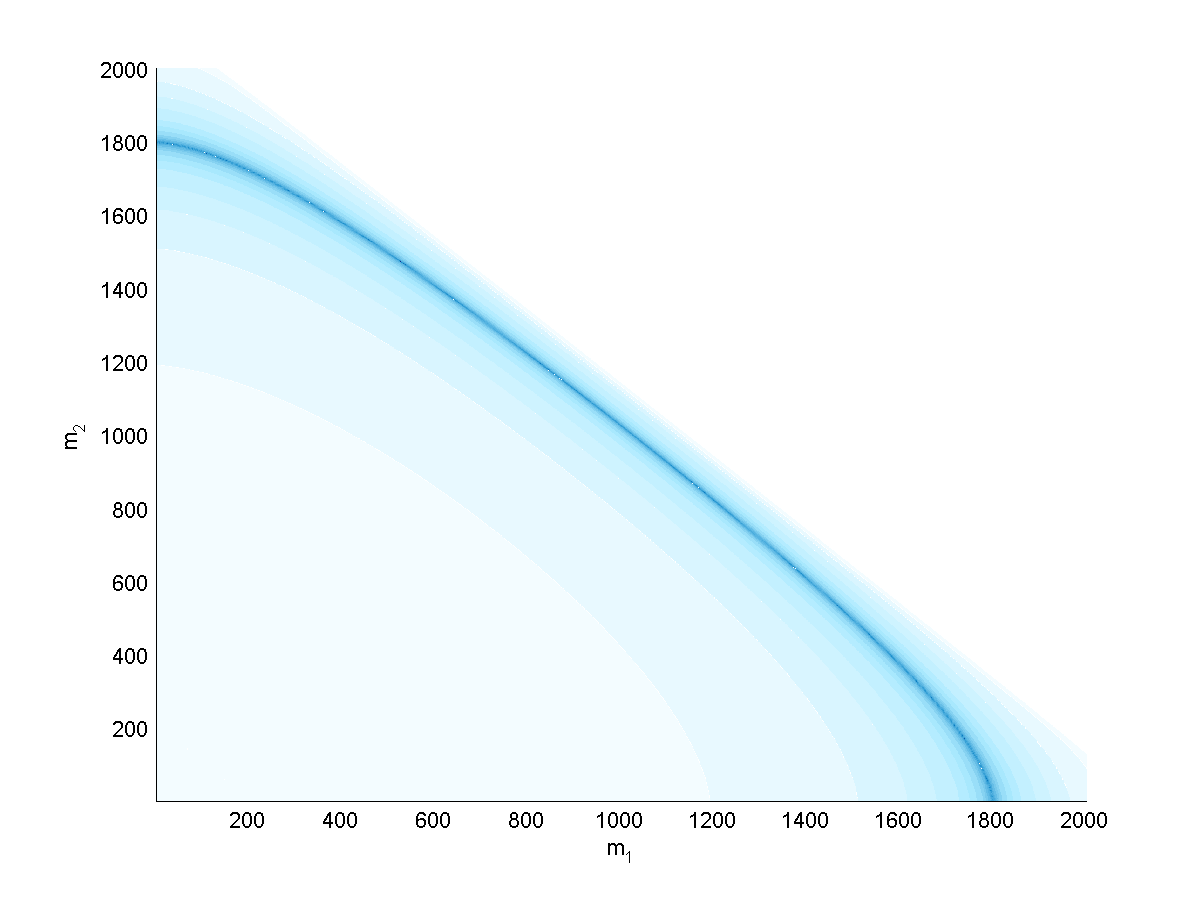} \\	
						\includegraphics[natwidth=1200bp, natheight=900bp, width=.48\textwidth]{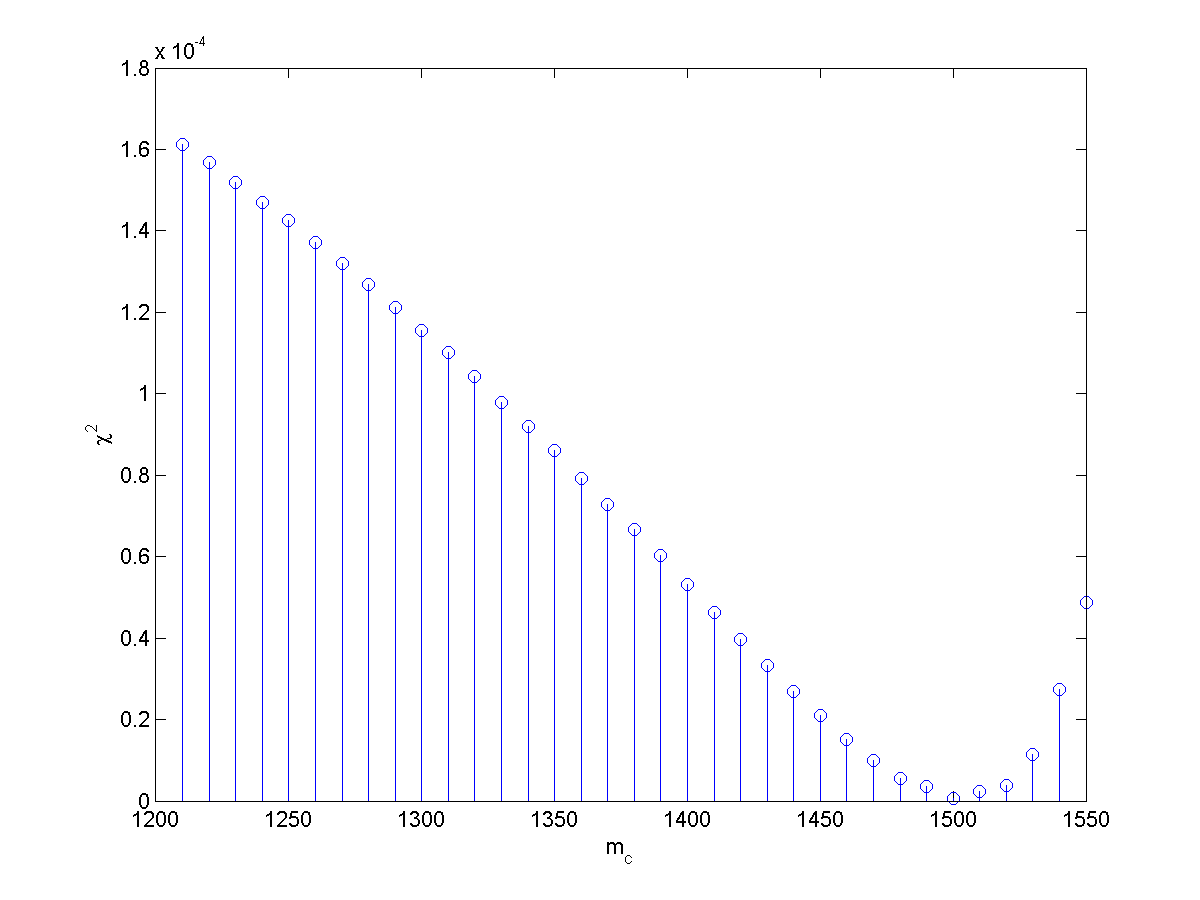} \hfill
						\includegraphics[natwidth=1200bp, natheight=900bp, width=.48\textwidth]{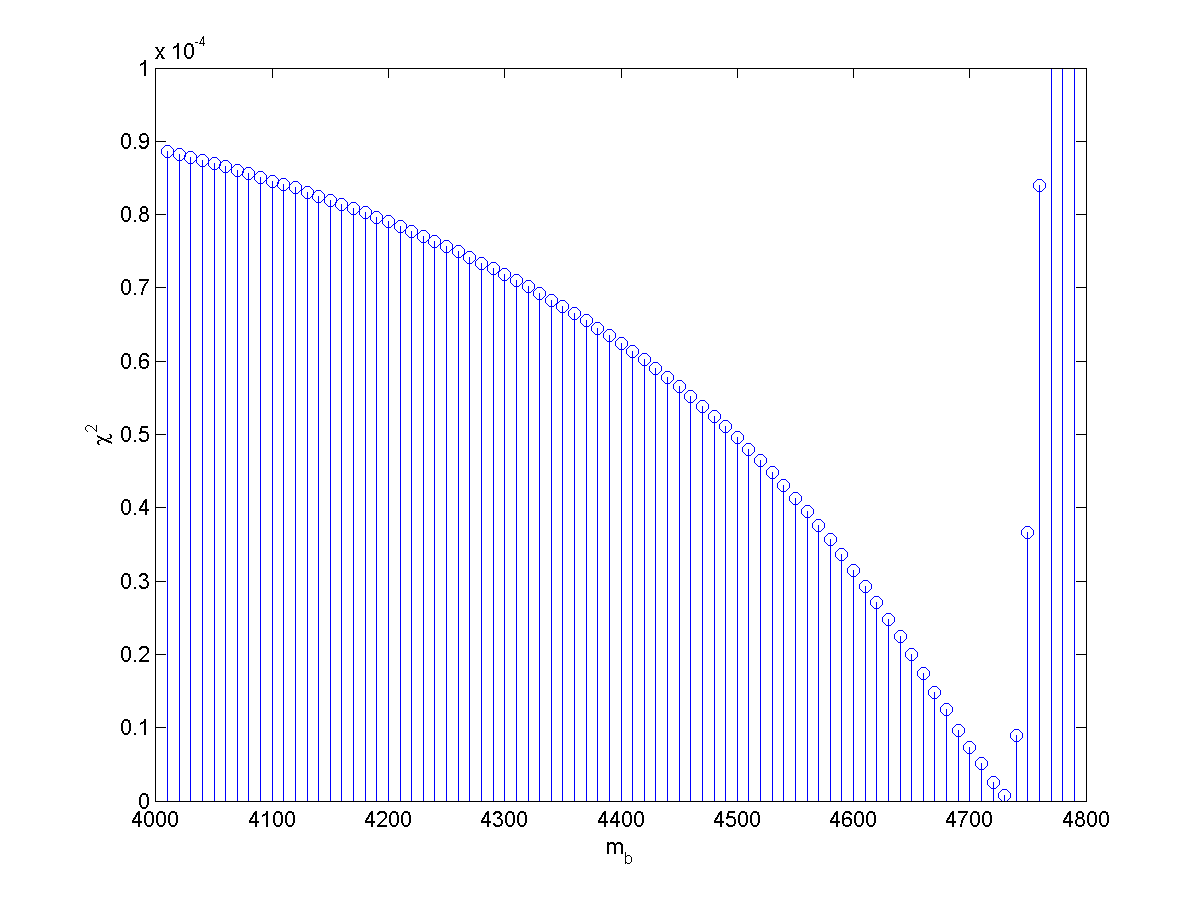}
						\caption{\label{fig:chi_j_heavy} Top: \(\chi^2\) as a function of two masses for the \(D\) (left) and \(D^*_s\) (right) trajectories. The coloring is based on a logarithmic scale, with the entire colored area having \(\chi^2_m/\chi^2_l < 1\). The minimum is \rchi{5\ten{-4}} for the \(D\), and \rchi{2\ten{-6}} for the \(D^*_s\). On the bottom are \(\chi^2\) as a function of \(m\) for the \((J,M^2)\) trajectory of the \(\Psi\) (left) and the same for the \(\Upsilon\) (right). In all plots, \(a\) and \(\alp\) are optimized for each choice of the endpoint masses.}
				\end{figure}
				
			There are three trajectories we analyze involving a charm quark. The first is of the \(D\), comprised of a light quark and a \(c\) quark, the second is the \(D^*_s\) with a \(c\) and an \(s\), and the third is \(\ccb\) - the \(\Psi\). All trajectories have only three data points.
			
			For the \(D\) meson, the optimal fit has \(m_c = 1640\), \(\mud = 80\) and \(\alp = 1.07\). In this case, unlike the result for the \(K^*\) trajectory, there is a preference for an imbalanced choice of the masses, although with four fitting parameters and three data points we can't claim this with certainty. The fit for the \(D^*_s\) has a good fit consistent with the previous \(s\) and \(c\) fits at \(m_c = 1580\), \(m_s = 400\), and \(\alp = 1.09\). The plots of \(\chi^2\) vs. the two masses (\(m_c\) and \(\mud\)/\(m_s\)) can be seen in figure (\ref{fig:chi_j_heavy}).
			
			In the same figure, we have \(\chi^2\) as a function of the single mass \(m_c\) for the \(\ccb\) \(\Psi\) trajectory. The minimum there is obtained at \(m_c = 1500\) MeV, where the slope is \(\alp = 0.98\) GeV\(^{-2}\).
			
			It is worth noting that while the linear fit results in values for \(\alp\) that are very far from the one obtained for the \(u\), \(d\), and \(s\) quark trajectories - \(0.42\), \(0.48\), and \(0.52\) for the \(\Psi\), \(D\), and \(D^*\) respectively - the massive fits point to a slope that is very similar to the one obtained for the previous trajectories.	This is also true, to a lesser extent, of the values of the intercept \(a\).
		
	\subsubsection{\texorpdfstring{$\bbb$}{b-bbar} mesons}
		The last of the \((J,M^2)\) trajectories is that of the \(\bbb\) \(\Upsilon\) meson, again a trajectory with only three data points.
		The fits point to an optimal value of \(m_b = 4730\), exactly half the mass of the lowest particle in the trajectory. The slope is significantly lower than that obtained for other mesons, \(\alp = 0.64\) at the optimum. The bottom plot in figure (\ref{fig:chi_j_heavy}) shows \(\chi^2\) for the \(\bbb\) trajectory.
		
		\begin{figure}[t!] \centering
					\includegraphics[natwidth=1200bp, natheight=900bp, width=.90\textwidth]{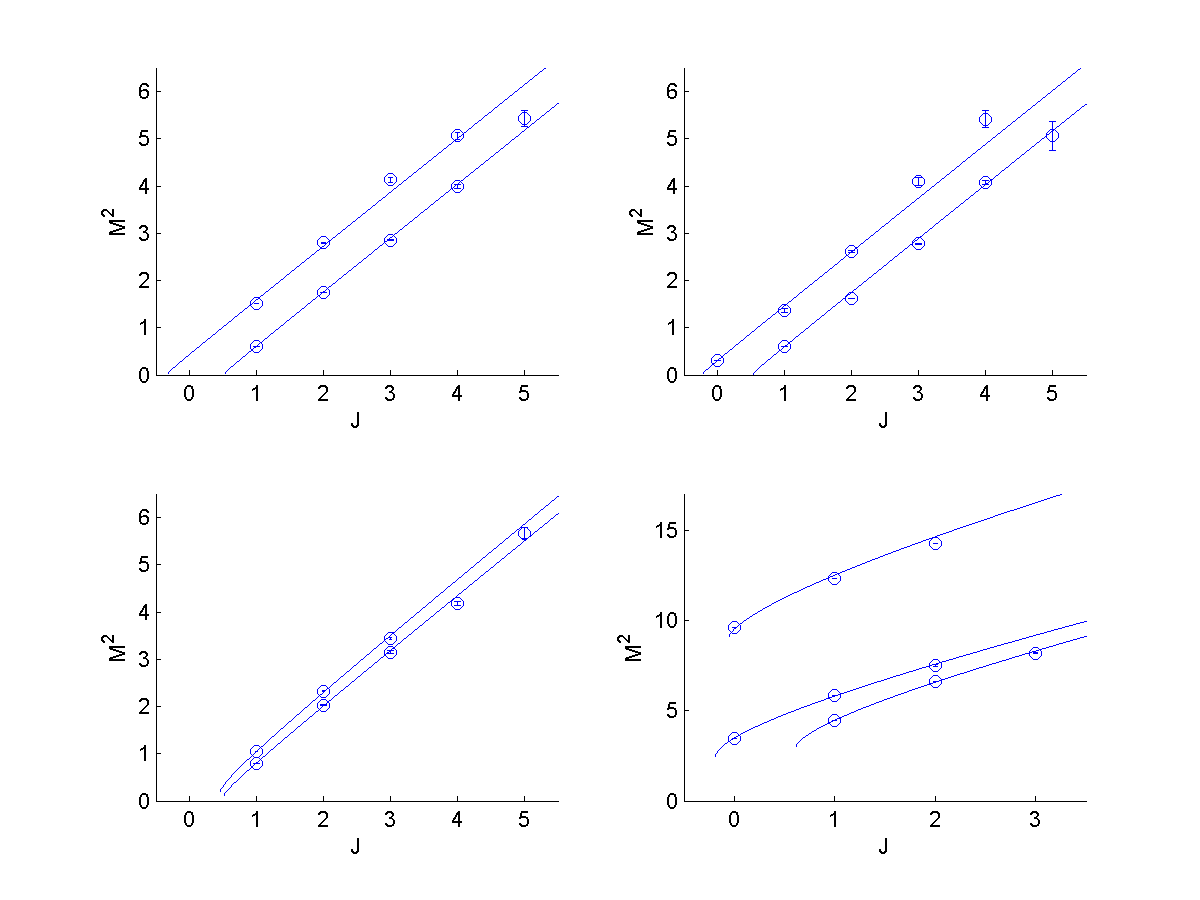}
					\caption{\label{fig:multiFit_j} Nine \((J,M^2)\) trajectories fitted using universal quark masses and slope (\(\mud = 60\), \(m_s = 220\), \(m_c = 1500\), and \(\alp = 0.884\)). Top left: \(\pi\) and \(\rho\), top right: \(\eta\) and \(\omega\), bottom left: \(K^*\) and \(\phi\), bottom right: \(D\), \(D^*_s\), and \(\Psi\).}
				\end{figure}
		
		\subsubsection{Universal fit} \label{sec:universal}
				
				Based on the combined results of the individual fits for the \((J,M^2)\) trajectories of the \(u\), \(d\), \(s\), and \(c\) quark mesons, we assumed the values
				\be m_{u/d} = 60, m_s = 220, m_c = 1500 \ee
				for the endpoint masses and attempted to find a fit in which the slope is the same for all trajectories. This wish to use a universal slope forces us to exclude the \(\bbb\) trajectory from this fit, but we can include the three trajectories involving a \(c\) quark. For these, with added endpoint masses (and only with added masses), the slope is very similar to that of the light quark trajectories.
								
				The only thing that was allowed to change between different trajectories was the intercept. With the values of the masses fixed, we searched for the value of \(\alp\) and the intercepts that would give the best overall fit to the nine trajectories of the \(\pi/b\), \(\rho/a\), \(\eta/h\), \(\omega/f\), \(K^*\), \(\phi\), \(D\), \(D^*_s\), and \(\Psi\) mesons. The best fit of this sort, with the masses fixed to the above values, was
				\be \alp = 0.884 \ee
				\[ a_\pi = -0.33 \qquad a_\rho = 0.52 \qquad a_\eta = -0.22 \qquad a_\omega = 0.53 \]
				\[ a_{K^*} = 0.50 \qquad a_\phi = 0.46 \qquad a_D = -0.19 \qquad a_{D^*_s} = -0.39 \qquad a_\Psi = -0.06\]
				and it is quite a good fit with \(\chi^2 = 13.13\ten{-4}\). The trajectories and their fits are shown in figure (\ref{fig:multiFit_j}). The values obtained for the masses vs. their experimental counterparts are in appendix \ref{app:universal}.
				
\subsection{Trajectories in the \texorpdfstring{$(n,M^2)$}{(n,M2)} plane}
	\subsubsection{Light quark mesons}
		In the light quark sector we fit the trajectories of the \(\pi\) and \(\pi_2\), the \(h_1\), the \(a_1\), and the \(\omega\) and \(\omega_3\).
		
		The \(h_1\) has a very good linear fit with \(\alp = 0.83\) GeV\(^{-2}\), that can be improved upon slightly by adding a mass of 100 MeV, with the whole range \(0-130\) MeV being nearly equal in \(\chi^2\).
		
		The \(a_1\) offers a similar picture, but with a higher \(\chi^2\) and a wider range of available masses. Masses between \(0\) and \(225\) are all nearly equivalent, with the slope rising with the added mass from \(0.78\) to \(0.80\) GeV\(^{-2}\).
		
		The \(\pi\) and \(\pi_2\) trajectories were fitted simultaneously, with a shared slope and mass between them and different intercepts. Again we have the range \(0\) to \(130\) MeV, \(\alp = 0.78-0.81\) GeV\(^{-2}\), with \(\mud = 100\) MeV being the optimum. The preference for the mass arises from non-linearities in the \(\pi\) trajectory, as the \(\pi_2\) when fitted alone results in the linear fit with \(\alp = 0.84\) GeV\(^{-2}\) being optimal.
		
		\begin{figure}[t!] \centering
						\includegraphics[natwidth=1200bp, natheight=900bp, width=.48\textwidth]{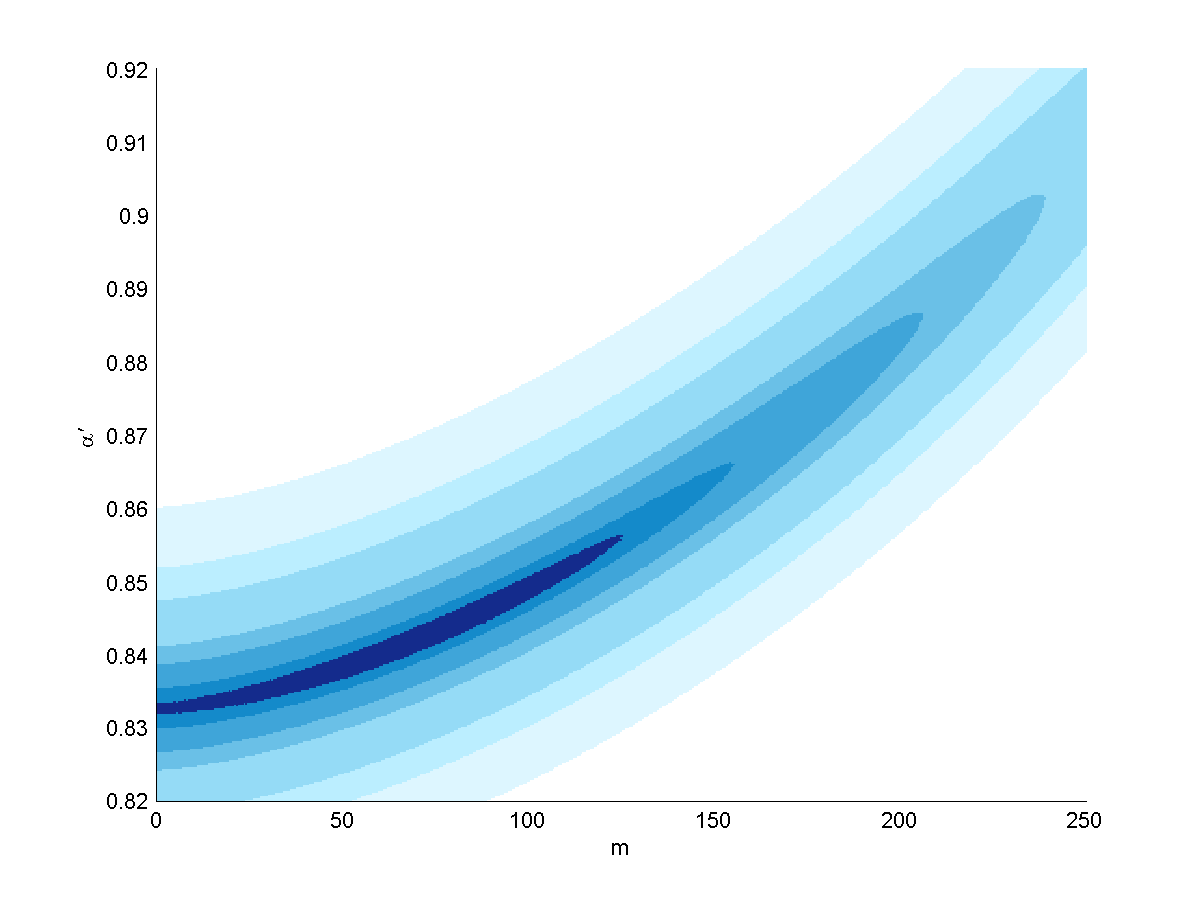}	 \hfill
						\includegraphics[natwidth=1200bp, natheight=900bp, width=.48\textwidth]{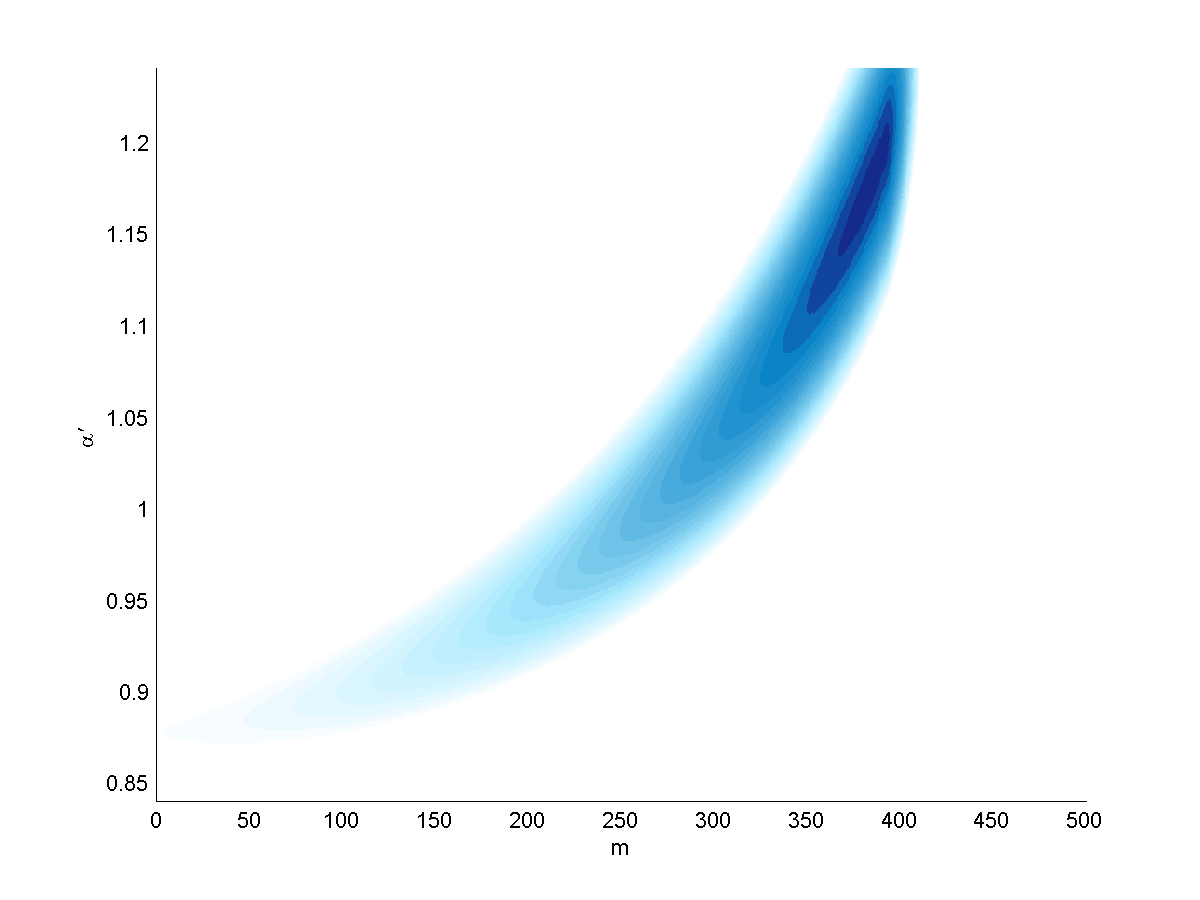}
						\caption{\label{fig:chi_n_h1_omg} \(\chi^2\) vs. \(\mud\) and \(\alp\) for the \(h_1\) (left) and \(\omega\) (\(J=1\) states alone, right) trajectories. For the \(h_1\) only the darkest area has \(\chi^2_m/\chi^2_l < 1\), while for the \(\omega\) the entire colored area offers better than linear fits and the minimum has \rchi{0.51}.}
				\end{figure}
		
		The \(\omega\) and \(\omega_3\) trajectories were also fitted simultaneously. Here again the higher spin trajectory alone resulted in an optimal linear fit, with \(\alp = 0.86\) GeV\(^{-2}\). The two fitted simultaneously are best fitted with a high mass, \(\mud = 340\), and high slope, \(\alp = 1.09\) GeV\(^{-2}\). Excluding the ground state \(\omega(782)\) from the fits eliminates the need for a mass and the linear fit with \(\alp = 0.97\) GeV\(^{-2}\) is then optimal. The mass of the ground state from the resulting fit is \(950\) MeV. This is odd, since we have no reason to expect the \(\omega(782)\) to have an abnormally low mass, especially since it fits in perfectly with its trajectory in the \((J,M^2)\) plane.
		
		The fit for the \(J^{PC} = 1^{--}\) \(\omega\) with the ground state included is shown in figure (\ref{fig:chi_n_h1_omg}), along with the fit for the \(h_1\), which has \(J^{PC} = 1^{+-}\).
							
	\subsubsection{\texorpdfstring{$\ssb$}{s-sbar} mesons}
		For the \(\ssb\) we have only one trajectory of three states, that of the \(\phi\). There are two ways to use these states. The first is to assign them the values \(n = 0,1,2\). Then, the linear fit with the slope \(\alp = 0.54\) GeV\(^{-2}\) is optimal.
		
		\begin{figure}[t!] \centering
						\includegraphics[natwidth=1200bp, natheight=900bp, width=.48\textwidth]{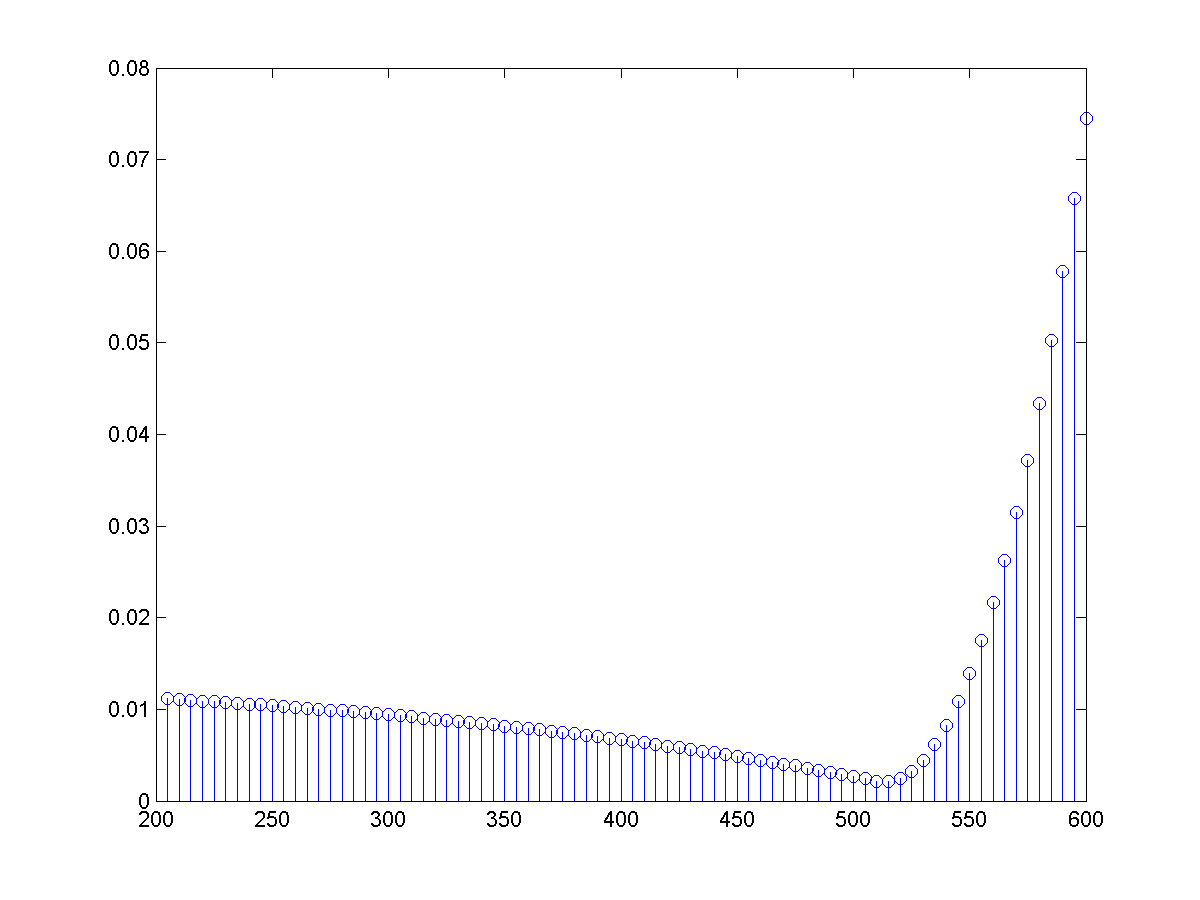}	 \hfill
						\includegraphics[natwidth=1200bp, natheight=900bp, width=.48\textwidth]{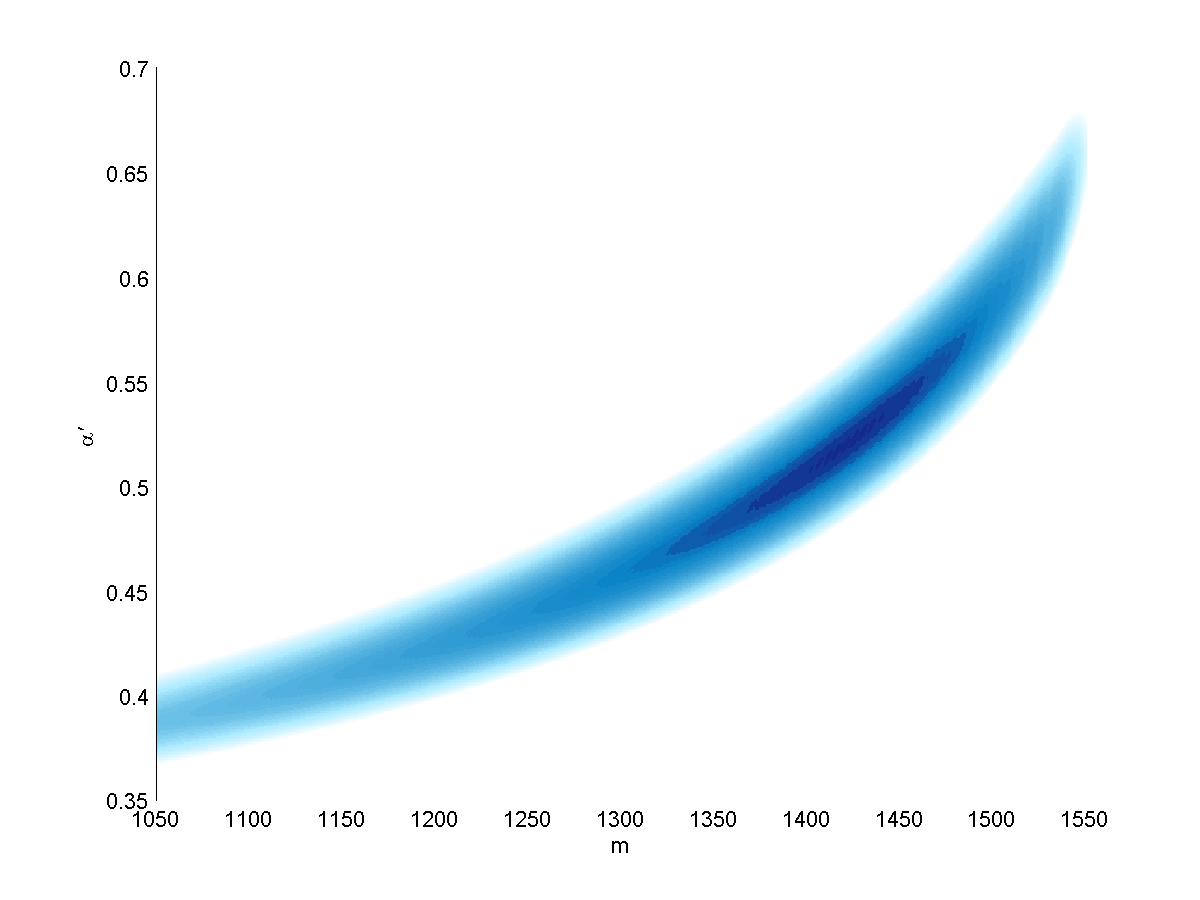}
						\caption{\label{fig:chi_n_sc} Left: \(\chi^2\) vs. \(m_s\) for the radial trajectory of the \(\ssb\) \(\phi\), with optimum at \(m_s = 515\). Right: Radial trajectory of the \(\ccb\) \(\Psi\) meson, \(\chi^2\) vs. \(\alp\) and \(m_c\).}
				\end{figure}
		
		Since this result in inconsistent both in terms of the low value of the slope, and the absence of a mass for the strange quark, we tried a different assignment. We assumed the values \(n = 0,1,\) and \(3\) for the highest state and obtained the values \(\alp = 1.10, m_s = 515\) for the optimal fit. These are much closer to the values obtained in previous fits.
		
		The missing \(n = 2\) state is predicted to have a mass of around \(1960\) MeV. Interestingly, there is a state with all the appropriate quantum numbers at exactly that mass - the \(\omega(1960)\), and that state lies somewhat below the line formed by the linear fit to the radial trajectory of the \(\omega\). Even if the \(\omega(1960)\) is not the missing \(\ssb\) (or predominantly \(\ssb\)) state itself, this could indicate the presence of a \(\phi\) state near that mass.
		
	\subsubsection{\texorpdfstring{$\ccb$}{c-cbar} mesons}
		Here we have the radial trajectory of the \(J/\Psi\), consisting of four states.
		
		The massive fits now point to the range \(1350-1475\) MeV for the \(c\) quark mass.	The biggest difference between the fits obtained here and the fits obtained before, in the \((J,M^2)\) plane is not in the mass, but in the slope, which now is in the range \(0.48-0.56\) GeV\(^{-2}\), around half the value obtained in the angular momentum trajectories involving a \(c\) quark - \(0.9-1.1\).
		
		It is also considerably lower than the slopes obtained in the \((n,M^2)\) trajectories of the light quark mesons, which would make it difficult to repeat the achievement of having a fit with a universal slope in the \((n,M^2)\) plane like the one we had in the \((J,M^2)\) plane.
	
	\subsubsection{\texorpdfstring{$\bbb$}{b-bbar} mesons}
	There are two trajectories we use for the \(\bbb\) mesons.
	\begin{figure}[t!] \centering
						\includegraphics[natwidth=1200bp, natheight=900bp, width=.48\textwidth]{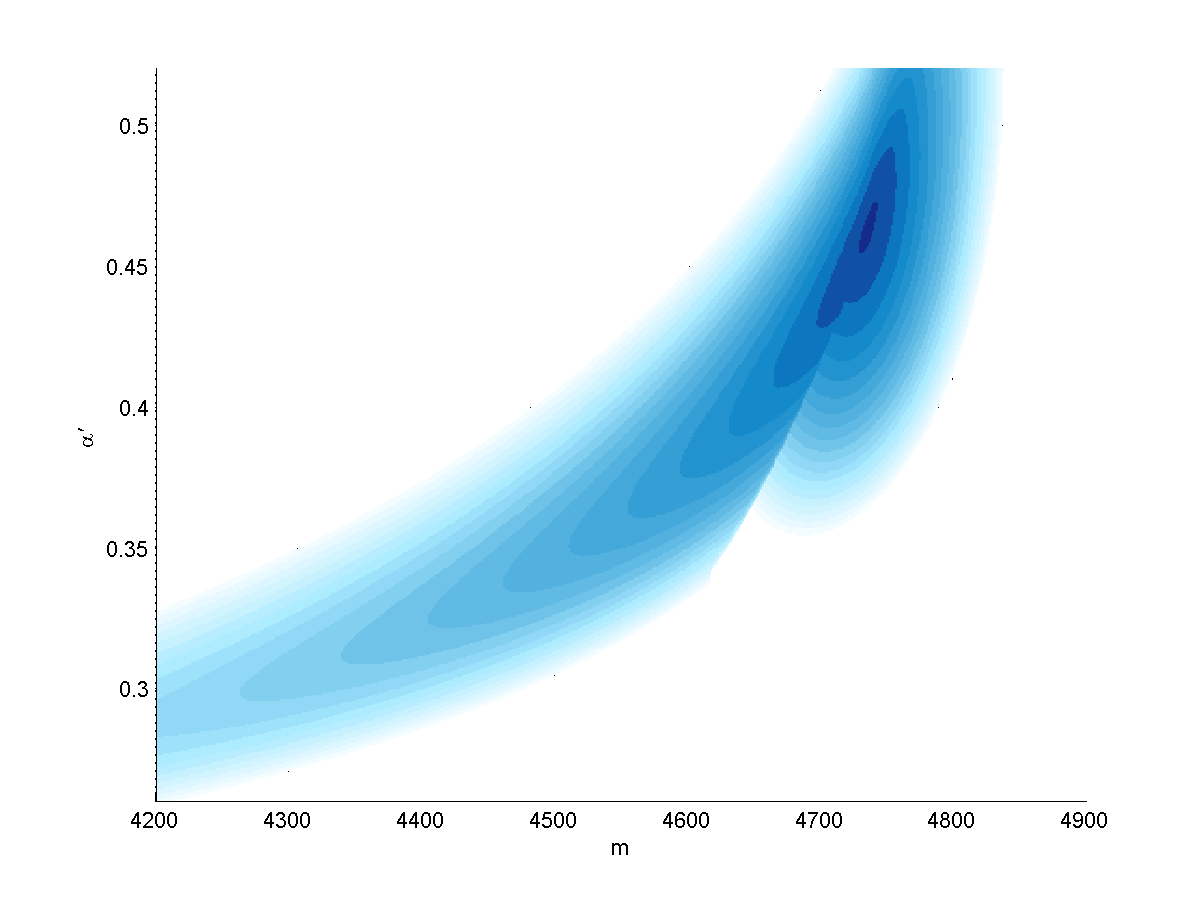}	 \hfill
						\includegraphics[natwidth=1200bp, natheight=900bp, width=.48\textwidth]{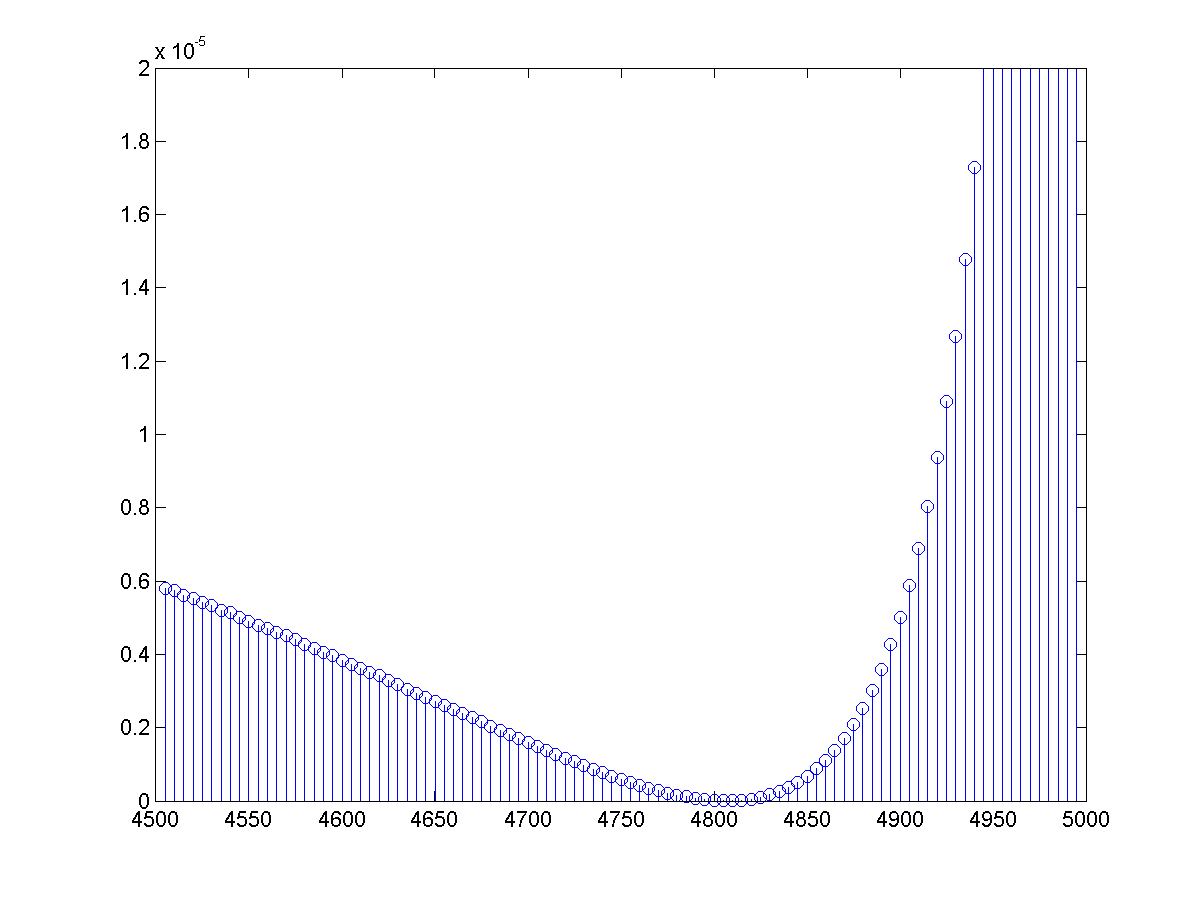}
						\caption{\label{fig:chi_n_bb} Left: \(\chi^2\) as a function of \(\alp\) and \(m_b\) for the \(\Upsilon\) radial trajectory. The discontinuity in the plot arises from the condition that the intercept \(a \leq 1\), otherwise the mass of the ground state is undefined. The two areas in the plot are then where \(a\) is still allowed to change (left) and where \(a\) is blocked from increasing further and is fixed at \(a = 1\) (oval shape on the right). Right: \(\chi^2\) as a function of \(m_b\) for the \(\chi_b\) trajectory.}
				\end{figure}
	
	The first is that of the \(\Upsilon\) meson, with six states in total, all with \(J^{PC} = 1^{--}\). For this trajectory we have an excellent fit with \(m_b = 4730\) MeV and the slope \(\alp = 0.46\) GeV\(^{-2}\). It is notable for having a relatively large number of states and still pointing clearly to a single value for the mass.
	
	The second \(\bbb\) trajectory is that of the \(\chi_b\) - \(J^{PC} = 1^{++}\). Here we have only three states and the best fit has a slightly higher mass for the \(b\) quark - \(m_b = 4800\) MeV - and a higher value for the slope \(\alp = 0.50\) GeV\(^{-2}\).

\subsection{WKB fits}

\begin{figure}[t!] \centering
						\includegraphics[natwidth=1200bp, natheight=900bp, width=.48\textwidth]{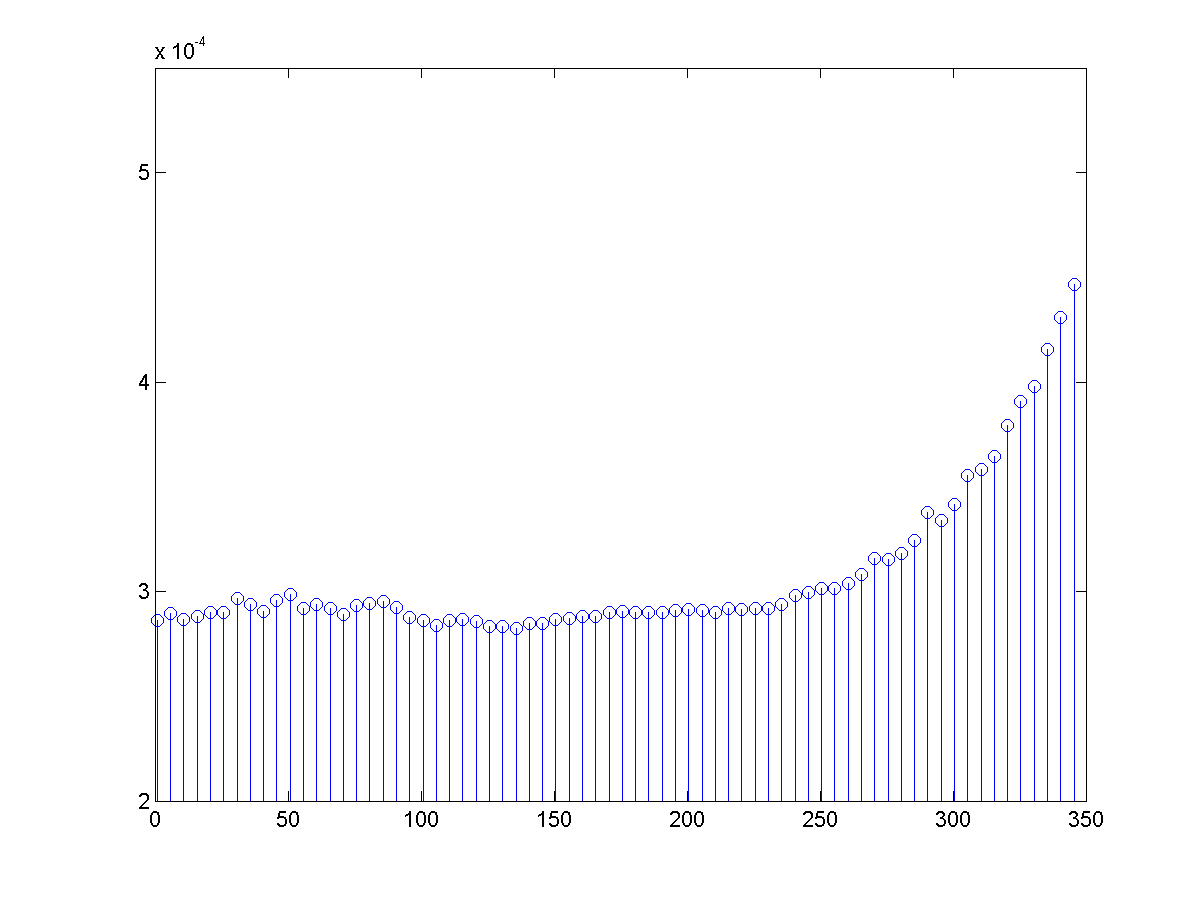}	 \hfill
						\includegraphics[natwidth=1200bp, natheight=900bp, width=.48\textwidth]{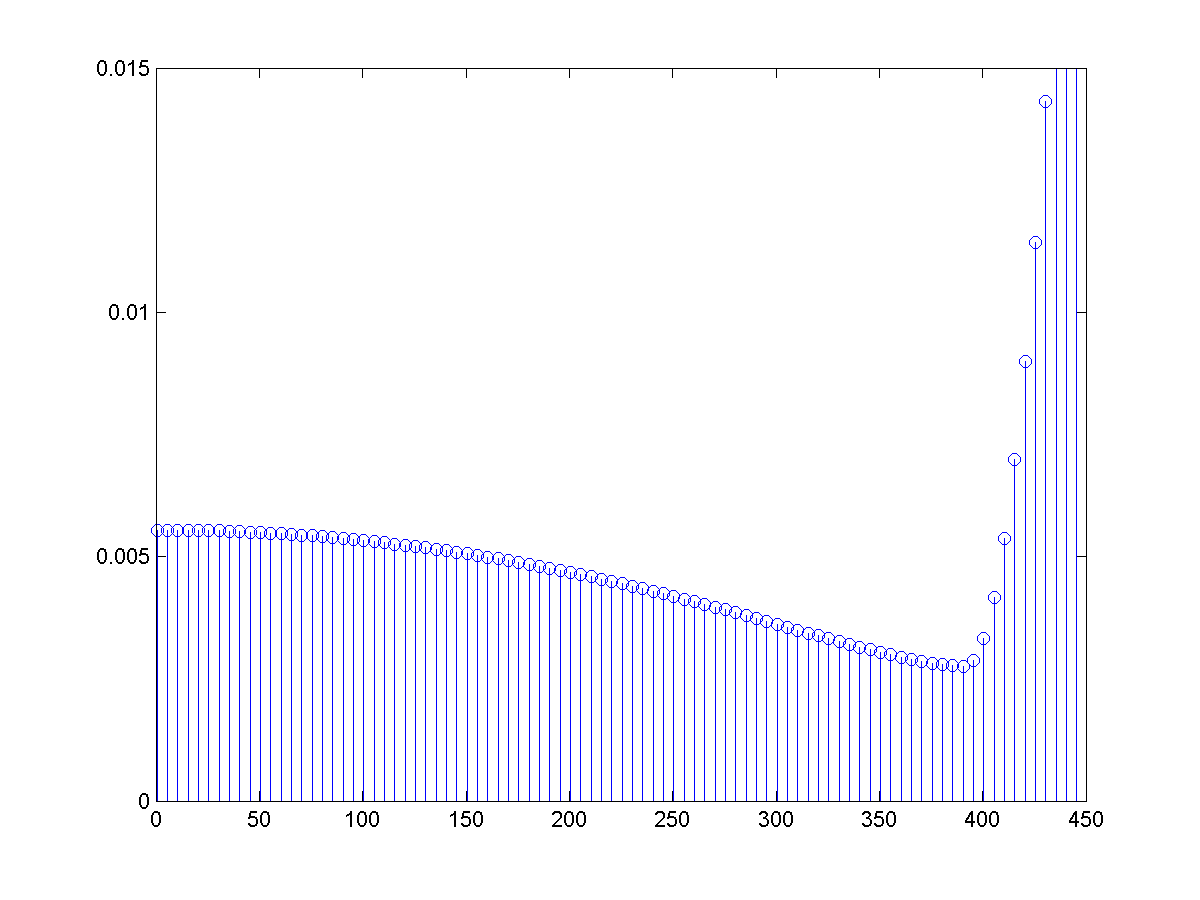}
						\caption{\label{fig:chi_w_ud} \(\chi^2\) as a function of \(\mud\) for the \(h_1\) trajectory (left) and for the \(\omega\) trajectory (right). \(\alp\) and \(a\) are always optimized.}
				\end{figure}
				
The WKB fits are all done in the \((n,M^2)\) plane. The biggest difference between the WKB fits in and the fits done using the expressions obtained from the classically rotating string is the way in which the angular momentum is included. In eq. \eqref{eq:massFitJ}, which was used for all the previous fits, we ultimately have a functional dependence of the form
	\be n + J - a = f(E;m,\alp) \ee
The contribution from the angular momentum, when fitting trajectories in the \((n,M^2)\) plane, amounts to nothing more than a shift of the \(n\) axis, and can be fully absorbed into the intercept \(a\). Eq. \eqref{eq:fitWKB}, on the other hand, carries out the contribution from the angular momentum in a different way.
The following fits are done assuming the angular momentum carried by the quarks, \(J_q\) in the notation of eq. \eqref{eq:fitWKB}, is the orbital angular momentum \(L\).

\begin{figure}[t!] \centering
						\includegraphics[natwidth=1200bp, natheight=900bp, width=.48\textwidth]{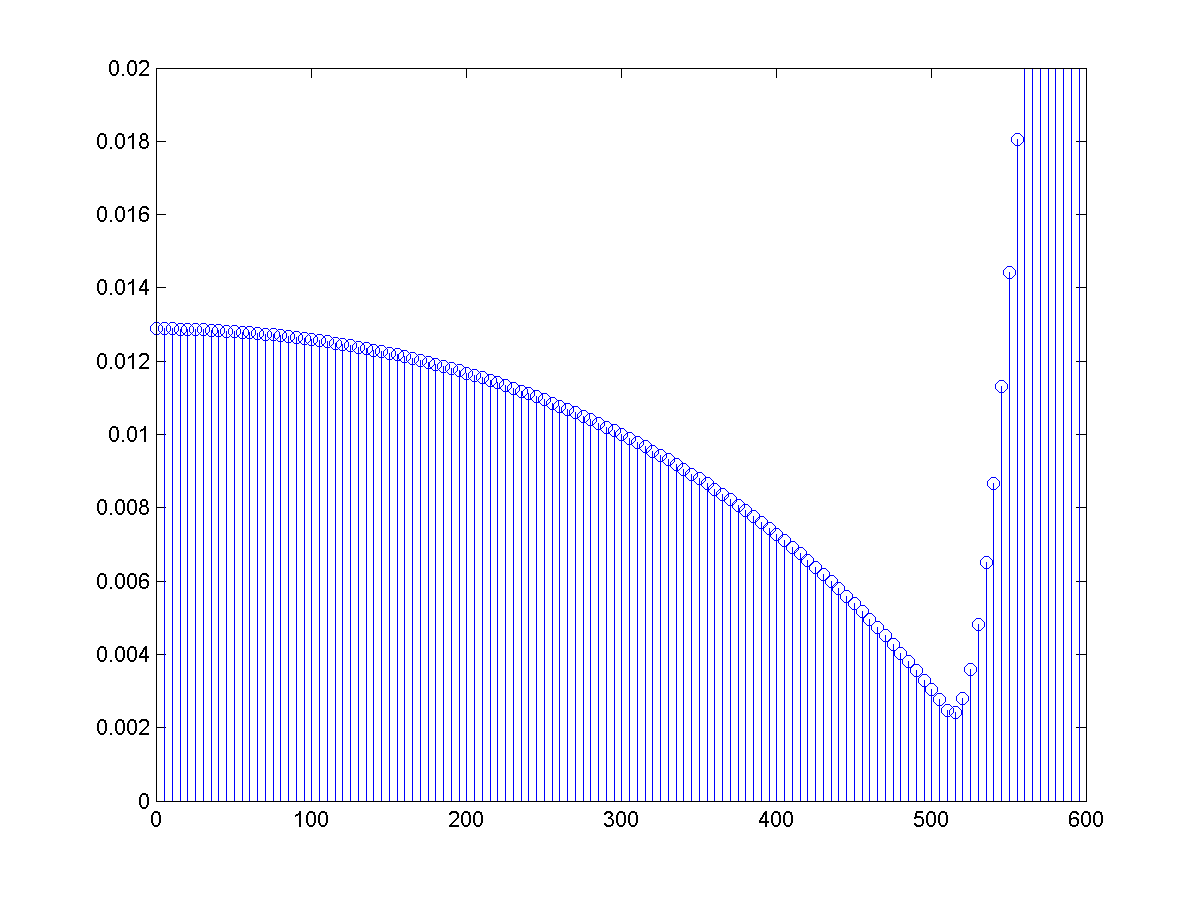}	 \hfill
						\includegraphics[natwidth=1200bp, natheight=900bp, width=.48\textwidth]{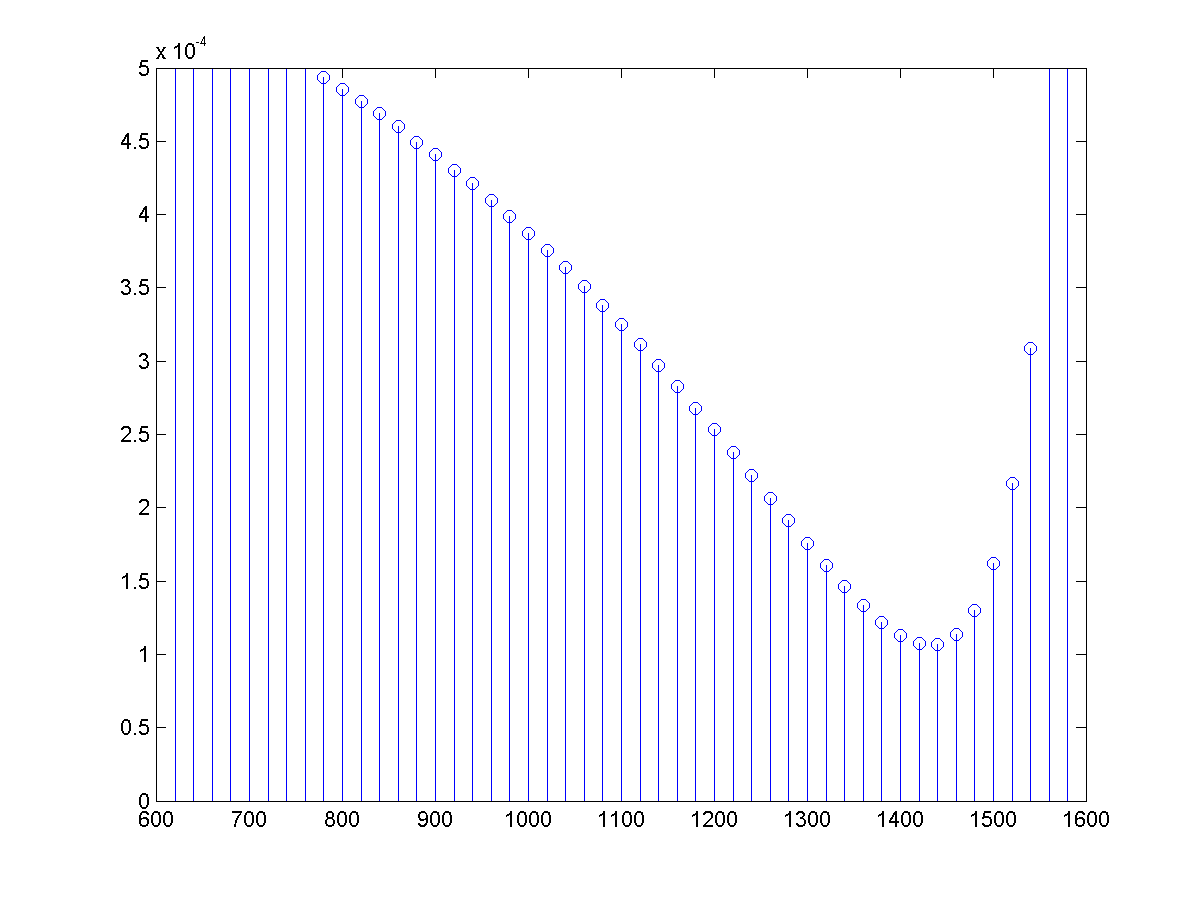} \\
						\includegraphics[natwidth=1200bp, natheight=900bp, width=.48\textwidth]{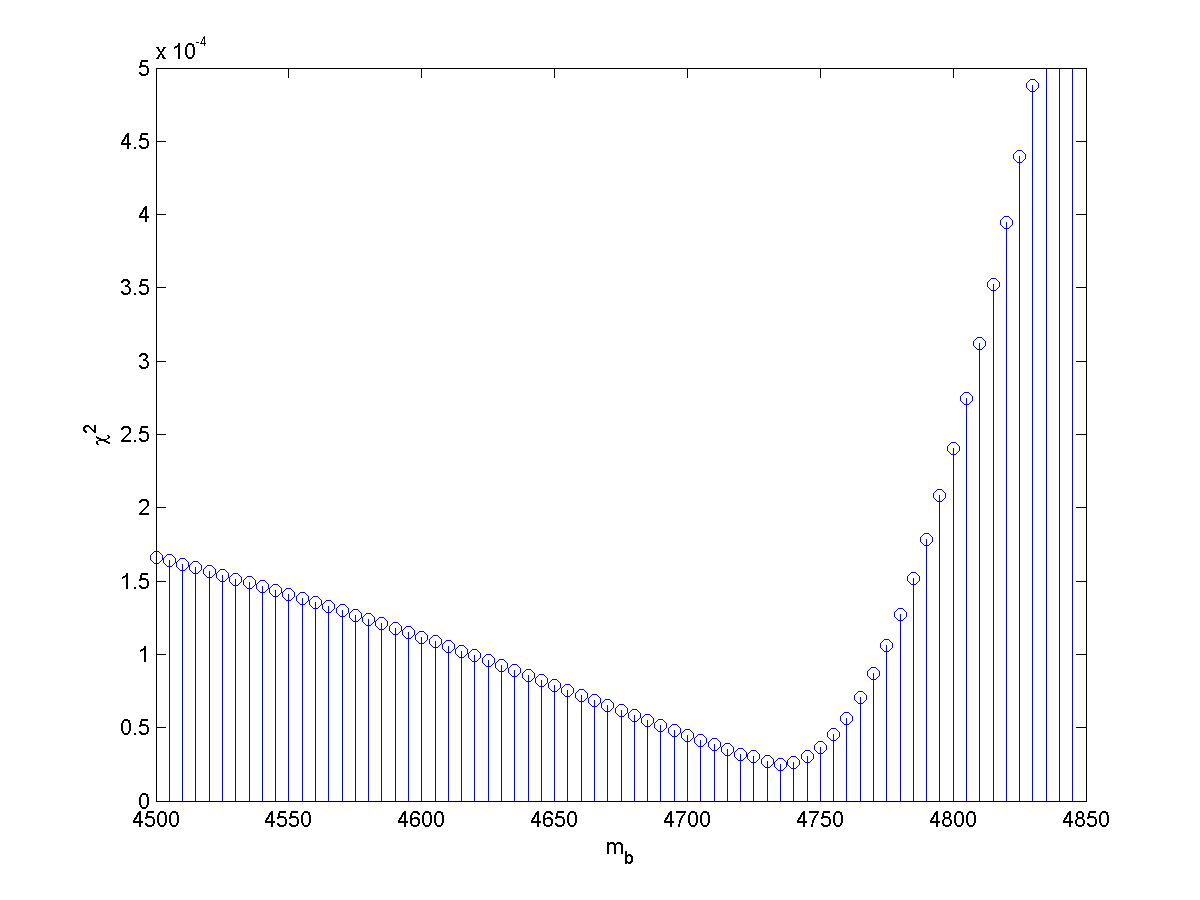}	 \hfill
						\includegraphics[natwidth=1200bp, natheight=900bp, width=.48\textwidth]{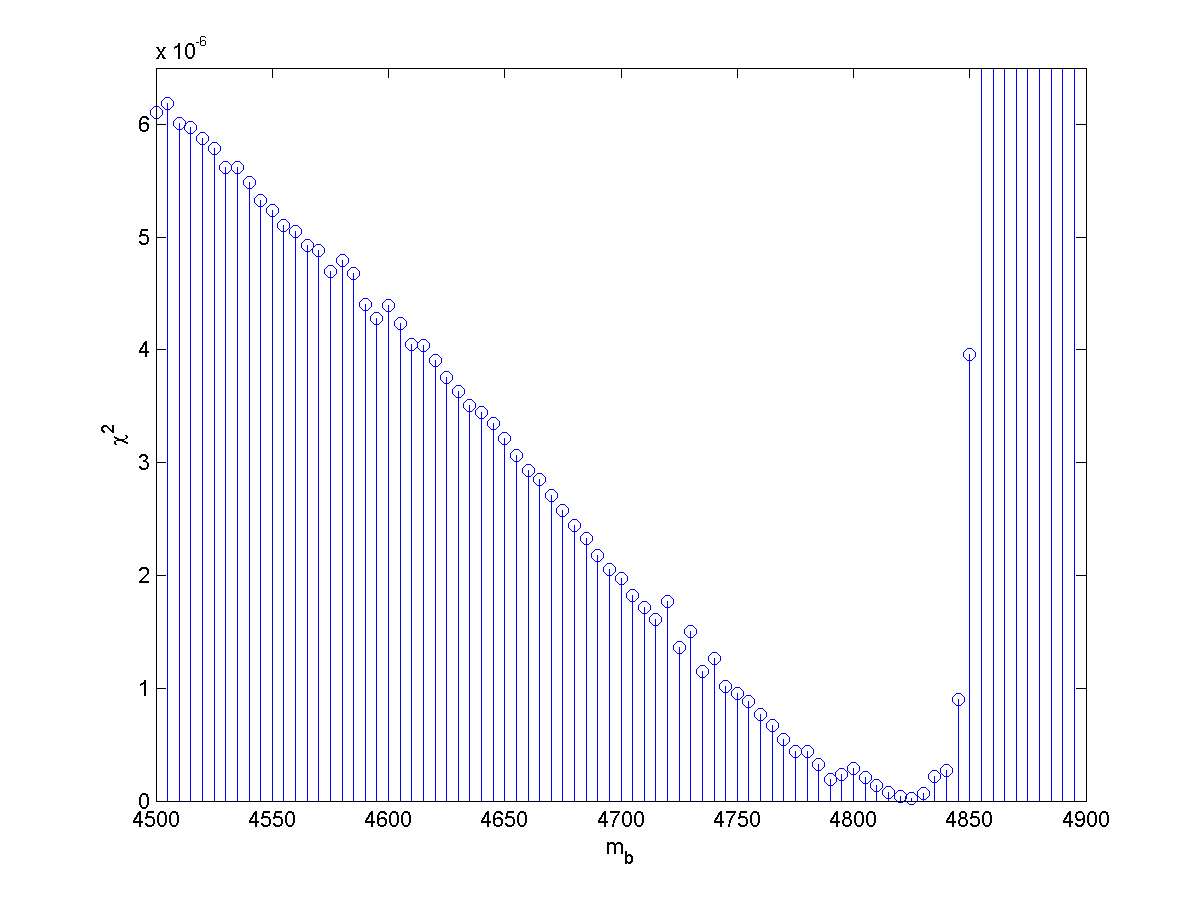}
						\caption{\label{fig:chi_w_scb} \(\chi^2\) as a function of \(m_s\) for the \(\phi\) radial trajectory (top left), as a function of \(m_c\) for the \(\Psi\) trajectory (top right), and as a function of \(m_b\) for the \(\Upsilon\) (bottom left) and \(\chi_b\) (bottom right) trajectories. The values obtained are \(m_s = 515\) MeV and \(m_c = 1500\) MeV. \(m_b = 4735\) or \(4825\) MeV. \(\alp\) and \(a\) are always optimized.}
				\end{figure}
Another point of difference between the two fits is in the values of the slope, which tend to be lower in the WKB fits. For the heavy quark trajectories we can understand this by comparing the heavy mass expansions in eqs. \eqref{eq:highMass} and \eqref{eq:highMassW}, and the ratio between the massive fit slopes and the WKB slopes is usually close to the ratio between the leading term coefficients of each of the two expansions (\(2\sqrt{3}/\pi \approx 1.1\)).
The WKB fits generally allow for higher masses for the light quarks, as can be seen in figure (\ref{fig:chi_w_ud}). For the \(\pi/\pi_2\) trajectories we actually obtain a minimum around \(\mud = 230\) MeV, where before it was less than half that value. The \(h_1\) trajectory now has an optimum at a mass of \(100-150\) MeV, with masses lower than \(100\) MeV now excluded. The \(\omega/\omega_3\) trajectory again has an optimum at the high mass of \(350\) MeV, and the \(a_1\) trajectory now has an even wider range of nearly equivalent mass than before, \(\mud = 0-250\) MeV.
				
For the heavier quark trajectories we obtain the same masses as before. The fits for the \(\ssb\) trajectory of the \(\phi\) result in a mass of \(515\) MeV for the \(s\) quark. The \(\Psi\) trajectory narrows down somewhat the mass of the \(c\) quark to the range \(m_c = 1390-1460\) MeV. The bottomonium trajectories of the \(\Upsilon\) and \(\chi_b\)	 indicate the value of the \(b\) quark mass to be \(4735\) or \(4825\) MeV, respectively. The values of \(\chi^2\) as a function of the mass for these four trajectories can be seen in figure (\ref{fig:chi_w_scb}).
		
\subsection{Summary of results for the mesons}

Table (\ref{tab:mes_j}) summarizes the results of the fits for the mesons in the \((J,M^2)\) plane. Tables (\ref{tab:mes_n}) and (\ref{tab:mes_w}) likewise summarize the results of the two types of fits for the \((n,M^2)\) trajectories, that of the rotating string and of the WKB approximation.
The higher values of \(\alp\) and \(a\) always correspond to higher values of the endpoint masses, and the ranges listed are those where \(\chi^2\) is within 10\% of its optimal value.
			\begin{table}[tp!] \centering
					\begin{tabular}{|c|c|cc|c|c|c|} \hline
					Traj. & \(N\) & \multicolumn{2}{|c|}{\(m\)} & \alp & \(a\) \\ \hline
					
					\(\pi/b\) & \(4\) & \multicolumn{2}{|c|}{\(\mud = 90-185\)} & \(0.808-0.863\) & \((-0.23)-0.00\) \\
					
					\(\rho/a\) & \(6\) & \multicolumn{2}{|c|}{\(\mud =0-180\)} & \(0.883-0.933\) & \(0.47-0.66\) \\
				
					\(\eta/h\) &  \(5\) & \multicolumn{2}{|c|}{\(\mud = 0-70\)} & \(0.839-0.854\) & \((-0.25)-(-0.21)\) \\
					
					\(\omega\) &  \(6\) & \multicolumn{2}{|c|}{\(\mud = 0-60\)}& \(0.910-0.918\) & \(0.45-0.50\) \\
					
					\(K^*\) &  \(5\) & \(\mud = 0-240\) & \(m_s = 0-390\) & \(0.848-0.927\) & \(0.32-0.62\) \\
					
					\(\phi\) &  \(3\) & \multicolumn{2}{|c|}{\(m_s = 400\)} & \(1.078\) & \(0.82\) \\
					
					\(D\) &  \(3\) & \(\mud = 80\) & \(m_c = 1640\) & \(1.073\) & \(-0.07\) \\
					
					\(D^*_s\) & \(3\) & \(m_s = 400\) & \(m_c = 1580\) & \(1.093\) & \(0.89\) \\
					
					\(\Psi\) &  \(3\) & \multicolumn{2}{|c|}{\(m_c = 1500\)} & \(0.979\) & \(-0.09\) \\
					
					\(\Upsilon\) &  \(3\) & \multicolumn{2}{|c|}{\(m_b = 4730\)} & \(0.635\) & \(1.00\) \\ \hline \end{tabular}
					
					\caption{\label{tab:mes_j} The results of the meson fits in the \((J,M^2)\) plane. For the uneven \(K^*\) fit the higher values of \(m_s\) require \(\mud\) to take a correspondingly low value. \(\mud + m_s\) never exceeds 480 MeV, and the highest masses quoted for the \(s\) are obtained when \(\mud = 0\). The ranges listed are those where \(\chi^2\) is within 10\% of its optimal value. \(N\) is the number of data points in the trajectory.}
				\end{table}
				
				\begin{table}[tp!] \centering
					\begin{tabular}{|c|c|c|c|cc|} \hline
					Traj. & \(N\) & \(m\) & \alp & \multicolumn{2}{|c|}{\(a\)} \\ \hline
					
					\(\pi\)/\(\pi_2\) & \(4+3\) & {\(\mud = 110-250\)} & \(0.788-0.852\) & \(a_0 = (-0.22)-0.00\) & \(a_2 = (-0.00)-0.26\) \\
					
					\(a_1\) & \(4\) & \(\mud = 0-390\) & \(0.783-0.849\) & \multicolumn{2}{|c|}{\((-0.18)-0.21\)} \\
					
					\(h_1\) & \(4\) & \(\mud = 0-235\) & \(0.833-0.850\) & \multicolumn{2}{|c|}{\((-0.14)-(-0.02)\)} \\
					
					\(\omega/\omega_3\) & \(5+3\) & \(\mud = 255-390\) & \(0.988-1.18\) & \(a_1 = 0.81-1.00\) & \(a_3 = 0.95-1.15\) \\
					
					\(\phi\) & \(3\) & \(m_s = 510-520\) & \(1.072-1.112\) & \multicolumn{2}{|c|}{\(1.00\)} \\
					
					\(\Psi\) & \(4\) & \(m_c = 1380-1460\) & \(0.494-0.547\) & \multicolumn{2}{|c|}{\(0.71-0.88\)} \\
					
					\(\Upsilon\) & \(6\) & \(m_b = 4725-4740\) & \(0.455-0.471\) & \multicolumn{2}{|c|}{\(1.00\)} \\
					
					\(\chi_b\) & \(3\) & \(m_b = 4800\) & \(0.499\) & \multicolumn{2}{|c|}{\(0.58\)} \\	\hline \end{tabular}
					
					\caption{\label{tab:mes_n} The results of the meson fits in the \((n,M^2)\) plane. The ranges listed are those where \(\chi^2\) is within 10\% of its optimal value. \(N\) is the number of data points in the trajectory.}
				\end{table}
				
				\begin{table}[tp!] \centering
					\begin{tabular}{|c|c|c|c|cc|} \hline
					Traj. & \(N\) & \(m\) & \alp & \multicolumn{2}{|c|}{\(a\)} \\ \hline
					
					\(\pi\)/\(\pi_2\) & \(4+3\) & \(\mud = 0-250\) & \(0.770-0.801\) & \(a_0 = (-0.34)-0.00\) & \(a_2 = (-1.53)-(-1.20)\) \\
					
					\(a_1\) & \(4\) & \(\mud = 0-380\) & \(0.777-0.862\) & \multicolumn{2}{|c|}{\((-0.89)-(-0.20)\)} \\
					
					\(h_1\) & \(4\) & \(\mud = 0-265\) & \(0.827-0.876\) & \multicolumn{2}{|c|}{\((-0.85)-(-0.71)\)} \\
					
					\(\omega/\omega_3\) & \(5+3\) & \(\mud = 240-345\) & \(0.937-1.000\) & \(a_1 = (-0.23)-(-0.04)\) & \(a_3 = (-1.54)-(-1.28)\) \\
					
					\(\phi\) & \(3\) & \(m_s = 505-520 \) & \(1.005-1.045\) & \multicolumn{2}{|c|}{\(0.00\)} \\
					
					\(\Psi\) & \(4\) & \(m_c = 1390-1465\) & \(0.464-0.514\) & \multicolumn{2}{|c|}{\((-0.27)-(-0.10)\)} \\
					
					\(\Upsilon\) & \(6\) & \(m_b = 4730-4740\) & \(0.417-0.428\) & \multicolumn{2}{|c|}{\(0.00\)} \\
					
					\(\chi_b\) & \(3\) & \(m_b = 4820\) & \(0.468\) & \multicolumn{2}{|c|}{\(-0.08\)} \\ \hline \end{tabular}
					\caption{\label{tab:mes_w} The results of the meson WKB fits, all in the \((n,M^2)\) plane. The ranges listed are those where \(\chi^2\) is within 10\% of its optimal value. \(N\) is the number of data points in the trajectory.}
				\end{table} \clearpage
				
\subsection{L vs. J and the values of the intercept}
Table (\ref{tab:l_vs_j}) offers a comparison between the values of the intercept when fitting \(M^2\) to \((n + L)\) instead of to \((n + J)\). In other words, they are the values obtained when identifying the \(J\) on the left hand side of eq. \eqref{eq:massFitJ} with the orbital, as opposed to total, angular momentum. The advantage of this choice is that the results are made more uniform between the different trajectories when doing the fits to \((n+L)\). With the exception of the \(\chi_b\) trajectory, all the trajectories have negative intercepts between \((-0.55)\) and zero, with the intercept being closer to zero as the endpoint masses grow heavier.
				\begin{table}[tp!] \centering
					\begin{tabular}{|c|c||c|c|c|} \hline
						
						\multicolumn{2}{|c||}{\(L,M^2\) trajectories} & \multicolumn{3}{|c|}{\(n,M^2\) trajectories} 		 \\ \hline
						Traj.     	& \(a_m\) 								& Traj. 			 & \(a_m\) 						 & \(a_w\) 							 \\ \hline
						
						\(\pi/b\)		& \((-0.23)-0.00\)			& \(\pi\)			& \((-0.22)-0.00\)	& \((-0.34)-0.00\)			 \\
						
						\(\rho/a\)	& \((-0.53)-(-0.34)\)		&	\(\pi_2\)		& \((-0.00)-0.26\)	  & \((-1.53)-(-1.20)\)		\\
						
						\(\eta/h\)	& \((-0.25)-(-0.21)\)		&	\(h_1\)			& \((-0.14)-(-0.02)\) & \((-0.85)-(-0.71)\)		\\
						
						\(\omega/f\)& \((-0.55)-(-0.50)\)		&	\(\omega\)	& \((-0.19)-0.00\)    &	 \((-0.23)-(-0.04)\)		\\
						\(K^*\)			& \((-0.68)-(-0.38)\)		 &	\(\omega_3\)& \((-0.05)-0.15\)		&	 \((-1.54)-(-1.28)\)		\\
						
						\(\phi\)		& \((-0.18)\)	&	\(a_1\)			& \((-0.18)-0.21\)    & \((-0.89)-(-0.20)\)		 \\
					
						\(D\)				& \((-0.07)\)						&	\(\phi\)		 & \(0.00\)    				 & \(0.00\)							 \\
						
						\(D^*_s\)		&	\((-0.11)\) & \(\Psi\)		& \((-0.29)-(-0.12)\) & \((-0.27)-(-0.10)\) 	 \\
						
						\(\Psi\) 		& \((-0.09)\)						&	 \(\chi_b\)	& \(0.58\)						 & \(-0.08\)	    				 \\
						
						\(\Upsilon\)& \(0.00\)							&	\(\Upsilon\)& \(0.00\)  					 & \(0.00\)  						 \\
						
					\hline \end{tabular}
					\caption{\label{tab:l_vs_j} The ranges of the intercept from tables (\ref{tab:mes_j})-(\ref{tab:mes_w}) adjusted to fits to \(n+L\). The right-most column is for the WKB fits and the other two the regular massive fits.}
					\end{table}

\subsection{The length of the mesonic strings}
Lacking the basic string theory of QCD, one may revert to an effective low energy  theory on long strings\cite{Aharony:2013ipa}. The effective theory is expanded in powers of $\frac{\sqrt{\alp}}{l}$. In such a framework, the  semi-classical approximation describes the system more faithfully the longer the string is. To examine the issue of how long are the rotating strings with massive endpoints that describe the mesons we have computed the length of the strings associated with various mesons.
Using eqs. \eqref{eq:massFitE} for the energy and the relation \eqref{eq:boundaryTwo} between \(q_1\) and \(q_2\) we extract the two velocities given the total mass \(E\) and the two endpoint masses. Then, again by using eq. \eqref{eq:boundaryTwo} and \(q_i = \omega l_i\), we have
	\be l_i = \frac{m_i}{T}\frac{q_i^2}{1-q_i^2} \ee
	with the total string length between the two masses being \(l_1 + l_2\).
	
	In table (\ref{tab:lengths}) we present the values of \((l_1+l_2)/\sqrt{\alp}\) for the fitted \((J,M^2)\) trajectories.
	
	We can see that for the \(u\), \(d\), and \(s\) mesons the lengths are not too small, with the ratio \(l/\sqrt{\alp}\) starting from \(2-3\) for the low spin mesons and increasing as \(J\) increases to values for which the string can be called a long string more confidently. For the mesons involving \(c\) quarks the lowest spin states are short strings, but the higher \(L\) states (the maximum we have for those is \(L = 2\)) are getting to be long enough. For the \(\bbb\) meson, the lowest state's string length tends to zero, and the highest spin state used (again with \(L = 2\)) has \(l/\sqrt{\alp}\) of only about 2.

\begin{table}[t!] \centering
	\begin{tabular}{|c|c|c|c|c|c|c|c|} \hline
	Traj. & \(L = 0\) & \(L = 1\) & \(L = 2\) & \(L = 3\) & \(L = 4\) & \(L = 5\) \\ \hline
	
	\(\pi/b\) & - & 3.6 & 5.2 & 6.4 & 7.4 & - \\
	
	\(\rho/a\) & 2.3 & 4.4 & 5.8 & 7.0 & 8.0 & 8.9 \\
	
	\(\eta/h\) & 1.8 & 4.3 & 5.8 & 7.0 & 8.0 & - \\
	
	\(\omega/f\) & 2.8 & 4.8 & 6.2 & 7.4 & 8.4 & 9.2 \\
	
	\(K^*\) & 2.4 & 4.3 & 5.7 & 6.9 & 7.9 & - \\
	
	\(\phi\) & 1.0 & 3.1 & 4.5 & - & - & - \\
	
	\(D\) & 0.6 & 3.0 & 4.3 & - & - & - \\
	
	\(D^*_s\) & 0.6 & 2.6 & 3.8 & - & - & - \\
	\(\Psi\) & 0.4 & 2.1 & 3.2 & - & - & - \\
	
	\(\Upsilon\) & 0.0 & 1.5 & 2.3 & - & - & - \\ \hline
	\end{tabular} \caption{\label{tab:lengths} \(l_s/\sqrt{\alp}\) for all the states used in the \((J,M^2)\)  trajectories (arranged here by their orbital angular momentum \(L\)). For each trajectory, the length was calculated at the mass in the midpoint of the range given in summary table (\ref{tab:mes_j}), except for the \(K^*\), where we used \(\mud = 60\), \(m_s = 220\).}
\end{table}
\section{Summary}
We have seen that the Regge trajectories of mesons involving the \(s\), \(c\), and \(b\) quarks are generally best fitted when introducing endpoint masses to the relativistic string. These masses help account for the deviations one can observe from the linear Regge trajectories.
The masses of the heaviest quarks, the \(c\) and the \(b\), as obtained from the fits, are the values one would assign to them as constituent quarks: around \(1500\) MeV for the \(c\) quark, and \(4730\) for the \(b\). This means that in our model the mass of the lowest \(\ccb\) and \(\bbb\) states is due only to the quark masses.
For the \(s\) quark we have a different picture, where the mass is somewhere between the QCD mass of 100 MeV and the constituent mass of around 500 MeV. The results for the \(s\) quark vary from 200-300 MeV in the best fits for the strange (i.e. \(s\bar{u}\) or \(s\bar{d}\)) \(K^*\) meson to 400 MeV for the \(\ssb\) states of the \(\phi\) trajectory. The radial trajectory for the \(\ssb\) gives the mass at an even higher value of 500 MeV. It is not clear if this discrepancy can be attributed to an actual physical feature of the mesons, that will result in different end-point masses for different physical configurations. We know, though, that we would have liked not to see a discrepancy between the mass obtained from the \((J,M^2)\) fit and the one obtained from the \((n,M^2)\) fit - the ground state in these two trajectories is the same, and naturally we don't expect the same physical state to have two different endpoint masses.

A similar discrepancy was found between the charmed \(D\) meson and the \(\ccb\) trajectories, but in this case the situation is reversed: the mass obtained when fitting the states with a single \(c\) quark was higher (1600-1700 MeV) than the 1400-1500 MeV mass of the \(c\) quark in the \(\ccb\) states. We also have the charmed/strange meson \(D^*_s\), which points towards the higher masses for the \(s\) and \(c\) quarks, of around \(400\) MeV for the \(s\) and \(1600\) MeV for the \(c\), where decreasing one of the masses would then require the other to increase even more.

One should note, however, that of all the relevant trajectories, involving \(s\) or \(c\) quarks, only two have more than three available data points. Those trajectories with only three points tend to pinpoint the mass at a very specific value, with a small margin of error, and it is hard to estimate the realistic value of the error in such a measurement. When we do not assume two equal endpoint masses, increasing the number of fitting parameters from three to four, we have to contend ourselves with the optimum lying along a curve in the \(m_1,m_2\) plane, rather than an accurate determination of both masses. This is also true of the 5-point \(K^*\) trajectory, where we cannot determine both \(\mud\) and \(m_s\) because of the near-equivalence (as far as Regge trajectories are concerned) of configurations with \(m_1^{3/2}+m_2^{3/2} = Const.\) for low \(m_1\) and \(m_2\). What we can do is check for consistency between the trajectories consisting of mesons of different flavor quarks, both in the quark masses and in the Regge slope.

The light quark (\(u\) and \(d\)) trajectories are the most problematic in terms of the mass, even though the light meson sector is the richest in data. In most cases, no real optimally fitting mass was found. A typically found range would have all the masses between 0 and 200 MeV as nearly equivalent. In those cases where an optimum is easily discernible from our fits, it is the massless linear fit.

For the radial trajectories, the masses obtained for the tend to be higher, as they seem to be more prone to deviations from the linear trajectories. These deviations sometimes result in optimal fits for relatively high masses, but this could be due to our simple model misinterpreting the more complex physical phenomena that are behind those non-linearities.

The slopes that were found, on the other hand, are quite uniform for the light quark trajectories. The \((J,M^2)\) trajectories have a slope in the range \(0.80-0.90\) GeV\(^{-2}\), and this slightly decreases to \(0.78-0.84\) GeV\(^{-2}\) in the \((n,M^2)\) plane fits. The slope for the strange meson is also in this range, while for the \(\ssb\) states the optimum is found with a higher slope of around \(1.1\) GeV\(^{-2}\), in both planes. The charmed and \(\ccb\) mesons are best fitted in the \((J,M^2)\) plane with a similar value of approximately \(1\) GeV\(^{-2}\) for the slope - and only when adding the appropriate mass for the \(c\) quark. This is what allows the universal slope fit in section \ref{sec:universal}, which had an optimum for the slope \(\alp = 0.884\) GeV\(^{-2}\).

This uniformity of the slope is then broken. First, the \(\bbb\) trajectory was excluded from the universal \((J,M^2)\) fit because its optimal slope is much lower, at \(0.64\) GeV\(^{-2}\), and in the \((n,M^2)\) plane both the \(\ccb\) and \(\bbb\) trajectories have a slope of \(0.42-0.50\) GeV\(^{-2}\). For the \(\bbb\) the difference between the \((J,M^2)\) slope and the \((n,M^2)\) slope is not too large, or different from what we have seen for the \(u/d\) and \(s\) quark trajectories, but for the \(\ccb\) the slope is nearly halved when moving from the trajectory in the \((J,M^2)\) plane to the trajectory beginning with the same ground state in the \((n,M^2)\) plane.

We can then divide the trajectories into four main groups based on the approximate value of their best fitting slope. In the \((J,M^2)\) plane we have \(\alp \approx 0.9\) GeV\(^{-2}\) for the \(u\), \(d\), \(s\), and \(c\) quark trajectories, and \(\alp \approx 0.6\) GeV\(^{-2}\) for the single \(\bbb\) trajectory. In \((n,M^2)\) we have lower values for the slope, around \(0.8\) GeV\(^{-2}\) for the \(u\), \(d\), and \(s\), and \(\alp \approx 0.5\) GeV\(^{-2}\) for the last group which now includes both the \(c\) and the \(b\). We had no a priori reason to anticipate a dependence of the slope (or equivalently, the string tension) on the mass, nor the behavior it seems to exhibit, with the slope being more or less constant for the lightest quark trajectories, and then dropping for the heavier mesons. The difference between the two slopes for the \(\ccb\) is especially puzzling.

The fact that the stringy description of mesons built from $b$ quarks refuses to unify with the one that is  associated with lighter quark mesons, is presumably related to the fact that for these mesons the string length, in units of the string basic length $\sqrt{\alp}$, is not really very long as we have seen in table (\ref{tab:lengths}). This is true also of the \(\ccb\) and \(\bbb\) in the \((n,M^2)\) plane, where the trajectories we have are comprised only of states with low angular momentum, and hence, short string lengths.

As for the intercept, the only assumption that was made regarding it in the fits was that it was constant (i.e. independent of \(J\), \(E\)and \(m\)).

The results appear at first glance to be quite scattered, with both positive and negative values appearing in our results.  This should not surprise us since the assumption is in fact not justified. It is clear that the intercept, that gets contributions from both the Polchinski-Strominger term and from the Casimir term, is some function of $\frac{m^2}{T}$ and not a constant.
In the picture of an effective string, the analog of expanding in $\frac{\sqrt{\alpha'}}{l}$ is expanding in $\frac{m}{\sqrt{TJ}}$\cite{AHSY}. In such an expansion the $J^0$ term of the intercept can for low values of $J$ be contaminated by terms of negative powers of $J$.

One thing that can make it easier to compare the values for the intercept between different trajectories is moving from the fits to \((n+J)\) to fits to \((n+L)\) - from the total to the orbital angular momentum. Then we get all the light and strange \((L,M^2)\) trajectories have \(a\) somewhere between \((-0.5)\) and \((-0.2\)), and this value increases to the upper limit of \(0\) for the heavier quarks. In the \((n,M^2)\) plane the picture is similar, but there the \(\ssb\) trajectory already has \(a = 0\).

The transition from the \((J,M^2)\) plane to the \((L,M^2)\) plane is easy to implement in practice, as it only requires the occasional shift of the intercept by one unit, but it requires us to do away completely with spin. Our model does not include spin in the first place, but it seems odd that ignoring it completely, by doing the fits to \(L\), should be rewarded with the added consistency in \(a\).

The WKB model was used in a way that distinguished between total and orbital momentum quite strongly, and the most significant change in the WKB fit results is in the values of the intercept. These are generally more negative, but there the values for the endpoint masses remain roughly the same in all trajectories, and there is always a small decrease in the values of the slope, relative to their values when fitting to the rotating string model. In terms of the goodness of the fits, the WKB model does not offer any significant improvement.

There are several questions and research directions that one can further investigate:
\begin{itemize}
\item
Our model does not incorporate spin degrees of freedom. It is well known that the spin and the spin-orbit interaction play an important role in the spectra of mesons. Thus the simple rotating string models that we are using have to be improved by introducing spin degrees of freedom to the endpoints. One way to achieve it is by replacing the spinless relativistic particle with one that carries spin or in the holographic framework associating spin to the vertical segments of the holographic string.
\item
Our model assumes chargeless massive endpoint particles. The endpoint of a string on a flavor brane carries a charge associated with the symmetry group of the flavor branes. Thus it is natural to add an interaction, for instance Abelian interaction, between the two string endpoints. It is easy to check that this change will introduce a modification of the intercept.
\item
In our WKB analysis we have used only the simplest linear potential. One obvious generalization, which probably will work better for heavy mesons, is the Cornell potential where a $\frac{1}{r}$ potential term is added to the linear one.
\item
As was discussed in the introduction, the models we are using are not the outcome of a full quantization of the system. We have been either using a WKB approximation for the spectra in the $(n,M^2)$ plane or using an ansatz of $J\rightarrow J+n-a$ for passing from the classical to the quantum model. In \cite{Hellerman:2013kba} the quantization of the rotating string without massive endpoints was determined. The quantum Regge trajectories associated with strings with massive endpoints require determining the contributions to the intercept to order $J^0$ from both the `` Casimir" term and the Polchinski-Strominger term\cite{AHSY}. Once a determination of the intercept as a function of $\frac{m^2}{T}$ is made, an improved fit and a re-examination of the deviations from a universal model should be made.
\item
%\subsection{Future work}
We have looked in the present work into only one feature of meson physics - the Regge trajectories of the spectra. One additional property that can be explored is the width of the decay of a meson into two mesons. The stringy holographic width was computed in \cite{Peeters:2005fq}. A detailed comparison with decay width of mesons can provide an additional way to extract string endpoint masses that can be compared to the one deduced from the spectra.
\item
There also remains the other sector of the hadronic spectrum - the baryons. As mentioned above the spectra of these hadrons could also be examined using a stringy model with or without massive endpoints. In addition, closed strings can be used to describe glueballs\cite{oai:arXiv.org:hep-th/0311190}, and using a stringy model we can search for evidence of glueballs among the observed flavorless baryon-less spectrum.
\item
Eventually we have in mind to perform ``precise comparisons" using holographic rotating string models instead of the model of rotating string with massive endpoints in flat space-time.
\end{itemize}
\acknowledgments{J.S. would like to thank Netanel Katz who took part in the early stages of the project. We are grateful to O. Aharony, S. Hellerman,  S. Nussinov,  M.E. Peskin, A. Soffer for useful discussions. We would like to thank O. Aharony and A. Soffer for their comments on the manuscript. This work is partially supported by the Israel Science Foundation (grant 1665/10).}
\clearpage
\appendix
\section{Individual trajectory fits} \label{app:individual}

\begin{table}[t!] \centering
	\begin{tabular}{|c|c|c|c|l|c|c|c|c|l|} \hline
		Traj. & \(I\) & \(J^{PC}\) & Status & State & Traj. & \(I\) & \(J^{PC}\) & Status & State \\ \hline
		
		\(\pi/b\) & 1&\(1^{+-}\) & \(\bullet\) & \(b_1(1235)\) 			     &	\(K^*\) &\(\frac{1}{2}\)&	\(1^-\) & \(\bullet\) & \(K^*(892)\)  \\
		          & &\(2^{-+}\) &  \(\bullet\) & \(\pi_2(1670)\)				            &&		& \(2^+\) & \(\bullet\) & \(K^*_2(1430)\) \\
		          & &\(3^{+-}\) & f. & \(b_3(2030)\)  				   &&& \(3^-\) & \(\bullet\) & \(K^*_3(1780)\) \\
							& &\(4^{-+}\) & f. & \(\pi_4(2250)\)				            &	&			& \(4^+\) & \(\bullet\) & \(K^*_4(2045)\) \\ \cline{1-5}
		\(\rho/a\)& 1&\(1^{--}\) & \(\bullet\) & \(\rho(770)\)  					          &	&			& \(5^-\) & & \(K^*_5(2380)\) \\ \cline{6-10}
							& &\(2^{++}\) & \(\bullet\) & \(a_2(1320)\) 		   &\(\phi/f'\) &0& \(1^{--}\) & \(\bullet\) & \(\phi(1020)\) \\
							& &\(3^{--}\) & \(\bullet\) &\(\rho_3(1690)\) &				            &	& \(2^{++}\) & \(\bullet\) & \(f_2'(1525)\) \\
							& &\(4^{++}\) & \(\bullet\) & \(a_4(2040)\) & 			            &	& \(3^{--}\) & \(\bullet\) & \(\phi_3(1850)\) \\ \cline{6-10}
							& &\(5^{--}\) && \(\rho_5(2350)\) &				    \(D\)   &\(\frac{1}{2}\)& \(0^-\) & \(\bullet\) & \(D^{0}(1865)\) \\
							& &\(6^{++}\) && \(a_6(2450)\) &				            &	& \(1^+\) & \(\bullet\) & \(D^0_1(2420)\) \\ \cline{1-5}
		\(\eta/h\)& 0&\(0^{-+}\) & \(\bullet\) & \(\eta(548)\) &				            &	& \(2^-\) & [a] & \(D_J(2740)\) \\ \cline{6-9}
							& &\(1^{+-}\) & \(\bullet\) & \(h_1(1170)\) &				  \(D^*_s\) & 0&\(1^-\) & \(\bullet\) & \(D^*_s{}^\pm(2112)\) \\
							& &\(2^{-+}\) & \(\bullet\) & \(\eta_2(1645)\) &				            &	& \(2^+\) & \(\bullet\) & \(D^*_{s2}(2573)\) \\
							& &\(3^{+-}\) & f. & \(h_3(2025)\) &				            &	& \(3^-\) & & \(D^*_{sJ}(2860)\) \\ \cline{6-9}
							& &\(4^{-+}\) & f. & \(\eta_4(2330)\) &					 \(\Psi\)& 0& \(1^{--}\) & \(\bullet\) & \(J/\Psi(1S)(3096)\) \\ \cline{1-5}
		\(\omega/f\)&0&\(1^{--}\)& \(\bullet\) & \(\omega(782)\) &				            &	& \(1^{++}\) & \(\bullet\) & \(\chi_{c1}(1P)(3510)\) \\
							& &\(2^{++}\) & \(\bullet\) & \(f_2(1270)\) &				            &	& \(1^{--}\) & \(\bullet\) & \(\Psi(3770)\) \\ \cline{6-10}
							& &\(3^{--}\) & \(\bullet\) & \(\omega_3(1670)\) &				  \(\Upsilon\) & 0 & \(1^{--}\)	& \(\bullet\) & \(\Upsilon(1S)(9460)\) \\
							& &\(4^{++}\) & \(\bullet\) & \(f_4(2050)\) &				            &	& \(2^{++}\) & \(\bullet\) & \(\chi_{b2}(1P)(9912)\) \\
							& &\(5^{--}\) & f. & \(\omega_5(2250)\) &				             & & \(2^{--}\) & & \(\Upsilon(1D)(10164)\) \\
							& &\(6^{++}\) && \(f_6(2510)\) &				            &	&  & & \\ \hline
	\end{tabular}
	\caption{\label{tab:states_j} The states used in the \((J,M^2)\) trajectory fits and their PDG status. States marked with a bullet are the established states appearing in the PDG summary tables, while those marked with an `f.' are the less established mesons classified as ``further states''. Unmarked states belong to the second tier of states omitted from the summary tables. [a] The \(D^J(2740)\) is a newly observed state, not yet listed by the PDG. See text for further explanation of the different state assignments.}
\end{table}
	In this appendix we present the results for the individual trajectory fits in detail. This includes a specification of the states used for each fit, the results for all fitting parameters (masses, slope, and intercept), and the values for \(\chi^2\) in each fit. The plots of all trajectories and their fits in the \((J,M^2)\) or \((n,M^2)\) planes are also presented here.

\subsection{The states used in the fits}
The experimental data is taken almost entirely from the Particle Data Group's (PDG) 2012 Review of Particle Physics \cite{PDG:2012}. Other sources are indicated where relevant. The observation of linear Regge behavior in the hadron spectrum dates back to the 1970s \cite{Collins:book} but has remained the subject of much more recent work as new states are continually discovered in experiment. The heavier sector of the meson spectrum in particular is getting richer and richer in data\cite{Eichten:2007qx}\cite{Brambilla:2010cs}.
	
\cite{Anisovich:2000kxa} and \cite{Gershtein:2006ng} are examples of relatively recent analyses of the meson spectra using Regge trajectories, as is the work of Ebert, Faustov, and Galkin, which covers the spectrum from light \cite{Ebert:2009ub}, to light-heavy \cite{Ebert:2009ua}, to heavy-heavy mesons \cite{Ebert:2011jc} using a relativistic quark model. The selection of trajectories was in part based on the later works cited here, but not before we have independently examined and selected states directly from the PDG review. Note that we have included in our analysis only those trajectories with three or more data points.
We begin by presenting, in tables (\ref{tab:states_j}) and (\ref{tab:states_n}), all the states we have used in our analysis. The two tables are for the trajectories in the \((J,M^2)\) and \((n,M^2)\) planes respectively. The two following subsections explain the selection of states and series of states for the fits, and mention some of those omitted from the fits.

\begin{table}[t!] \centering
	\begin{tabular}{|c|c|c|c|l|c|c|c|c|l|} \hline
		Traj. & \(I(J^{PC})\) & \(n\) & Status & State & Traj. & \(I(J^{PC})\) & \(n\) & Status & State \\ \hline
		
		\(\pi\) & \(1(0^{-+})\) &1 & \(\bullet\) & \(\pi(1300)\) 			     &	\(\omega_3\) &\(0(3^{--})\) &	0 & \(\bullet\) & \(\omega_3(1670)\)  \\

		          & &2 & \(\bullet\) & \(\pi(1800)\)				            					&&		& 1 & f. & \(\omega_3(1950)\) \\

		          & &3 & f. & \(\pi(2070)\)  				   										&&& 2 & f. & \(\omega_3(2255)\) \\ \cline{6-10}

							& &4 & f. & \(\pi(2360)\)				            & \(\phi\)	&	\(0(1^{--})\)	& 0 & \(\bullet\) & \(\phi(1020)\) \\ \cline{1-5}

		\(\pi_2\)& \(1(2^{-+})\) & 0 & \(\bullet\) & \(\pi_2(1670)\)  &	&			& 1 & \(\bullet\) & \(\phi(1680)\) \\

							& &1 & f. & \(\pi_2(2005)\) 		   & && 3 & \(\bullet\) & \(\phi(2170)\) \\ \cline{6-10}

							& &2 & f. &\(\pi_2(2285)\) &				          \(\Psi\)&	\(0(1^{--})\) & 0 & \(\bullet\) & \(J/\Psi(1S)(3097)\) \\ \cline{1-5}

		\(a_1\) & \(1(1^{++})\) &0 & \(\bullet\) & \(a_1(1260)\) & 			            &	& 1 & \(\bullet\) & \(\Psi(2S)(3686)\) \\

							& &1 & & \(a_1(1640)\) &				       && 2 & \(\bullet\) & \(\Psi(4040)\) \\

							& &2 & f. & \(a_1(2095)\) &				            &	& 3 & \(\bullet\)  & \(\Psi(4415)\) \\ \cline{6-10}

							& &3 & f. & \(a_1(2270)\) &	          \(\Upsilon\) & \(0(1^{--})\)& 0 & \(\bullet\) & \(\Upsilon(1S)(9460)\) \\ \cline{1-5}

		\(h_1\)		& \(0(1^{+-})\) & 0 & \(\bullet\) & \(h_1(1170)\) &				  & & 1 & \(\bullet\) & \(\Upsilon(2S)(10023)\) \\

							& &1 & & \(h_1(1595)\) &				            &	& 2 & \(\bullet\) & \(\Upsilon(3S)(10355)\) \\

							& &2 & f. & \(h_1(1965)\) &				            &	& 3 & \(\bullet\) & \(\Upsilon(4S)(10579)\) \\

							& &3 & f. & \(h_1(2215)\) &					 & & 4& \(\bullet\) & \(\Upsilon(10860)\) \\ \cline{1-5}

		\(\omega\)&\(0(1^{--})\) &0 & \(\bullet\) & \(\omega(782)\) &				            &	& 5 & \(\bullet\) & \(\Upsilon(11020)\) \\ \cline{6-10}

							& &1 & \(\bullet\) & \(\omega(1420)\) &				         \(\chi_{b1}\) &	\(0(1^{++})\) & 0 & \(\bullet\) & \(\chi_{b1}(1P)(9893)\) \\

							& &2 & \(\bullet\) & \(\omega(1650)\) &				   & & 1 & \(\bullet\) & \(\chi_{b1}(2P)(10255)\) \\

							& &3 & f. & \(\omega(1960)\) &				            &	& 2 & & \(\chi_{b}(3P)(10530)\) \\

							& &4 & f. & \(\omega(2290)\) &				             & & & & \\ \hline

	\end{tabular}

	\caption{\label{tab:states_n} The states used in the \((n,M^2)\) trajectory fits. Note that we assign \(n = 0\) to the ground state rather than \(n = 1\). States marked with a bullet are the established states appearing in the PDG summary tables, while those marked with an `f.' are the less established mesons classified as ``further states''. Unmarked states belong to the second tier of states omitted from the summary tables. See text for further explanation of the different state assignments.}

\end{table}

\subsubsection{The \texorpdfstring{$(J,M^2)$}{(J,M2)} trajectories}
The classification of the states into trajectories in the \((J,M^2)\) plane is relatively straightforward. We expect the usual relation between spin, orbital angular momentum, and a meson's parity and \(C\)-parity, to hold:
\be P = (-1)^{L+1} \qquad C = (-1)^{L+S} \ee
For states belonging to a trajectory in the \((J,M^2)\) plane, all quantum numbers except the orbital angular momentum are fixed. Therefore \(P\) and \(C\) have alternating values across the trajectory. Furthermore, we fitted only primary Regge trajectories in the \((J,M^2)\) plane, fitting states with no quantum excitations - \(n = 0\). The states we pick then for the trajectories are always the lightest known states with the appropriate quantum numbers.

Our interest is naturally drawn to states with high values of \(J\), where, as explained in the text, the long string approximation is expected to work best. Unfortunately, these states are not typically characterized by great experimental certainty regarding their properties. The PDG broadly divides the known mesons into three tiers.\footnote{We thank the referee for bringing this issue up for us.} The best established states are those included in the summary tables. These are the well defined states that have been observed in multiple experiments. Next are states with their own listings that are omitted from the summary tables. These are resonances that, depending on interpretation, may still move or disappear entirely. The third tier is of the mesons which the PDG classifies as ``further states''. These typically include states observed only in one experiment and considered for the present unconfirmed. Our fits include states belonging to all three tiers.

The \(\rho/a\) trajectory is the best of the light meson \((J,M^2)\) trajectories in terms of the availability and reliability of experimental data. We can confidently use all the six states from \(J = 1\) to \(6\). For the \(\omega/f\) trajectory, again of six states, the \(\omega_5(2250)\) is considered an unestablished state. Our decision to include it in the analysis does not alter the fit results significantly, and a fit done without the \(\omega_5\) predicts its mass to be around 2230 MeV.

For the \(\pi/b\) trajectory, we must include the two unconfirmed states \(b_3(2030)\) and \(\pi_4(2250)\) if we want to have enough data for our analysis. Without those two there are only two other states we can use (after excluding the pion ground state, whose low mass we cannot account for in our simple model). The \(b_3\) and \(\pi_4\) were both observed by the same group, and there is no reason to favor one with an inclusion and not the other.

The \(\eta/h\) trajectory is similar, but there we can choose to include the pseudo-scalar ground state. If we exclude the \(\eta\) we are left with only two states, and we include again both unconfirmed higher states, the \(h_3(2025)\) and \(\eta_4(2330)\) for the fit. If we include the \(\eta\) we may do a fit with only the first three states. The results for the mass in that fit are not altered, but the resulting slope is higher at \(0.89\) GeV\(^{-2}\), and the \(3^{+-}\) and \(4^{-+}\) states are predicted to be lower than the observed states: they should then be at 1910 MeV and 2180 MeV respectively. In the paper we present the analysis of the full five state trajectory. We aim to include as many high spin states as are available, since it is for those states that we expect our model to be most valid, but we will not be surprised if the \(J = 3\) and \(4\) states turn out to be lower than the states currently given.

Of the light-strange mesons we only fit the \(K^*\) trajectory, with the states with \(J = 1\) to \(4\) are in the summary tables, and the state with \(J = 5\) in the second tier of confirmed states not in the summary tables. We did not find a suitable trajectory to use with the \(S = 0\) \(K\) mesons. The \(\ssb\) trajectory of the \(\phi\) includes three states, all considered well established.

For the heavier mesons we begin to make some assignments of our own for the higher \(J\) states. In the trajectory of the charmed \(D\) meson, beginning with the \(D^0\) and the \(D^1_0\) we include a state not yet listed by the PDG as the third \(J = 2\) state - the \(D_J(2740)\). The last state was only recently observed and has been assigned the values \(J^P = 2^-\) \cite{Aaij:2013sza}\cite{Wang:2013tka}. For the charmed-strange \(D^*_s\) we identify the state \(D^*_{sJ}(2860\) as the \(J^P = 3^-\) state, to follow the \(D^{*\pm}_s\) and \(D^*_{s2}\).

In the last two sectors, of the \(\ccb\) and \(\bbb\), there are no confirmed states with \(J\) higher than \(2\). For the \(\ccb\) \(\Psi\) trajectory we then use states with \(J = 1\) but with increasing orbital angular momentum. The spin-orbit splitting between the \(J = 1\) states and the higher \(J\) states with the same orbital momentum is small. For the \(L = 1\) state the difference between the \(\chi_{c1}(1P)\) (\(J^P = 1^+\)) and the \(\chi_{c2}(1P)\) is 45 MeV. From the three \(L = 2\) states, only the state with \(J = 1\) was observed - the \(\Psi(3770)\). The \(J = 3\) state is expected to lie \(30-60\) MeV above the \(\Psi(3770)\)\cite{Eichten:2007qx}. This is again small compared with the masses of the mesons involved. For the trajectory of the \(\bbb\) \(\Upsilon\) we similarly use \(\Upsilon(1D)\) with \(J = 1\) in place of the \(J = 3\) state. The splitting in mass between the different \(J\) states is even less significant for the \(\bbb\) mesons, as can be seen by looking at the the \(L = 1\) \(\chi_{bJ}(1P)\) states: the mass differences due to spin-orbit splitting are 20-30 MeV, and they are completely negligible when compared with the \(\bbb\) mesons' mass.

\subsubsection{The \texorpdfstring{$(n,M^2)$}{(n,M2)} trajectories}
A trajectory in the \((n,M^2)\) plane is constructed by taking multiple states with the same observed quantum numbers and assigning them values of the quantum excitation number \(n\).

In assigning the \(n\) values of the light quark mesons, we began by assuming that the states belong to linear trajectories in the \((n,M^2)\) plane. Our massive model was only to check for small corrections following the assignment of the states into linear trajectories, knowing from the analysis of the better defined trajectories in the \((J,M^2)\) plane that the masses of the light quarks are indeed small.

Of the seven \(\omega\) meson states (\(J^{PC} = 1^{--}\) listed in the PDG, we select five. The first three, \(\omega(782)\), \(\omega(1420)\), and \(\omega(1650)\), are listed in the summary tables. Next there are four \(\omega\) resonances listed as further states. We select the two among them that best continue the linear trajectory formed by the first three states: the \(\omega(1960)\) and \(\omega(2290)\). Remaining are \(\omega(2205)\) and \(\omega(2330)\). The former is just a little too low to serve as the fifth state and too high to be the fourth. When the latter, the \(\omega(2330)\), is used as the \(n = 4\) state instead of the \(\omega(2290)\) we get no significant change. The trajectory of the higher spin \(\omega_3\) states (\(J^{PC} = 3^{--}\)) starts with the well established \(\omega_3(1450)\). Then, from the three remaining further states we find two that lie on a linear trajectory parallel to the trajectory of the lower spin states.

The case of the \(\pi\) is similar. For the \(J^{PC} = 0^{-+}\) trajectory we use all PDG listed states except the pion ground state. Moving on to the \(2^{-+}\) states, the \(\pi_2(1670)\) is the established lowest state, and the states that follow belong to the linear trajectory parallel to that of the \(0^{-+}\) \(\pi\) mesons. We omit the \(\pi_2(1880)\) (not in further states, but not in summary tables) which is too low to follow \(\pi_2(1670)\) in its trajectory.

We also examine the trajectories of the \(a_1\) and \(h_1\) mesons. Both cases are similar - the lowest state is listed in the PDG summary tables, while the third and fourth states are taken from the further states listings. From the \(h_1\) trajectory we omit second tier state \(h_1(1380)\), whose proximity to the \(h_1(1170)\) would give the resulting trajectory an unreasonably high slope. The rest of the \(h_1\) states are all included as they are located on a linear trajectory. As for the \(a_1\), we have a well established ground state, the next lowest state (second tier) is also included, and we pick two states out of three from the further states table to complete the trajectory.

There are some light meson series which we have left out altogether, most notably the \(\rho\) and the \(\eta\) of which the PDG lists 8 and 11 states respectively, all in various degrees of quality. The assignment of these states tends to be more ambiguous than that of the previous series, and it is harder to find trajectories that will be useful for our purpose of checking the massive string model against experimental data. One possible assignment for the \(\rho\) mesons has \(\rho(770)\), \(\rho(1450)\), \(\rho(1900)\) (not in summary table), and \(\rho(2270)\) (further states) in the leading trajectory. The trajectory formed by these states is \(0.66\) GeV\(^{-2}\). Another possibility is to take \(\rho(770)\), \(\rho(1700)\), \(\rho(2000)\) (further states), and \(\rho(2150)\) (not in summary table) as the \(n = 0\), \(2\), \(3\), and \(4\) states and get a trajectory with \(\alp = 0.92\) GeV\(^{-2}\). The missing \(n = 1\) state is predicted to be then at \(1300\) MeV. The linear fit works better for the first option, with the lower slope.

For the \(\eta\) states the assignment into trajectories seems again problematic, because there are so many states in a relatively small mass span. In the PDG summary tables there are five \(\eta\) meson states - from the \(548\) MeV ground state to the \(\eta(1475)\). If we are to have a linear trajectory with a consistent value of the Regge slope, we can do this only by choosing two of these, and then completing the trajectory using higher states (second and third tier). This can lead to a few possible assignments, none of which offers particularly illuminating results.

There are also many \(f_0\) and \(f_2\) states which we omit here. These have the right quantum numbers to be (or contain) glueballs - \(I = 0\) and \(J^{PC} = 0^{++}\) or \(2^{++}\) - and we therefore leave them out until a separate analysis is made.

For the heavier quark mesons we use, with one sole exception, only well established states, included in the PDG summary tables. There are three \(\phi\) states (\(\ssb\)), for which we assign \(n = 0\), \(1\), and for the highest state \(3\). The \(\omega(1960)\) (further states) could be interpreted as the missing \(n = 2\) state, but we leave it its original classification as an \(\omega\) and maintain that there should be another state with the same quantum numbers near it.\footnote{The two states will probably not be a pure \(\phi\) and a pure \(\omega\), but rather a mixture of the two.}

The charmonium sector is quite rich in data, with many observed states with low spin. In particular, there are many states with \(J^{PC} = 1^{--}\), of which we pick four for our trajectory, starting with the well known \(J/\Psi\) meson. Other well established states, the \(\chi_c\) and \(\eta_c\), have to be omitted simply because there are not enough data points to complete a trajectory for them. Of the states in this mass region, there are also some which are potential exotics, most famously the \(X(3872)\)\cite{Brambilla:2010cs} - these will also be interesting to examine once we generalize our stringy model to include non \(q\bar{q}\) states.

The bottomonium sector again offers us many low spin states. The \(\Upsilon\) (\(J^{PC} = 1^{--}\)) trajectory in particular uses six of them, all of them being summary table states. We also analyze the trajectory of the \(\chi_b\) (\(1^{++}\)), with three states. The \(\eta_b\) is left out because we have only two such states.
	\subsection{Trajectories in the \texorpdfstring{$(J,M^2)$}{(J,M2)} plane}
			\subsubsection{Light quark mesons}
			
			The states in this section are all comprised of \(u\) and \(d\) quarks only. We assume in our analysis that the two lightest quarks are equal in mass, and make no attempt to differentiate between them.
				
\begin{figure}[tbp] \centering
						\includegraphics[natwidth=1200bp, natheight=900bp, width=.48\textwidth]{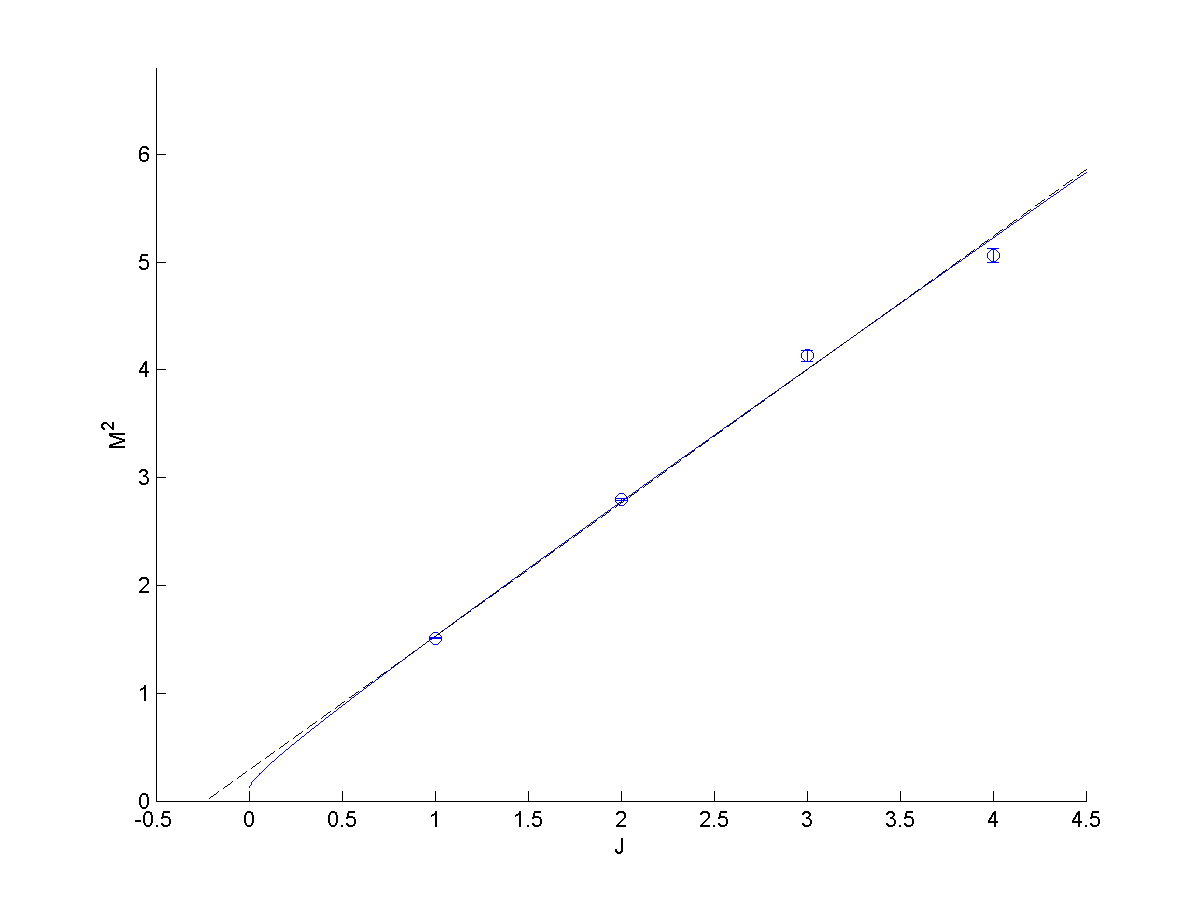}	 \hfill
						\includegraphics[natwidth=1200bp, natheight=900bp, width=.48\textwidth]{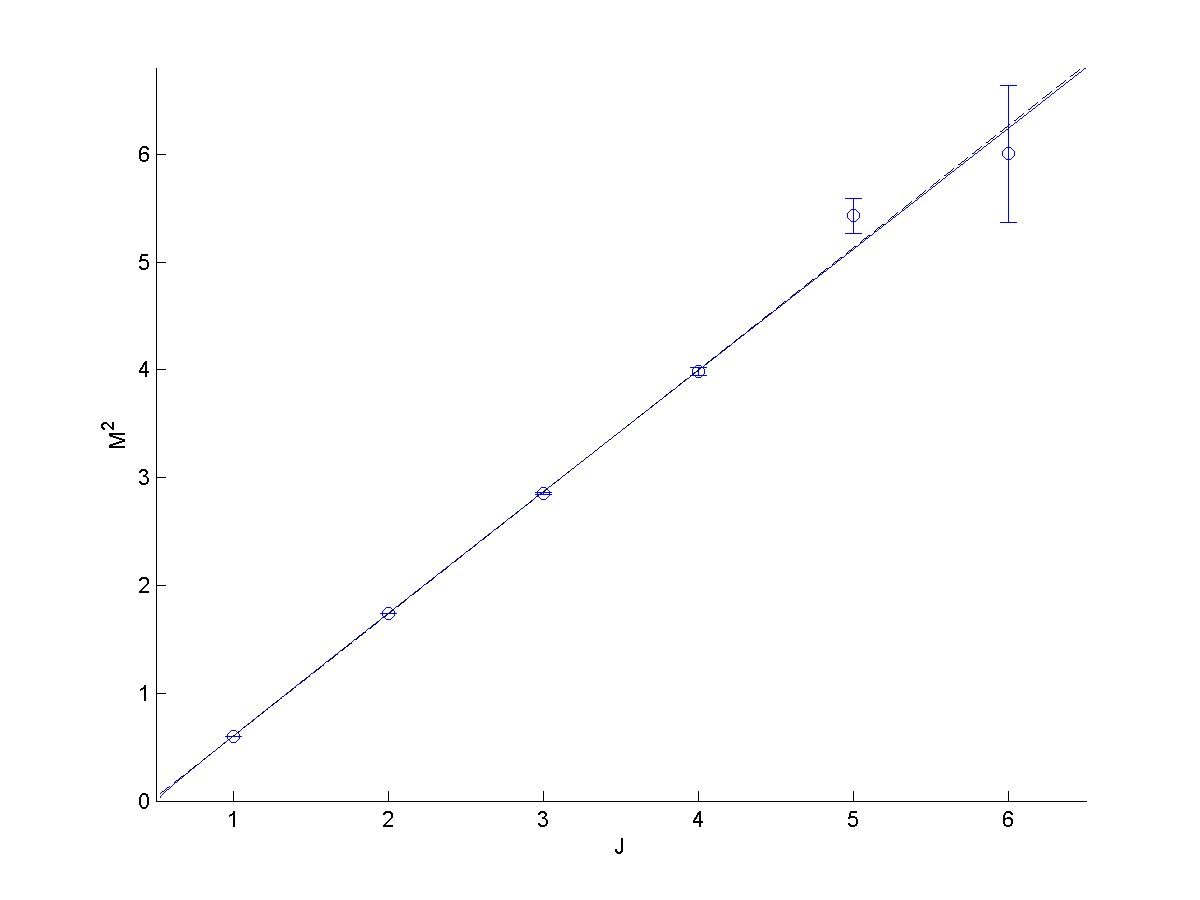}
						\caption{\label{fig:pi_rho} The \(I = 1\) light quark trajectories. Left: the \(\pi/b\) and optimal massive fit with \(m = 170\). The red marker represents the pion, not used in the analysis. The plot on the right is the \(\rho/a\) trajectory and fit with \(m = 65\). The blue lines are the trajectories with massive endpoints, dashed black lines are the linear fits.}
				\end{figure}

				\paragraph{\(I = 1\). The \(\pi/b\) trajectory:} The trajectory depicted in the left of figure (\ref{fig:pi_rho}) is comprised of the states \(b_1(1235) 1^{+-}, \pi_2(1670) 2^{-+}, b_3(2030) 3^{+-},\) and \(\pi_4(2250) 4^{-+}\). The lowest state in this trajectory is actually the pion, but we exclude it from our fits due to its abnormally low mass. The corresponding fits show a relatively large range of available masses, from \(m = 0\) to \(185\) MeV, with the optimum being at \(m = 170\) MeV. The linear fit,
				\[ \alp = 0.808, a = -0.23 \]
				has \(\chi^2_l = 7.99\ten{-4}\). The mass \(m = 185\) is where we get the nearly equal value with \rchi{0.99} before \(\chi^2_m\) starts growing higher and surpasses \(\chi^2_l\). The optimum is
				\[ m = 170, \alp = 0.844, a = 0.00 \]
				and it has \rchi{0.86}.
				
				\paragraph{\(I = 1\). The \(\rho/a\) trajectory:} The plot on the right of figure (\ref{fig:pi_rho}) is that of the \(\rho\) trajectory. The states are \(\rho(770)1^{--}, a_2(1320)2^{++}, \rho_3(1690)3^{--}, a_4(2040)4^{++}, \rho_5(2350)5^{--},\) and \(a_6(2450)6^{++}\). The linear fit is
				\[ \alp = 0.882, a = 0.47 \]
				with \(\chi^2_l = 9.90\ten{-4}\).	The massive fits exhibit a very weak dependence of \(\chi^2\) on the endpoint mass. All the masses in the range \(m = 0-125\) appear to be nearly equivalent. There is a very indistinct optimum with \rchi{0.99} at
				\[ m = 65, \alp = 0.896, a = 0.52 \]				
				but masses up to \(m = 160\) are still offer reasonable fits with \rchi{1.04} at that mass.				
				
\begin{figure}[tbp] \centering
						\includegraphics[natwidth=1200bp, natheight=900bp, width=.48\textwidth]{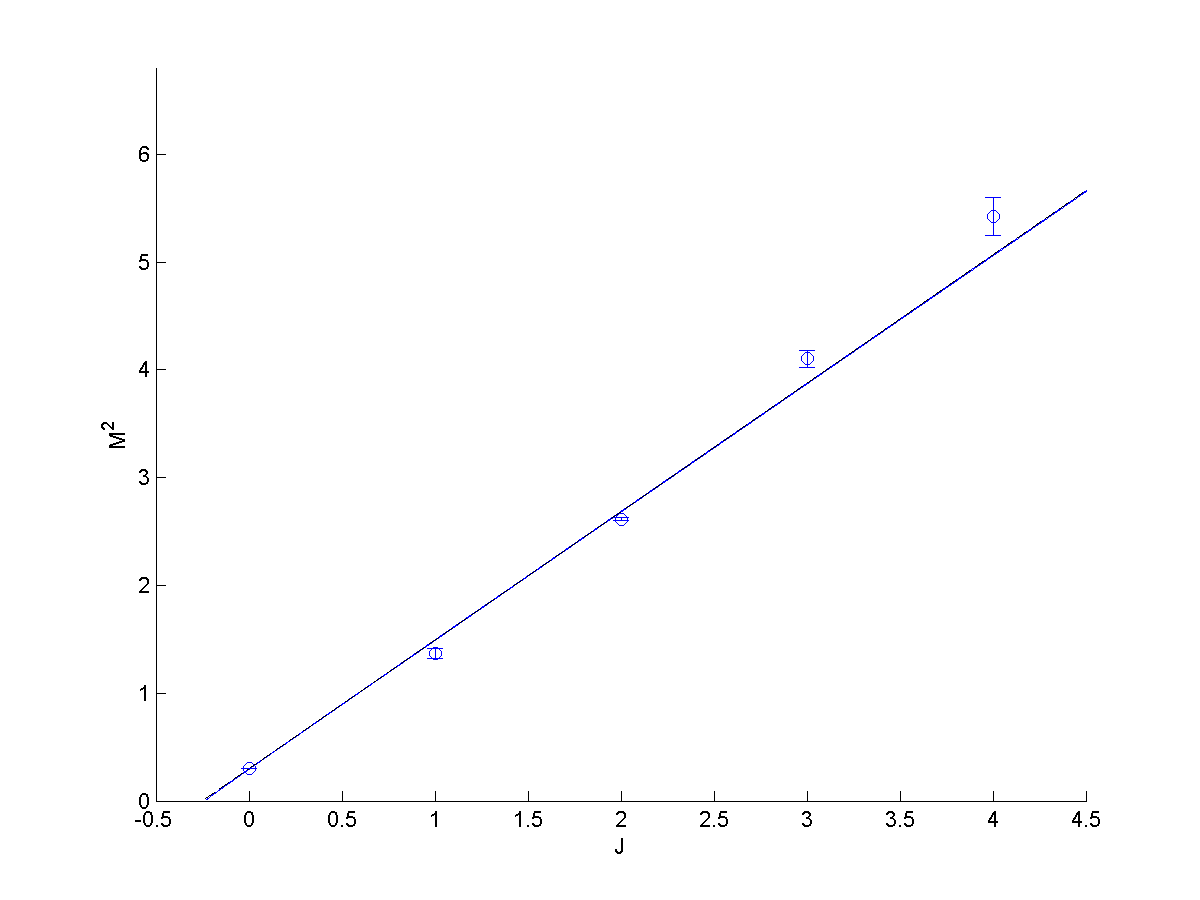}	 \hfill
						\includegraphics[natwidth=1200bp, natheight=900bp, width=.48\textwidth]{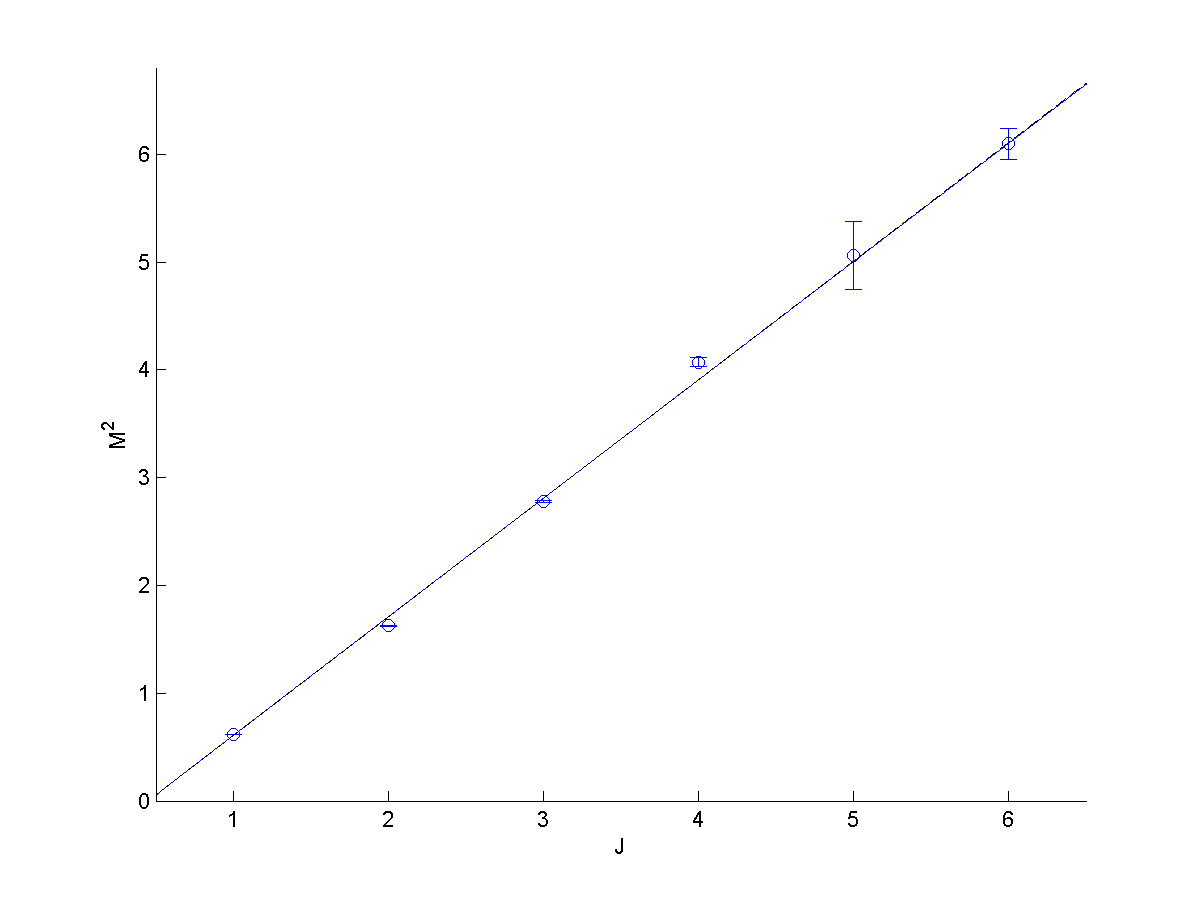}
						\caption{\label{fig:eta_omega} The \(I = 0\) light quark trajectories. The \(\eta/h\) are on the left, \(\omega/f\) on the right. For both trajectories the linear fit (dashed black line) is optimal, but we also plot a massive fit with \(m = 60\) for both trajectories (blue line).}
				\end{figure}

				\paragraph{\(I = 0\). The \(\eta/h\) trajectory:} The \(\eta/h\) trajectory is depicted on the right of figure (\ref{fig:eta_omega}). The states used were \(\eta(548) 0^{-+}, h_1(1170)1^{+-}, \eta_2(1645)2^{-+}, h_3(2025)3^{+-},\) and \(\eta_4(2330)4^{-+}\). The ground state is the scalar \(\eta\) meson, and we should consider excluding it from our analysis as we did the pion. With the \(\eta\) included, the linear fit gives
				\[ \alp = 0.839, a = -0.25 \]
				and \(\chi^2_l = 39.63\ten{-4}\). Here the linear fit is optimal. We need only go to \(m = 60\) to get \rchi{1.10}, and \(\chi^2\) only keeps on rising with the mass. If we exclude the \(\eta\) ground state, the linear fit is again optimal, but it is changed considerably. The new values are
				\[ \alp = 0.745, a = -0.01 \]
				and \(\chi^2_l = 4.57\ten{-4}\), a much better value. In terms of the endpoint masses, though, the picture is largely unchanged. The linear fit is optimal, and \(\chi^2\) rises somewhat quicker to give \rchi{1.11} at \(m = 45\).
				
				\paragraph{\(I = 0\). The \(\omega/f\) trajectory:} The right side of figure (\ref{fig:eta_omega}) depicts the \(\omega/f\) trajectory. It includes the states \(\omega(782)1^{--}, f_2(1270)2^{++}, \omega_3(1670)3^{--}, f_4(2050)4^{++}, \omega_5(2250)5^{--},\) and \(f_6(2510)6^{++}\). The linear fit is
				\[ \alp = 0.909, a = 0.45 \]
				with \(\chi^2_l = 8.85\ten{-4}\). It is optimal. We can place the limit on the mass at \(m = 60\), where \rchi{1.10}.
				
			\subsubsection{Strange and \texorpdfstring{$\ssb$}{s-sbar} mesons}
\begin{figure}[tbp] \centering
						\includegraphics[natwidth=1200bp, natheight=900bp, width=.48\textwidth]{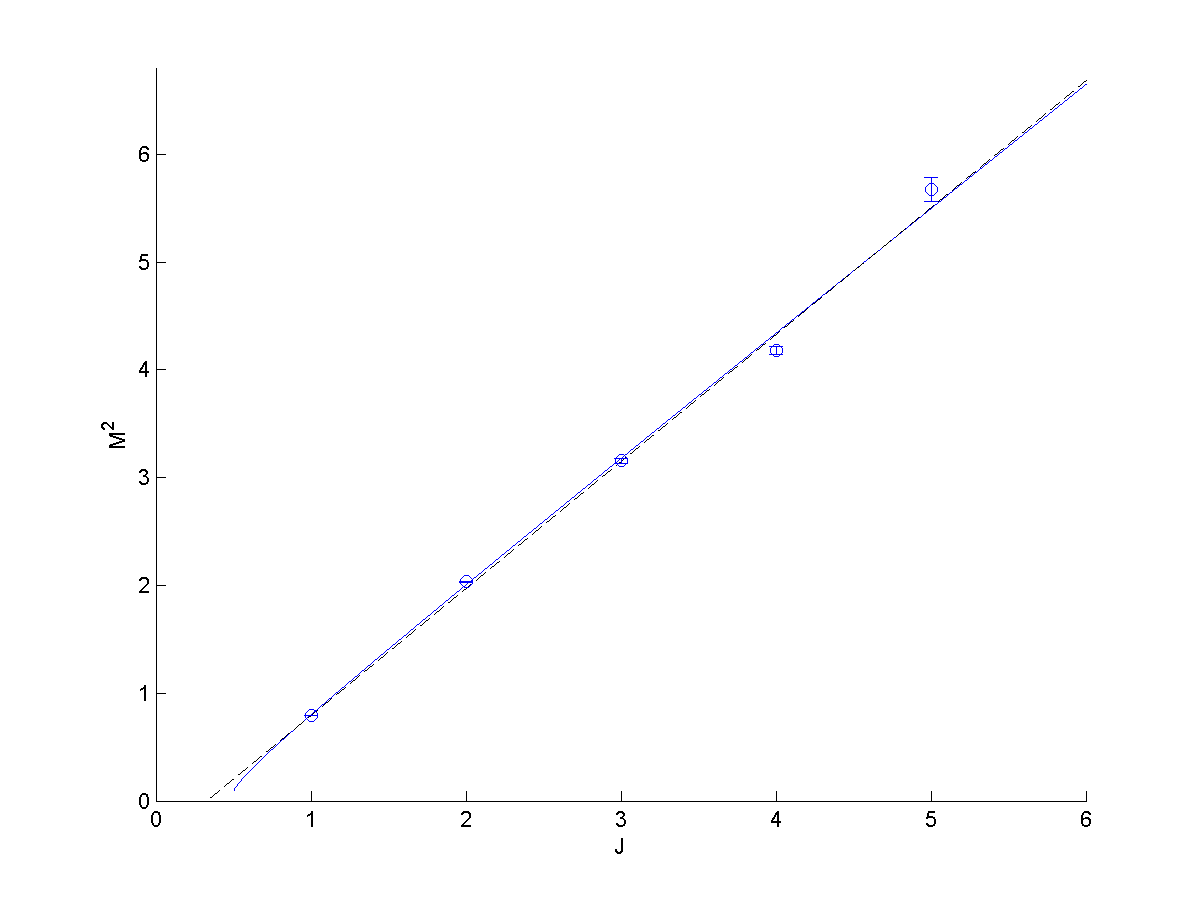}	 \hfill
						\includegraphics[natwidth=1200bp, natheight=900bp, width=.48\textwidth]{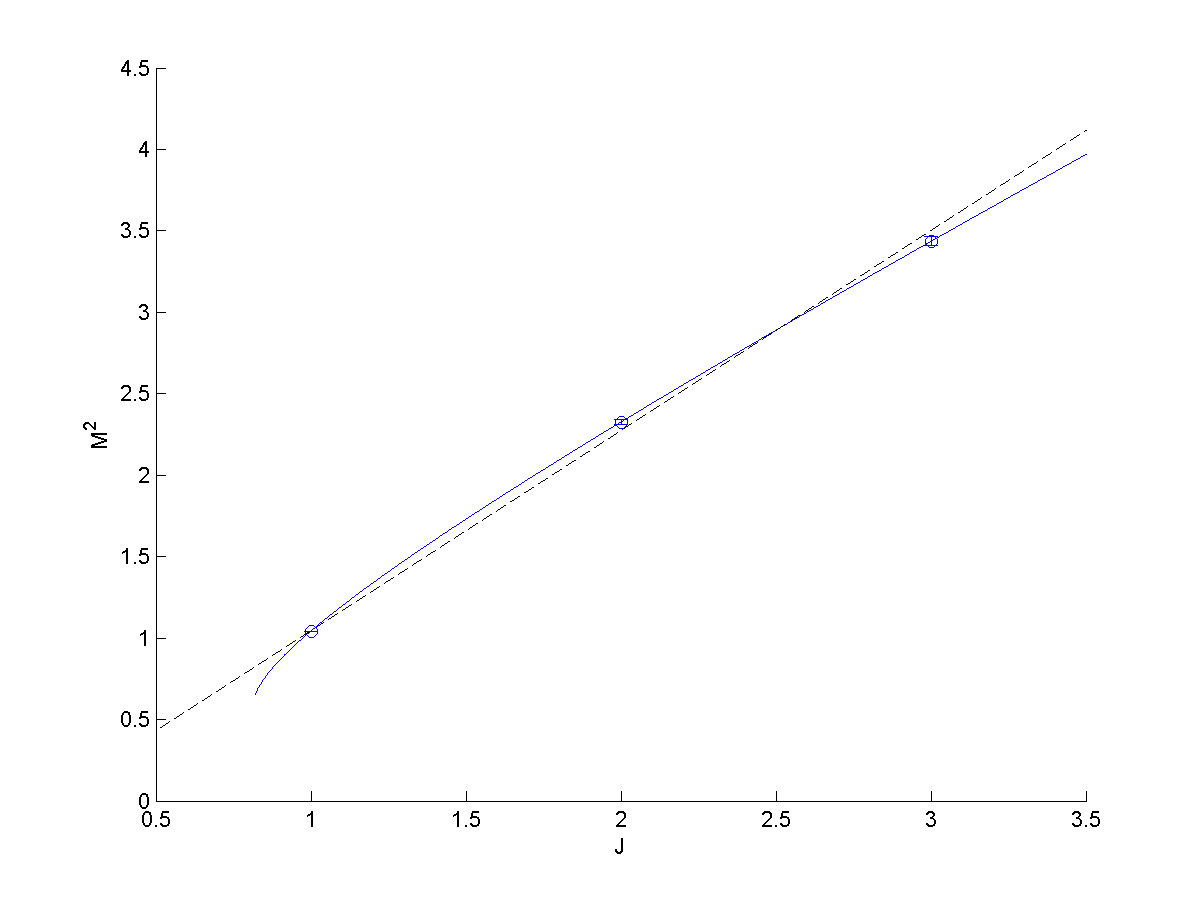}
						\caption{\label{fig:k_phi} The trajectories involving an \(s\) quark. On the left is \(K^*\), with the fit where \(m_{u/d} = 60\) and \(m_s = 220\). On the left is \(\phi\) with its optimal fit of \(m_s = 400\). The blue lines are the trajectories with massive endpoints, dashed black lines are the linear fits.}
				\end{figure}

				\paragraph{Strange. The \(K^*\) trajectory:} This is the trajectory depicted on the left of figure (\ref{fig:k_phi}). The states are \(K^*(892)1^-\), \(K^*_2(1430)2^+\), \(K^*_3(1780)3^-\), \(K^*_4(2045)4^+,\) and \(K^*_5(2380)5^-\). These are comprised of one light \(u\) or \(d\) quark and one \(s\) quark. Since we expect a difference between the mass of the \(s\) quark and that of the light quarks we fit to a formula with two different masses, \(m_s > m_{u/d}\).
				
				The linear fit has
				\[ \alp = 0.849, a = 0.33 \]
				and \(\chi^2_l = 7.15\ten{-4}\). The optimal massive fits have \(m_{u/d} + m_s \approx 300\), but there is no way to determine the masses separately from these fits. It is not possible even to distinguish between the symmetric case where the two masses are equal and the other extreme where one of the endpoints is massless, or nearly massless, and the other is not. The optimal fits are obtained on a curve on the \((m_{u/d},m_s)\) plane - \(\mud^{3/2}+m_s^{3/2} = 2\times(162)^{3/2}\). We can list some of the values along the curve:
				\[m_{u/d} = 60, m_s = 220, \alp = 0.885, a = 0.50\]
				\[m_{u/d} = 100, m_s = 180, \alp = 0.881, a = 0.49\]
				\[m_{u/d} = 140, m_s = 180, \alp = 0.889, a = 0.52\]
				These are all nearly equal, with \rchi{0.932-0.935}. Higher masses are also possible, with \(\mud + m_s \approx 360\) still having \rchi{1} or less.
				
				\paragraph{\(\ssb\). The \(\phi/f'\) trajectory:} The trajectory on the right of figure (\ref{fig:k_phi}) is that of the \(\ssb\). Here we have only three states: \(\phi(1020)1^{--}, f'_2(1525)2^{++}, \phi_3(1850)3^{--}\). The best linear fit is
				\[ \alp = 0.814, a = 0.15 \]
				with \(\chi^2_l = 4.43\ten{-4}\). The massive fits point to a very distinct minimum at
				\[ m_s = 400, \alp = 1.078, a = 0.82 \]
				which has \rchi{0.01}. Fits with a mass closer to that which the \(K^*\) trajectory fits imply for the \(s\) quark still offer a significant improvement when compared to the linear fit. For example,
				\[ m_s = 200, \alp = 0.882, a = 0.41 \]
				has \rchi{0.60}.
				
			\subsubsection{Charmed, Charmed/Strange, and \texorpdfstring{$\ccb$}{c-cbar} mesons}
\begin{figure}[tbp] \centering
						\includegraphics[natwidth=1200bp, natheight=900bp, width=.48\textwidth]{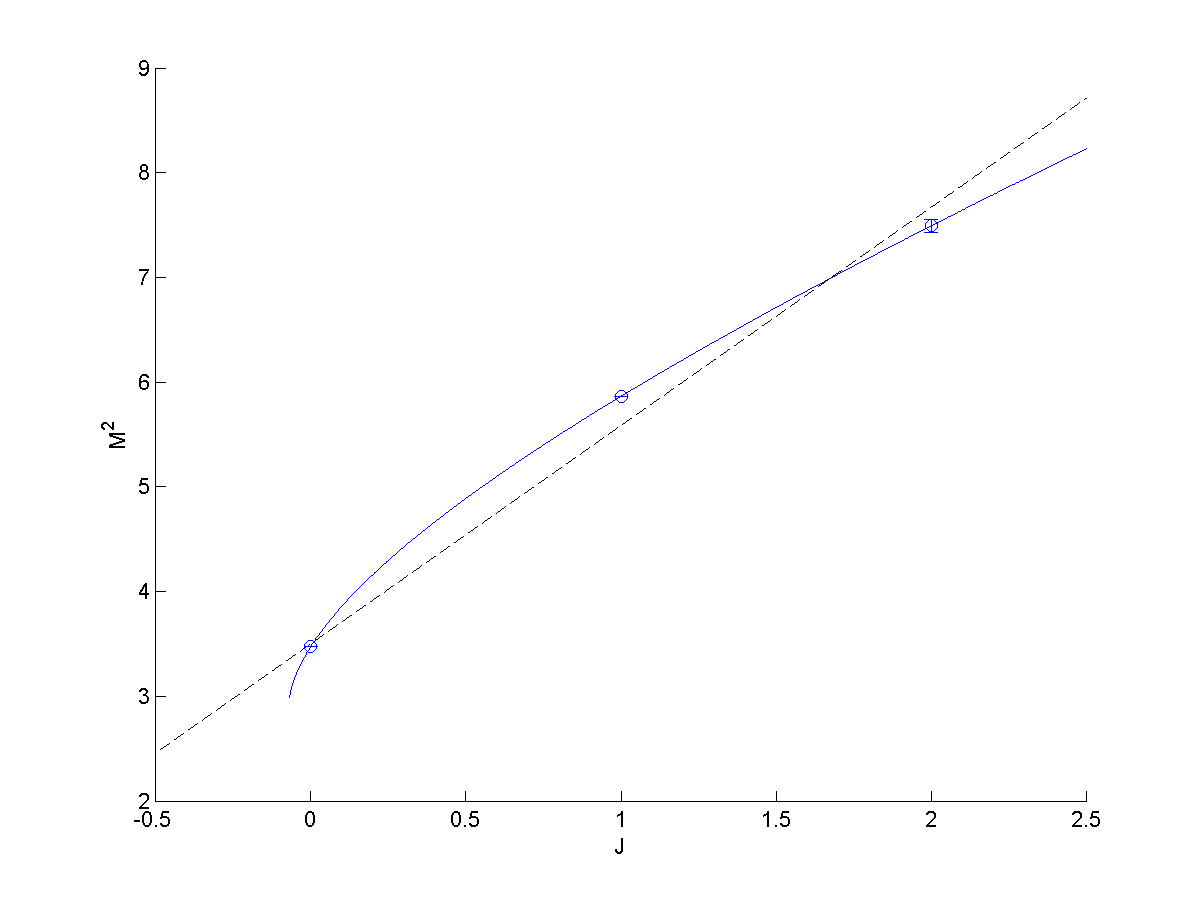}	 \hfill
						\includegraphics[natwidth=1200bp, natheight=900bp, width=.48\textwidth]{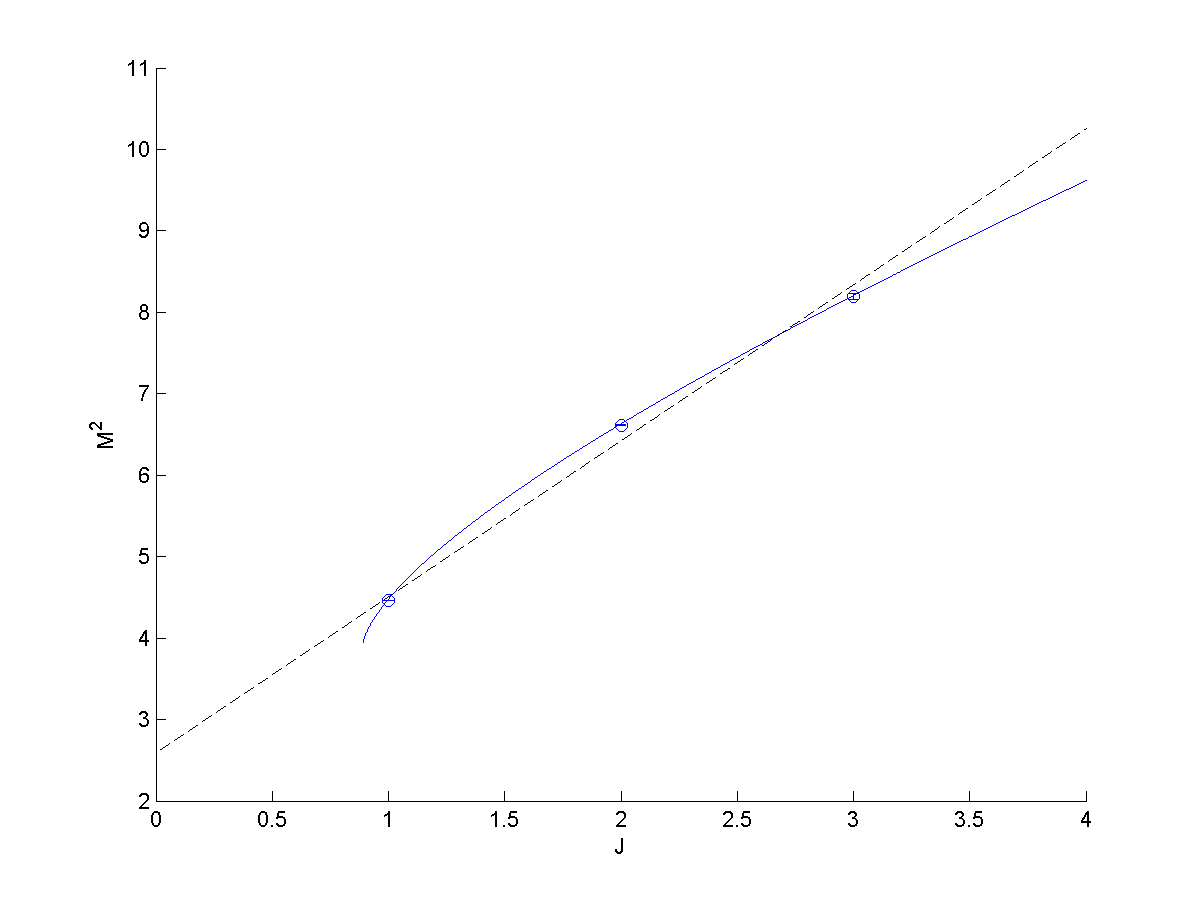}
						\caption{\label{fig:d_ds} The charmed meson trajectories. On the left is \(D\) with the massive fit \(m_{u/d} = 80\) and \(m_c = 1640\), and on the right is the charmed/strange \(D^*_s\), with its fit of \(m_s = 400\), \(m_c = 1580\). The blue lines are the trajectories with massive endpoints, dashed black lines are the linear fits.}
				\end{figure}

				 \paragraph{Charmed. The \(D\) trajectory:} On the left side of figure (\ref{fig:d_ds}) is the trajectory of the charmed \(D\) mesons, comprised of one \(u/d\) quark and one \(c\) quark. Here we used the	states \(D^0(1865)0^-, D_1(2420)^01^+,\) and \(D_J(2740)2^-\). The last state, not yet listed by the PDG, was only recently observed and given the assignment \(J^P = 2^-\) \cite{Aaij:2013sza}\cite{Wang:2013tka}.
			The linear fit to the trajectory is
			\[ \alp = 0.480, a = -1.69 \]
			and it has \(\chi^2_l = 13.92\ten{-4}\). The massive fits here show a preference for one light quark and one heavy quark, with the optimal fit being
			\[ m_c = 1640, m_{u/d} = 80, \alp = 1.073, a = -0.07 \]
			with \(\chi^2_m = 5\ten{-8}\) (\rchi{3\ten{-5}}). We can still shift some of the mass from one endpoint to another:
			\[ m_c = 1500, m_{u/d} = 300, \alp = 1.021, a = -0.03\]
			has \(\chi^2_m = 3\ten{-7}\), but not to the point where the two masses are equal. If we assume the symmetric case, the best fit we get is
			\[ 2m = 1840, \alp = 0.933, a = -0.01 \]
			with \(\chi^2_m = 7\ten{-6}\).
			
				\paragraph{Charmed/Strange. The \(D^*_s\) trajectory}. On the right of figure (\ref{fig:d_ds}) is the \(D^*_s\) trajectory. These contain an \(s\) quark and a \(c\) quark. We use the states \(D^*_s{}^\pm(2112)1^-\), \(D^*_{s2}(2573)2^+\), and take \(D^*_{sJ}(2860)\) to be the \(J^P = 3^-\) state. The linear fit
				\[ \alp = 0.522, a = -1.35 \]
				has \(\chi^2_l = 6.44\ten{-4}\). The fits don't point to a specific value of the two masses, nor does the optimum lie along a simple curve like they did for the \(K^*\). We have, for example
				\[ m_s = 200, m_c = 1720, \alp = 1.133, a = 0.88 \]
				with \(\chi^2_m = 5\ten{-9}\), or
				\[ m_s = 400, m_c = 1580, \alp = 1.093, a = 0.89 \]
				with \(\chi^2_m = 4\ten{-9}\). The best symmetric fit (which maximizes \(m_1 + m_2\)) is
				\[ 2m = 2020, \alp = 1.028, a = 0.93\]
				with \(\chi^2_m = 16\ten{-9}\).
				
				\begin{figure}[tbp] \centering
						\includegraphics[natwidth=1200bp, natheight=900bp, width=.48\textwidth]{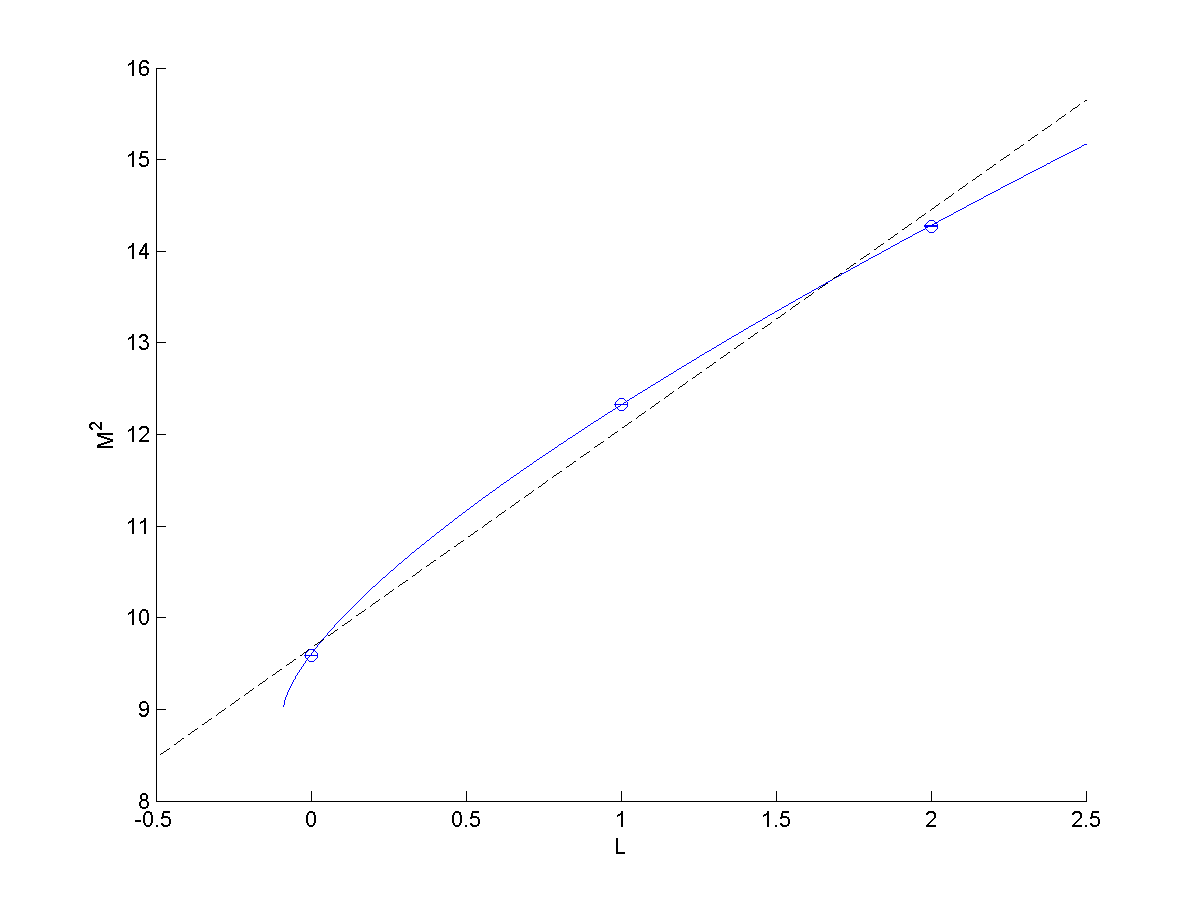}	 \hfill
						\includegraphics[natwidth=1200bp, natheight=900bp, width=.48\textwidth]{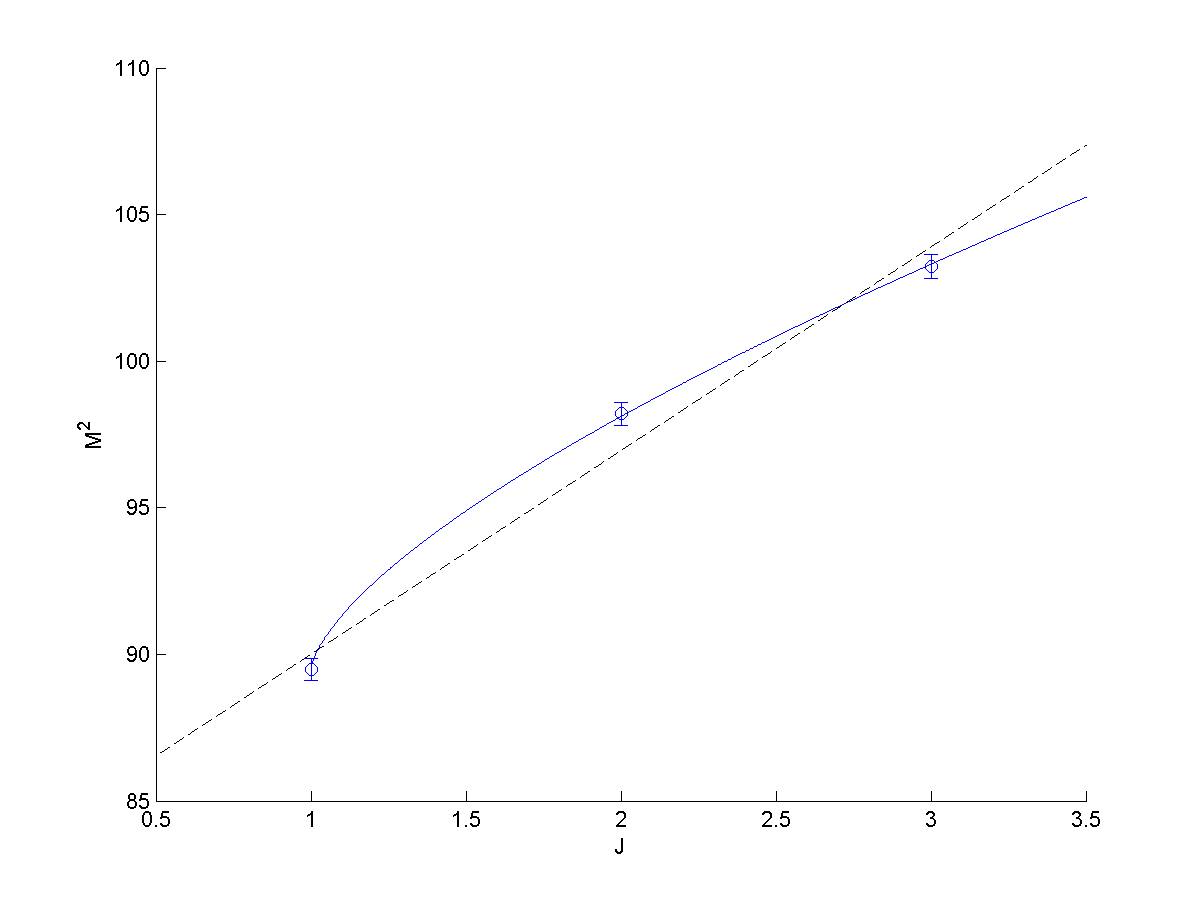}	 \hfill
						\caption{\label{fig:psi_ups} The \(\ccb\) \(\Psi\) (left) and \(\bbb\) \(\Upsilon\) (right) trajectories with the massive fits \(m_c = 1500\) and \(m_b = 4730\) respectively. The blue line is the fit with massive endpoints, dashed black line the linear fit.}
				\end{figure}
			
				\paragraph{\(\ccb\). The \(\Psi\) trajectory:} The left side of figure (\ref{fig:psi_ups}) depicts the \(\Psi\) trajectory. Here we use the states \(J/\Psi(1S)(3097)1^{--}, \chi_{c1}(1P)(3510)1^{++},\) and \(\Psi(3770)1^{--}\). Since no \(J = 3\) state has been observed, we use three states with \(J = 1\), but with increasing orbital angular momentum (\(L = 0,1,2\)) and do the fit to \(L\) instead of \(J\). To give an idea of the shifts in mass involved, the \(J^{PC} = 2^{++}\) state \(\chi_{c2}\) has a mass of \(3556\) MeV, and the \(J^{PC} = 3^{--}\) state is expected to lie \(30-60\) MeV above the \(\Psi(3770)\)\cite{Eichten:2007qx}.
				
				The best linear fit is
				\[ \alp = 0.418, a = -4.04 \]
				with \(\chi^2_l = 3.41\ten{-4}\), but the optimal fit is far from the linear, with endpoint masses in the range of the constituent \(c\) quark mass:
				\[ m_c = 1500, \alp = 0.979, a = -0.09 \]
				with \(\chi^2_m = 5\ten{-7}\) (\rchi{0.002}). Aside from the improvement in \(\chi^2\), by adding the mass we also get a value for the slope (and to a lesser extent, the intercept) that is much closer to that obtained in fits for the light meson trajectories.
				
			\subsubsection{\texorpdfstring{$\bbb$}{b-bbar} mesons}
				\paragraph{\(\bbb\). The \(\Upsilon\) trajectory:} The right side of figure (\ref{fig:psi_ups}) shows the \(\Upsilon\) trajectory, comprised of the three states \(\Upsilon(1S)(9460)1^{--}, \chi_{b2}(1P)(9910)2^{++},\) and \(\Upsilon(1D)(10160)2^{--}\). The actual third state in this trajectory, with \(J^{PC} = 3^{--}\), should be a little higher in mass compared with the \(\Upsilon(1D)\). We can estimate the difference in mass between the \(J = 2\) and \(3\) states based on the splitting of the three \(\chi_b\) states. These have \(^{2S+1}\! L_J =\) \(^3P_0\), \(^3P_1,\) and \(^3P_2\) and the differences between masses of the different \(J\) states is around \(20-30\) MeV. This results in a difference of less than one percent in their masses squared, so we can safely assume that using the \(2^{--}\) state in place of the \(3^{--}\) won't affect our fits significantly.
				The linear fit for this trajectory is
				\[ \alp = 0.144, a = -11.96 \]
				and it has \(\chi^2_l = 1.20\ten{-4}\). The massive fit gives an optimum when the endpoint masses are equal to the constituent mass. It is
				\[ m = 4730, \alp = 0.635, a = 1.00 \]
				with \(\chi^2_m = 8\ten{-7}\) (\rchi{0.007}).
				
		\subsection{Trajectories in the \texorpdfstring{$(n,M^2)$}{(n,M2)} plane}
			\subsubsection{Light quark mesons}
\begin{figure}[tbp] \centering
						\includegraphics[natwidth=1200bp, natheight=900bp, width=.48\textwidth]{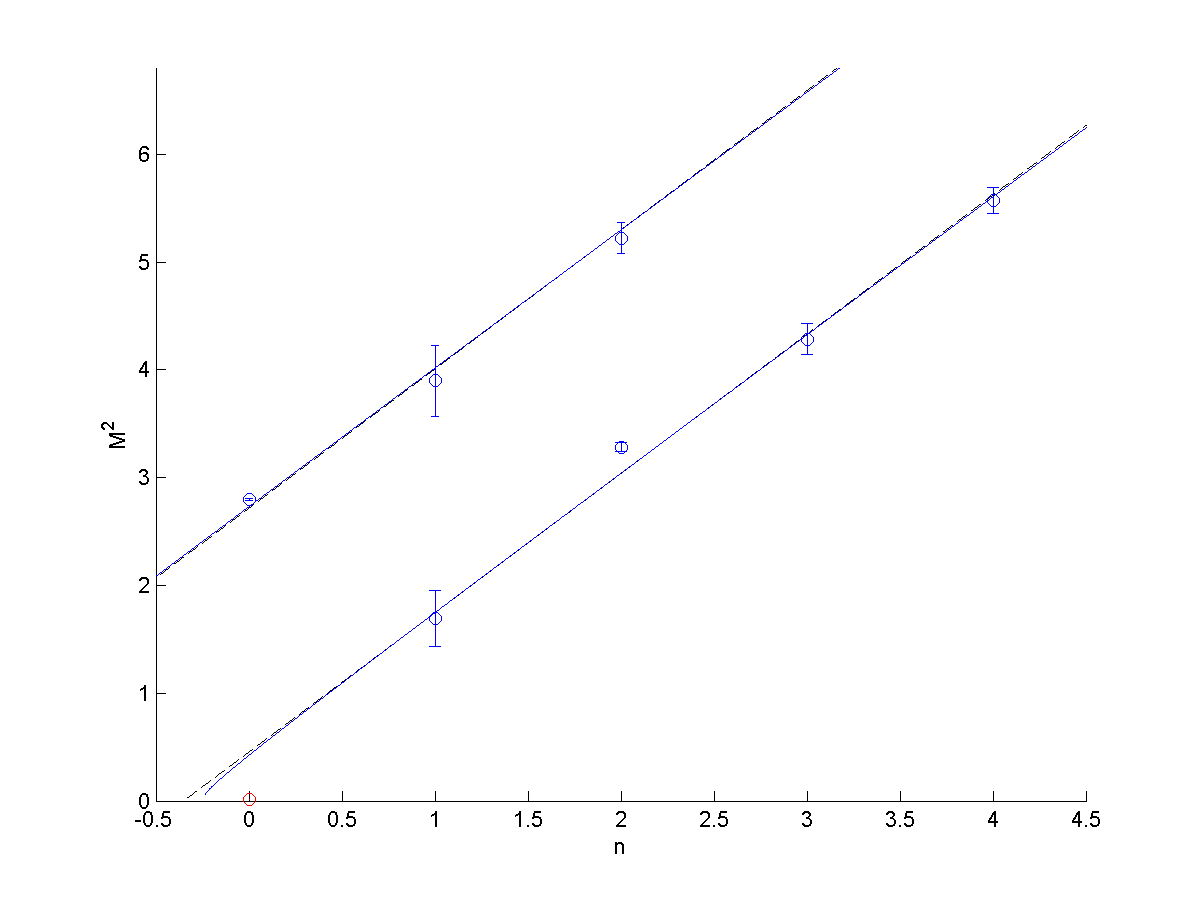}	 \hfill
						\includegraphics[natwidth=1200bp, natheight=900bp, width=.48\textwidth]{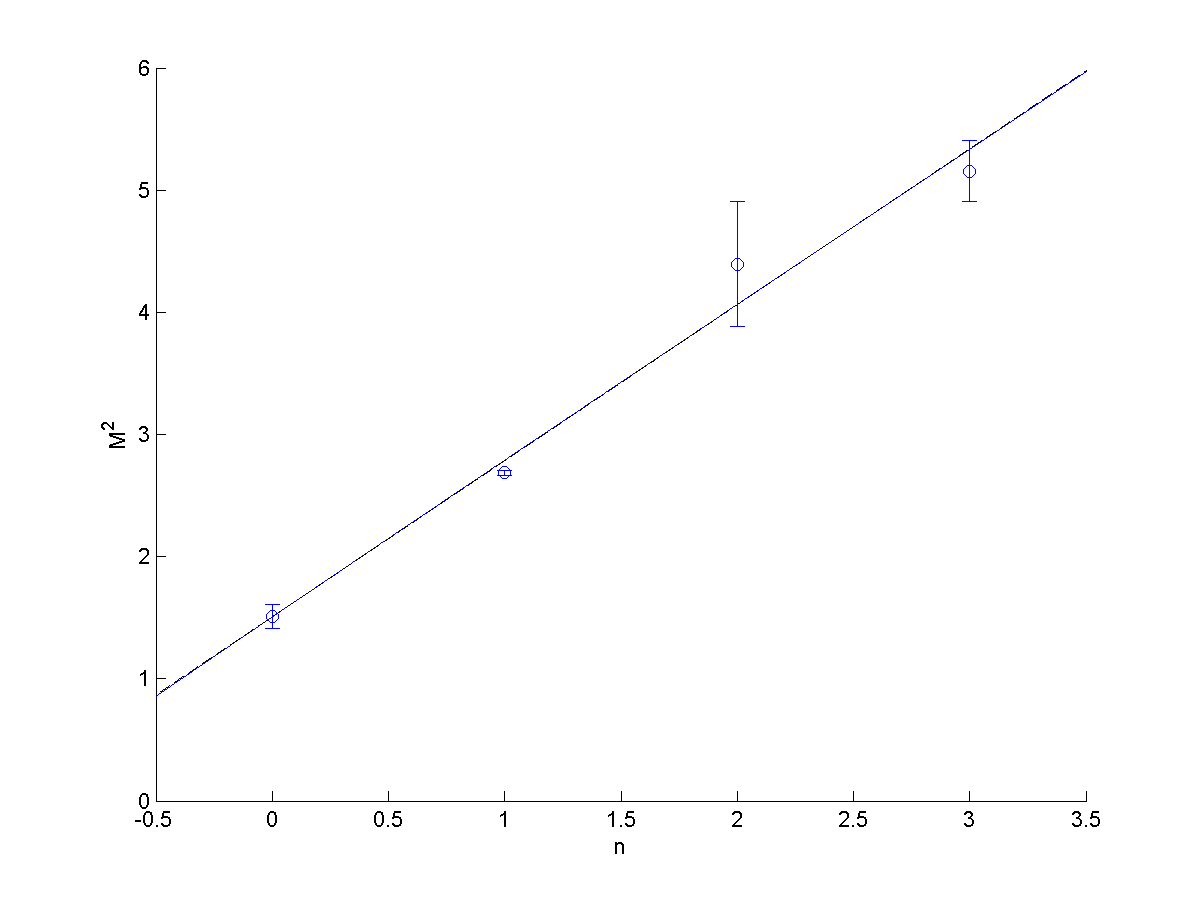}
						\caption{\label{fig:pi_a1} Left: the \(\pi\) and \(\pi_2\) with the massive fit \(\mud = 105\). Right: the \(a_1\) with its fit for \(\mud = 100\).}
				\end{figure}

				\paragraph{\(I = 1\). The \(\pi\) trajectory:} The left of figure (\ref{fig:pi_a1}) depicts the two \(\pi\) trajectories. Here we use the states \(\pi(1300)\), \(\pi(1800)\), \(\pi(2070)\), and \(\pi(2360)\) with \(J^{PC} = 0^{-+}\), and \(\pi_2(1670)\), \(\pi_2(2005)\), and \(\pi_2(2285)\) with \(J^{PC} = 2^{-+}\). The pion ground state is again excluded from the analysis.

				The fits are done simultaneously to the \(J = 0\) and \(J = 2\) states, with the same mass and slope for both trajectories. We do allow, though, a difference in the intercept. The best linear fit is
				\[ \alp = 0.774, a_0 = -0.35, a_2 = -0.10 \]
				with \(\chi^2_l = 14.56\ten{-4}\). The massive fits are better than t
he linear fit for masses up to 250 MeV, with the optimum being
				\[ m = 225, \alp = 0.823, a_0 = 0.00, a_2 = 0.26 \]
				with \rchi{0.87}. The optimum of the WKB fits is at a slightly higher mass. It is
				\[ m_w = 235, \alp = 0.789, a_0 = 0.00, a_2 = -1.20 \]
				and it has \rchi{0.86}. The big difference between the two values of the intercept is because now \(a_2\) includes a shift originating in the angular momentum.
				
				When fitting the \(\pi_2\) states alone, the linear fit
				\[ \alp = 0.840, a_2 = -0.33 \]
				is optimal (\(\chi^2_l = 2.61\ten{-4}\)).
				
				\paragraph{\(I = 1\). The \(a_1\) trajectory:} Depicted in figure (\ref{fig:pi_a1}), these are states with \(J^{PC} = 1^{++}\). They are: \(a_1(1260), a_1(1640), a_1(2095),\) and \(a_1(2270)\). The linear fit is
				\[ \alp = 0.783, a = -0.18 \]
				and it has \(\chi^2_l = 27.82\ten{-4}\). The massive fits here have an remarkably weak dependence of \(\chi^2\) on \(m\). The entire range \(m = 0-225\) MeV has values of \(\chi^2_m\) within \(1\%\) of that of the linear fit. For example,
				\[ m = 100, \alp = 0.787, a = -0.14\]
				has \(\chi^2_m = 27.82\ten{-4}\), and doubling the mass
				\[ m = 200, \alp = 0.796, a = -0.07\]
				only has the effect of changing \(\chi^2_m\) to \(27.84\ten{-4}\). The WKB fits likewise point to a large range - \(0-250\) MeV, and again there is nothing to distinguish any particular value of the mass. For the two masses quoted above, we have here the fits
				\[ m = 100, \alp = 0.787, a = -0.82 \]
				\[ m = 200, \alp = 0.803, a = -0.64 \]
				with \(\chi^2_w = 27.84\ten{-4}\) and \(27.79\ten{-4}\) respectively.
			
				\begin{figure}[tbp] \centering
						\includegraphics[natwidth=1200bp, natheight=900bp, width=.48\textwidth]{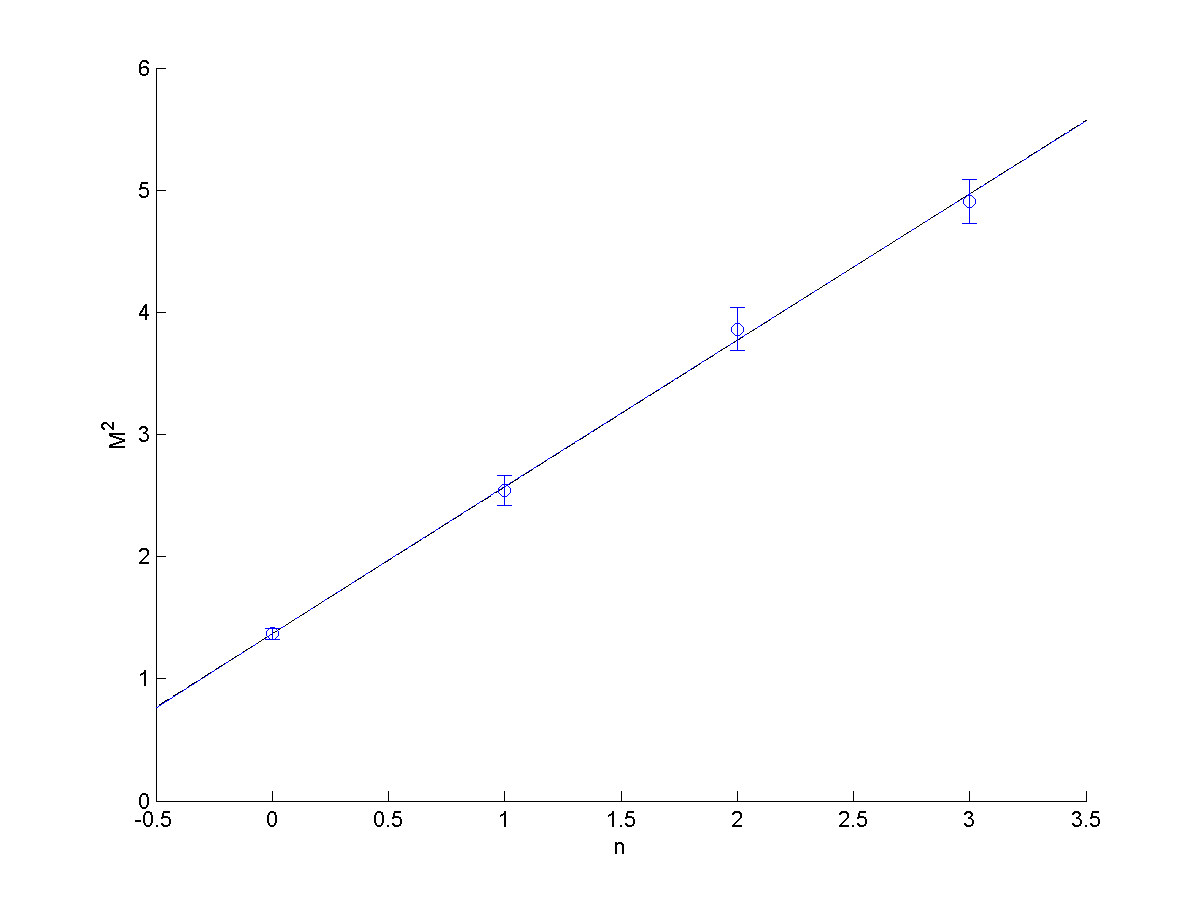}	 \hfill
						\includegraphics[natwidth=1200bp, natheight=900bp, width=.48\textwidth]{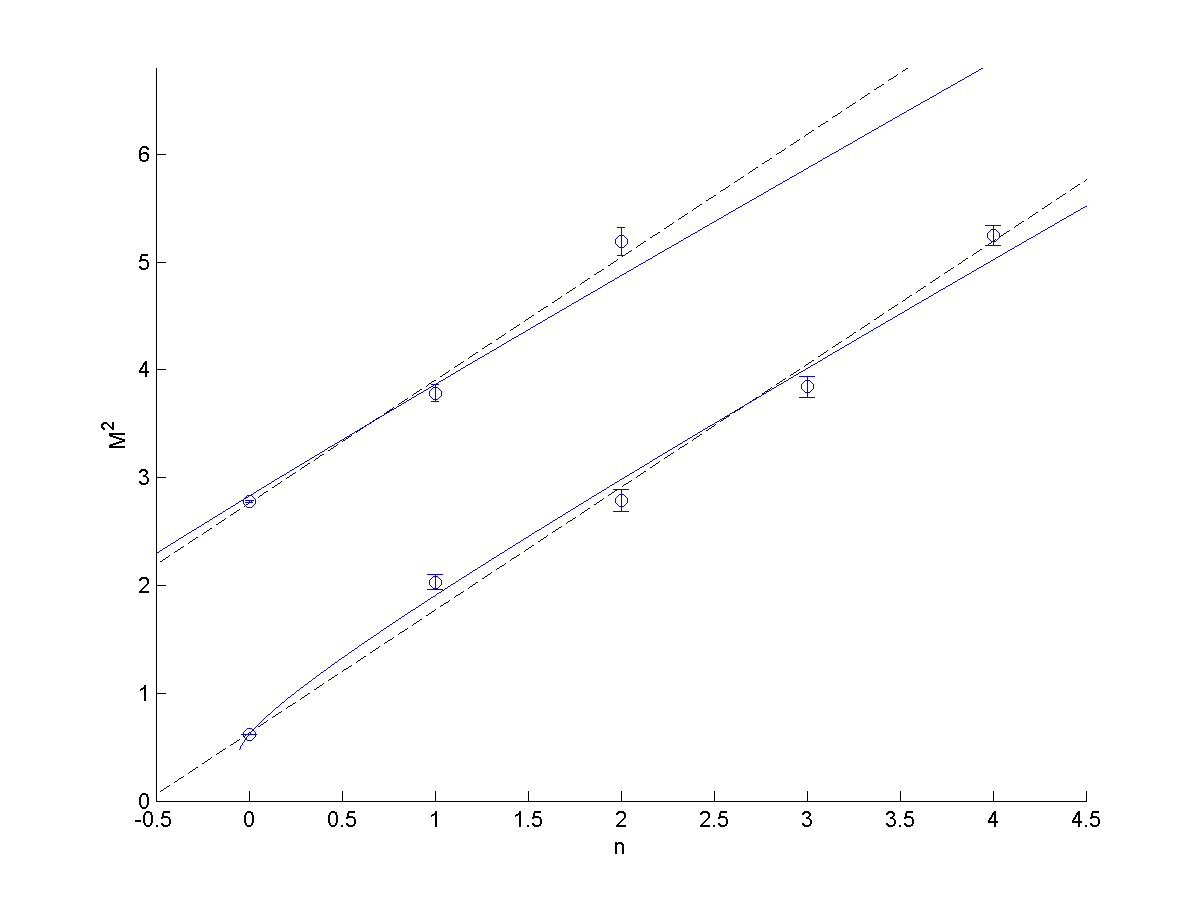}
						\caption{\label{fig:h1_omega} Left: the \(h_1\) with the massive fit \(\mud = 75\). Right: the \(\omega\) and \(\omega_3\) with their optimal fit of \(\mud = 305\).}
				\end{figure}
				\paragraph{\(I = 0\). The \(h_1\) trajectory:} In figure (\ref{fig:h1_omega}) we have the states \(h_1(1170), h_1(1595), h_1(1965), h_1(2215)\). They have \(J^{PC} = 1^{+-}\). The optimal linear fit is
				\[ \alp = 0.833, a = -0.14 \]
				It has \(\chi^2_l = 2.854\ten{-4}\). The massive fits are nearly all equivalent in the range \(m = 0-130\) MeV, with no clear optimum. The highest mass which gives a fit that is better than the linear is
				\[ m = 105, \alp = 0.850, a = -0.02 \]
				with \(\chi^2_m = 2.848\ten{-4}\). The best WKB fits are in the range \(m = 100-150\) MeV, with the minimum being
				\[ m_w = 135, \alp = 0.840, a = -0.71 \]
				with \(\chi^2_w = 2.826\ten{-4}\).
				
				\paragraph{\(I = 0\). The \(\omega\) trajectories:} Also in figure (\ref{fig:h1_omega}) are the states \(\omega(782), \omega(1420), \omega(1650), \omega(1960)\), and \(\omega(2290)\), with \(J^{PC} = 1^{--}\), and \(\omega_3(1670), \omega_3(1950),\) and \(\omega_3(2255)\) with \(J^{PC} = 3^{--}\).
				The best linear fit has
				\[ \alp = 0.877, a_1 = 0.45, a_3 = 0.58 \]
				and \(\chi^2_l = 34.30\ten{-4}\). Due to deviations of some of the states from the linear trajectory, we have a large range of masses that improve on it, up to \(400\) MeV. The optimum is with a very high mass
				\[ \mud = 340, \alp = 1.085, a_1 = 0.95, a_3 = 1.10 \]
				with \rchi{0.70}. The WKB fit is similar, with the optimum at a high mass:
				\[ m_w = 310, \alp = 0.979, a_1 = -0.10, a_3 = -1.38 \]
				and \rchi{0.75}.
				
				Fitting the \(\omega_3\) states alone, we get that the linear fit
				\[ \alp = 0.860, a_3 = 0.64 \]
				is optimal (\(\chi^2_l = 8.80\ten{-4}\)).
				
				Excluding the \(\omega(782)\) ground state and redoing the fits (for both the \(\omega\) and \(\omega_3\) states) results in the linear fit
				\[ \alp = 0.973, a_1 = 0.12, a_3 = 0.25 \]
				being optimal (\(\chi^2_l = 22.69\ten{-4}\)).
				
			\subsubsection{\texorpdfstring{$\ssb$}{s-sbar} mesons}
				\paragraph{\(\ssb\). The \(\phi\) trajectory:} There are three \(\phi\) states with \(J^{PC} = 1^{--}\): \(\phi(1020), \phi(1680)\), and \(\phi(2170)\).
				
				\begin{figure}[tbp] \centering
						\includegraphics[natwidth=1200bp, natheight=900bp, width=.48\textwidth]{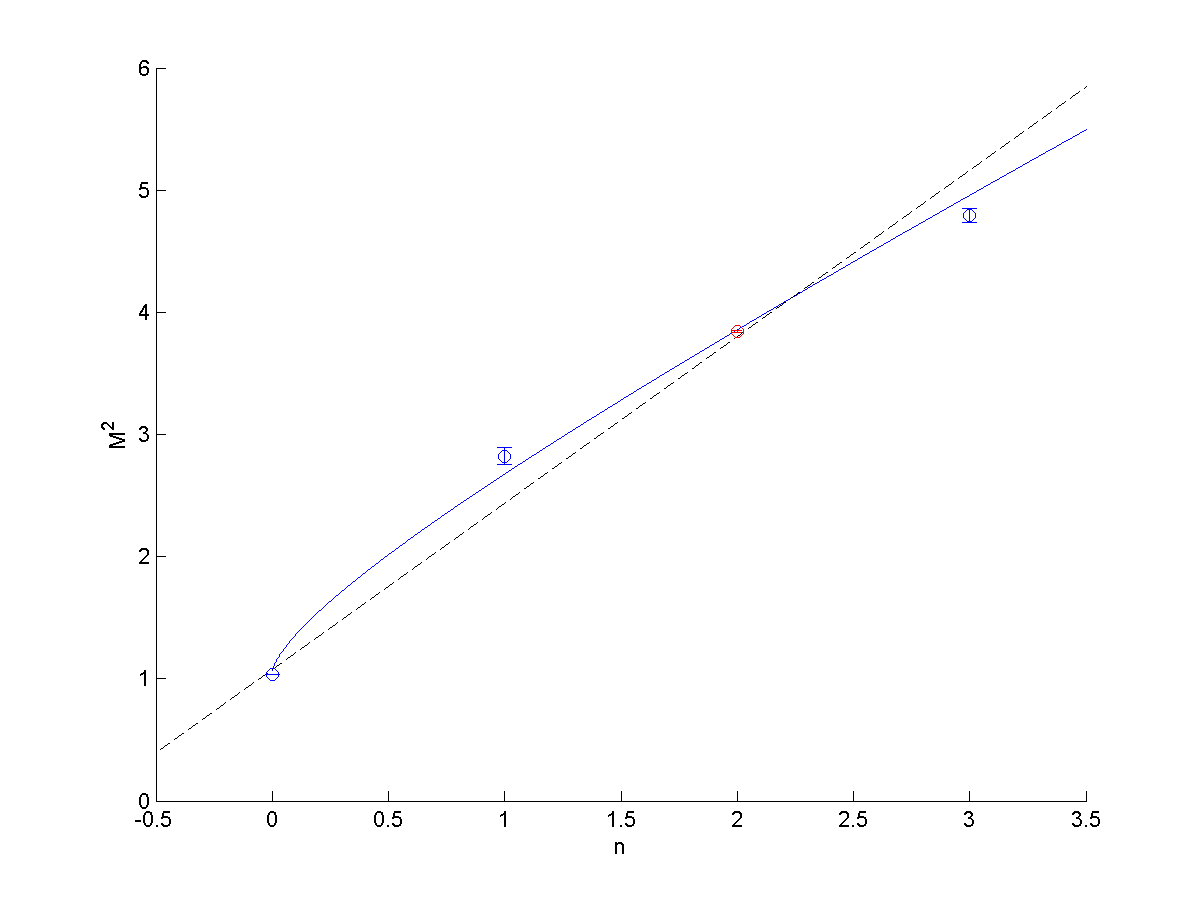}	 \hfill
						\includegraphics[natwidth=1200bp, natheight=900bp, width=.48\textwidth]{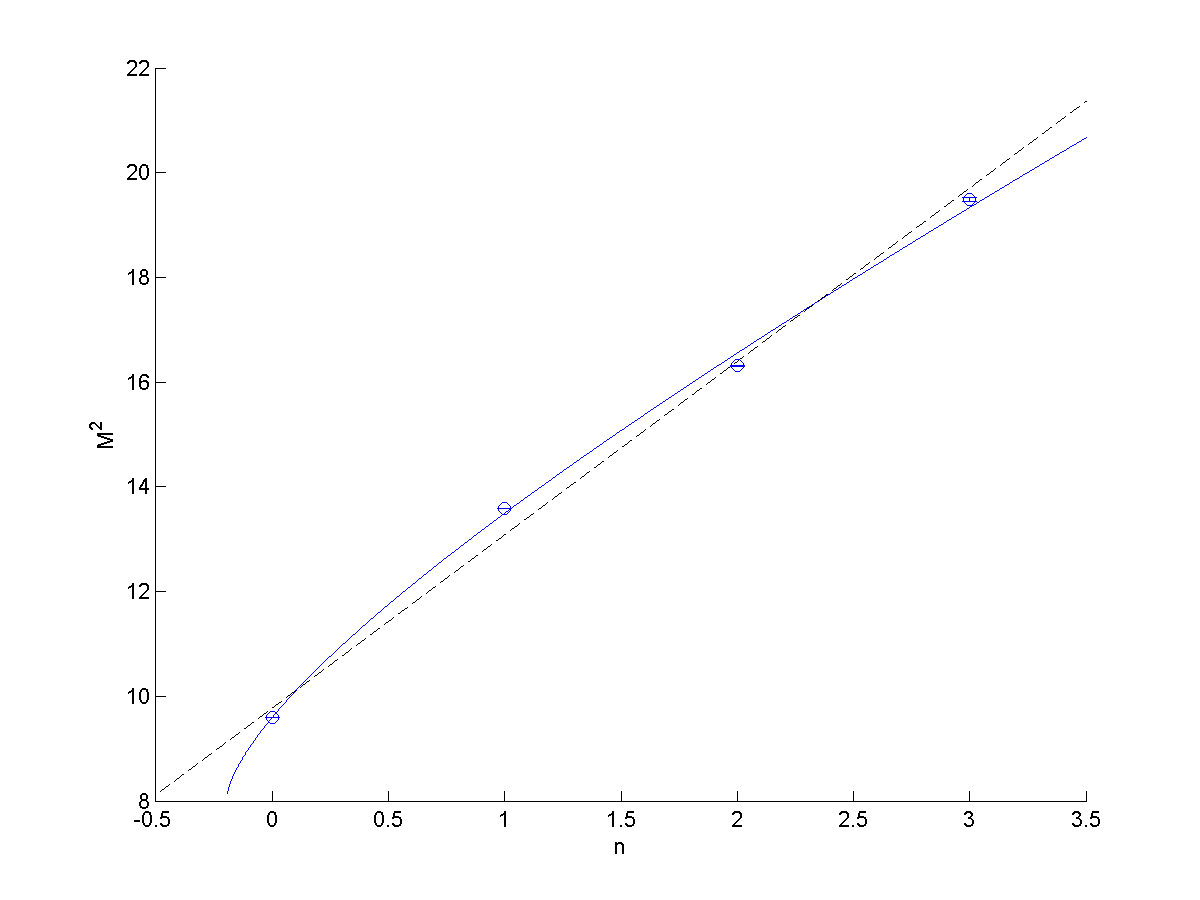}
						\caption{\label{fig:phi_psi} Left: the \(\ssb\) \(\phi\) radial trajectory with the massive fit \(m_s = 515\). The red marker is the state \(\omega(1960)\), which has the right quantum numbers and mass for the fit but is classified a (primarily) \(u/d\) state, and not \(\ssb\). Right: the \(\ccb\) \(\Psi\) trajectory with an optimal massive fit of \(m_c = 1425\).}
				\end{figure}
				
				Assuming the assignment \(n = 0, 1, 2\) for the three states we get the best fit is the linear fit with
				\[ \alp = 0.543, a = 0.44 \]
				It has \(\chi^2_l = 3.11\ten{-4}\). This is inconsistent with previous results, both in the resulting slope and \(s\) quark mass. We can get a fit with parameters closer to what we expect them to be if we make the assignment \(n = 0,1,3\) for the three states. The fits with this assignment are what is depicted in figure (\ref{fig:phi_psi}).
				
				The linear fit then is
				\[ \alp = 0.724, a = 0.21 \]
				with \(\chi^2_l = 129\ten{-4}\). The best massive fit has
				\[ m_s = 515, \alp = 1.098, a = 1.00 \]
				with \rchi{0.16}. We can fit with a lower mass as well. The fit
				\[ m_s = 400, \alp = 0.909, a = 0.84 \]
				has \rchi{0.52}. The best WKB fit, using this same assignment for \(n\), is
				\[ m_w = 515, \alp = 1.027, a = 0.00 \]
				
				The mass of the missing \(n = 2\) state is predicted to be in the range \(M = 1949-1963\) MeV. Interestingly, there is a state with the appropriate quantum numbers \(I^G(J^{PC}) = 0^-(1^{--})\) at that exact mass - the \(\omega(1960)\). If we add this state in this trajectory, the fits don't change much:
				\[ \alp = 0.730, a = 0.22 \]
				is the optimal linear fit - \(\chi^2_l = 86\ten{-4}\), and the optimal massive fit is still at the same mass
				\[ m_s = 515, \alp = 1.100, a = 1.00 \]
				with \(\chi^2_m = 13.85\ten{-4}\) (\rchi{0.16}).
				
			\subsubsection{\texorpdfstring{$\ccb$}{c-cbar} mesons}
				\paragraph{\(\ccb\). The \(\Psi\) trajectory:} The right plot in figure (\ref{fig:phi_psi}) shows the trajectory formed by the states	\(J/\Psi(1S)(3097), \Psi(2S)(3686), \Psi(4040),\) and \(\Psi(4415)\), all with \(J^{PC} = 1^{--}\). The best fitting linear trajectory is
				\[\alp = 0.299, a = -1.91\]
				with \(\chi^2_l = 6.23\ten{-4}\). The massive fit does not show a clear preference for a single value for the mass as it did for the \(\Psi\) trajectory in the angular momentum plane. Instead we find that the optimal mass is in the range \(m_c = 1350-1475\), where the best fit is
				\[ m_c = 1425, \alp = 0.482, a = 0.81 \]
				with \rchi{0.17}. The WKB offers similar results for the mass. The best fits are in the range \(1390-1460\) MeV with the optimum at
				\[ m_w = 1435, \alp = 0.488, a = -0.17 \]
				
				\begin{figure}[tbp] \centering
						\includegraphics[natwidth=1200bp, natheight=900bp, width=.48\textwidth]{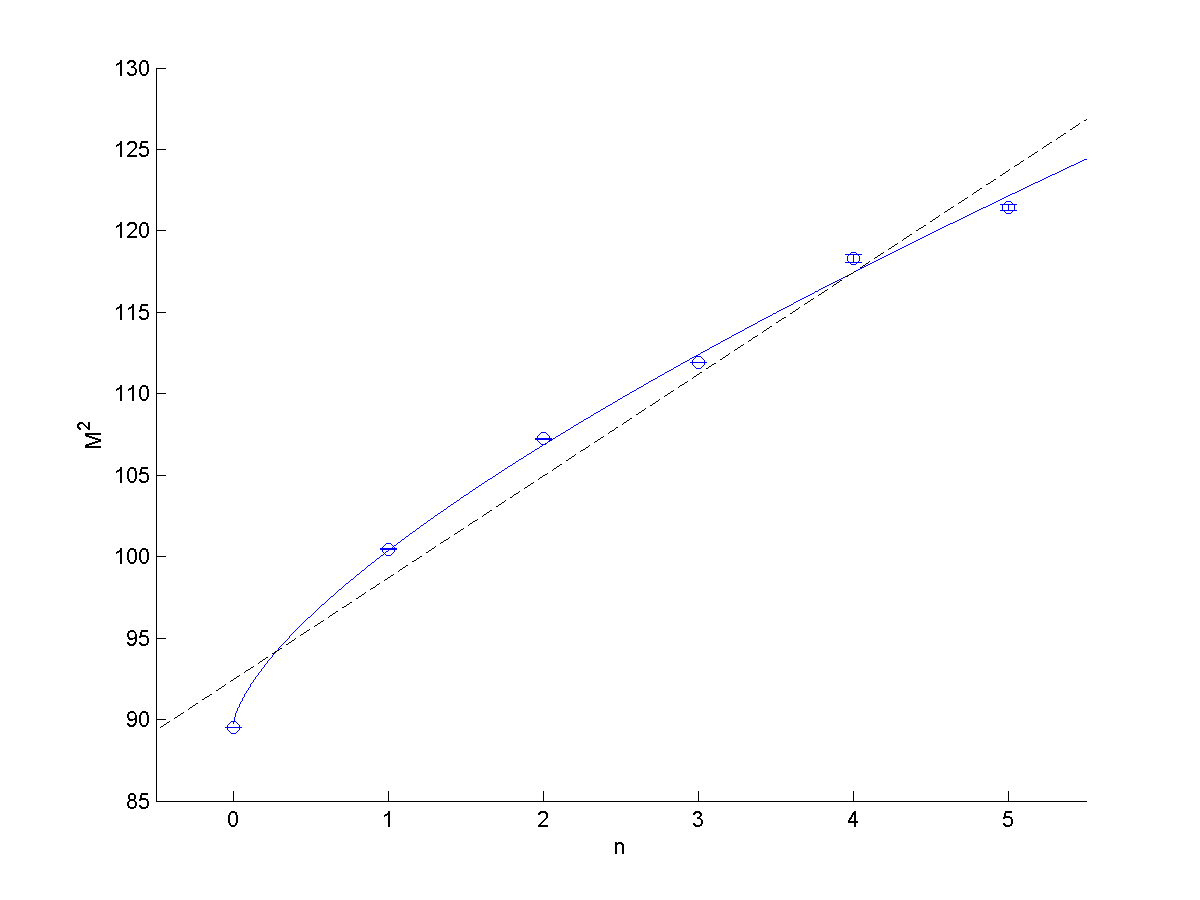}	 \hfill
						\includegraphics[natwidth=1200bp, natheight=900bp, width=.48\textwidth]{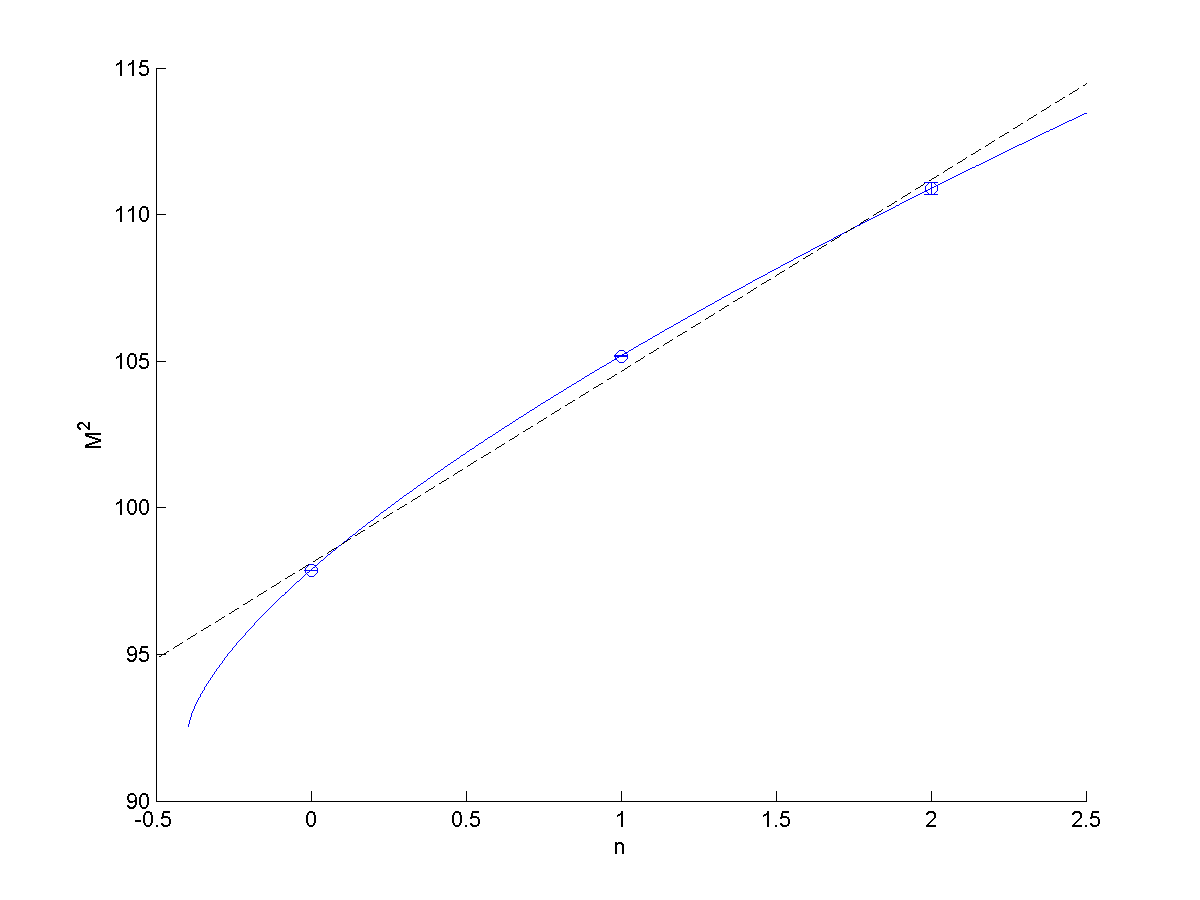}
						\caption{\label{fig:ups_chib} The \(\bbb\) trajectories and massive fits. Left: \(\Upsilon\) with its optimal fit of \(m_b = 4730\). Right: the \(\chi_b\) with \(m_b = 4800\).}
				\end{figure}

			\subsubsection{\texorpdfstring{$\bbb$}{b-bbar} mesons}
				\paragraph{\(\bbb\). The \(\Upsilon\) trajectory:} Depicted in the left side of figure (\ref{fig:ups_chib}) is the \(\Upsilon\) radial trajectory. It consists of six states: \(\Upsilon(1S)(9460), \Upsilon(2S)(10023), \Upsilon(3S)(10355) ,\Upsilon(4S)(10579), \Upsilon(10860),\) and \(\Upsilon(11020)\), with \(J^{PC} = 1^{--}\). The best linear fit for it is
				\[ \alp = 0.157, a = -13.46 \]
				which has \(\chi^2_l = 4.50\ten{-4}\). The best massive fit points to the constituent mass for the \(b\) quark again, with
				\[ m_b = 4730, \alp = 0.458, a = 1.00 \]
				being the optimum with \(\chi^2_m = 0.26\ten{-4}\) (\rchi{0.06}). The optimal WKB fit is similar with
				\[ m_b = 4735, \alp = 0.422, a = 0.00 \]
				and it has \(\chi^2_w = 0.25\ten{-4}\). The best fit with \(J_q = J\) is
				\[ m_w = 4625, \alp = 0.357, a = 0.00 \]
				which has \(\chi^2_w = 1.04\ten{-4}\).				
				
				\paragraph{\(\bbb\). The \(\chi_b\) trajectory:} The other \(\bbb\) trajectory, on the right side of figure (\ref{fig:ups_chib}) is that of the \(\chi_{b1}\). It consists of three states with \(J^{PC} = 1^{++}\): \(\chi_{b1}(1P)(9893), \chi_{b1}(2P)(10255),\) and \(\chi_b(3P)(10530)\). The linear fit for them is
				\[ \alp = 0.153, a = -13.01 \]
				with \(\chi^2_l = 0.19\ten{-4}\). The optimum is located at
				\[ m_b = 4800, \alp = 0.499, a = 0.58 \]
				and it has \(\chi^2_m = 4\ten{-8}\) (\rchi{0.002}). The best WKB fit is
				\[ m_w = 4825, \alp = 0.473, a = -0.06 \]
				and it has \(\chi^2_w = 2\ten{-8}\) (\rchi{0.001}).
				
\begin{figure}[tbp] \centering
						\includegraphics[natwidth=1200bp, natheight=900bp, width=.48\textwidth]{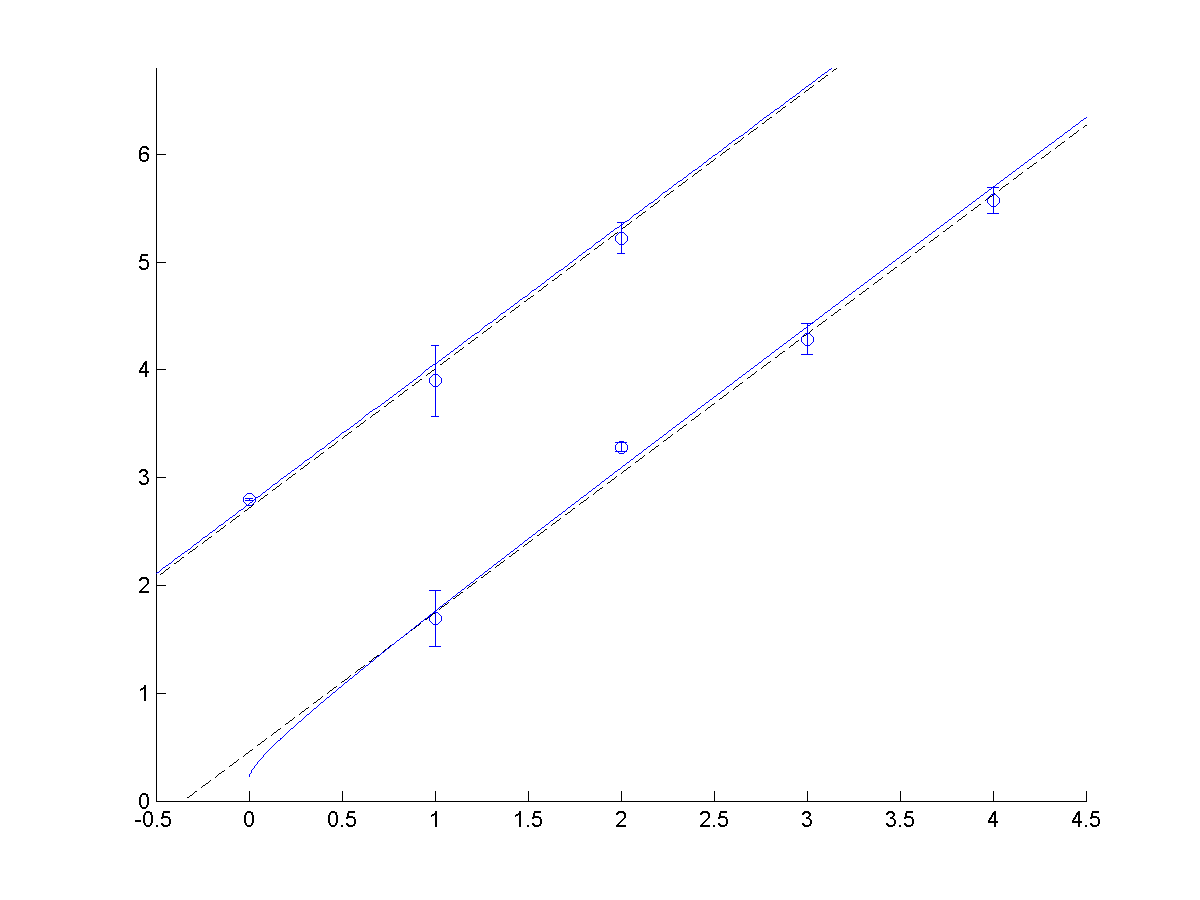}	 \hfill
						\includegraphics[natwidth=1200bp, natheight=900bp, width=.48\textwidth]{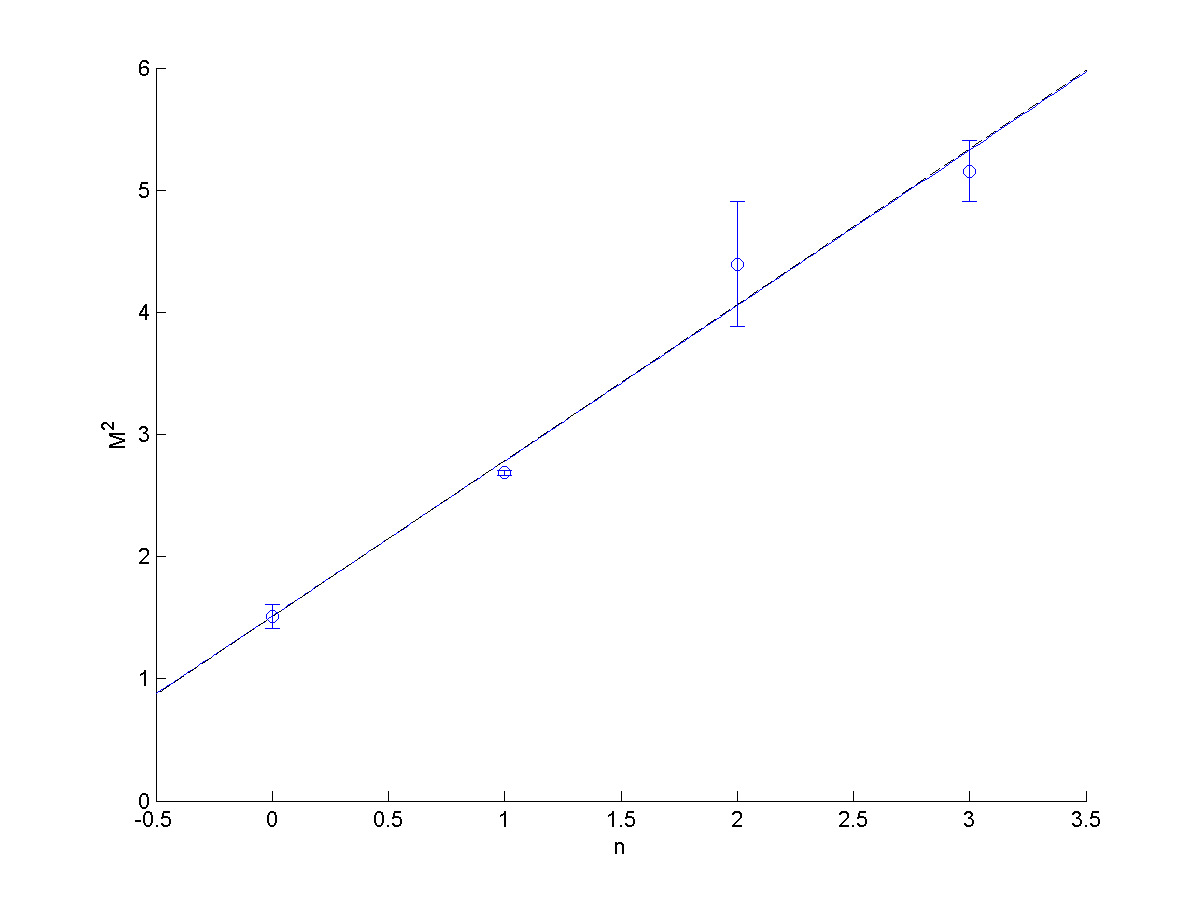} \\
						\includegraphics[natwidth=1200bp, natheight=900bp, width=.48\textwidth]{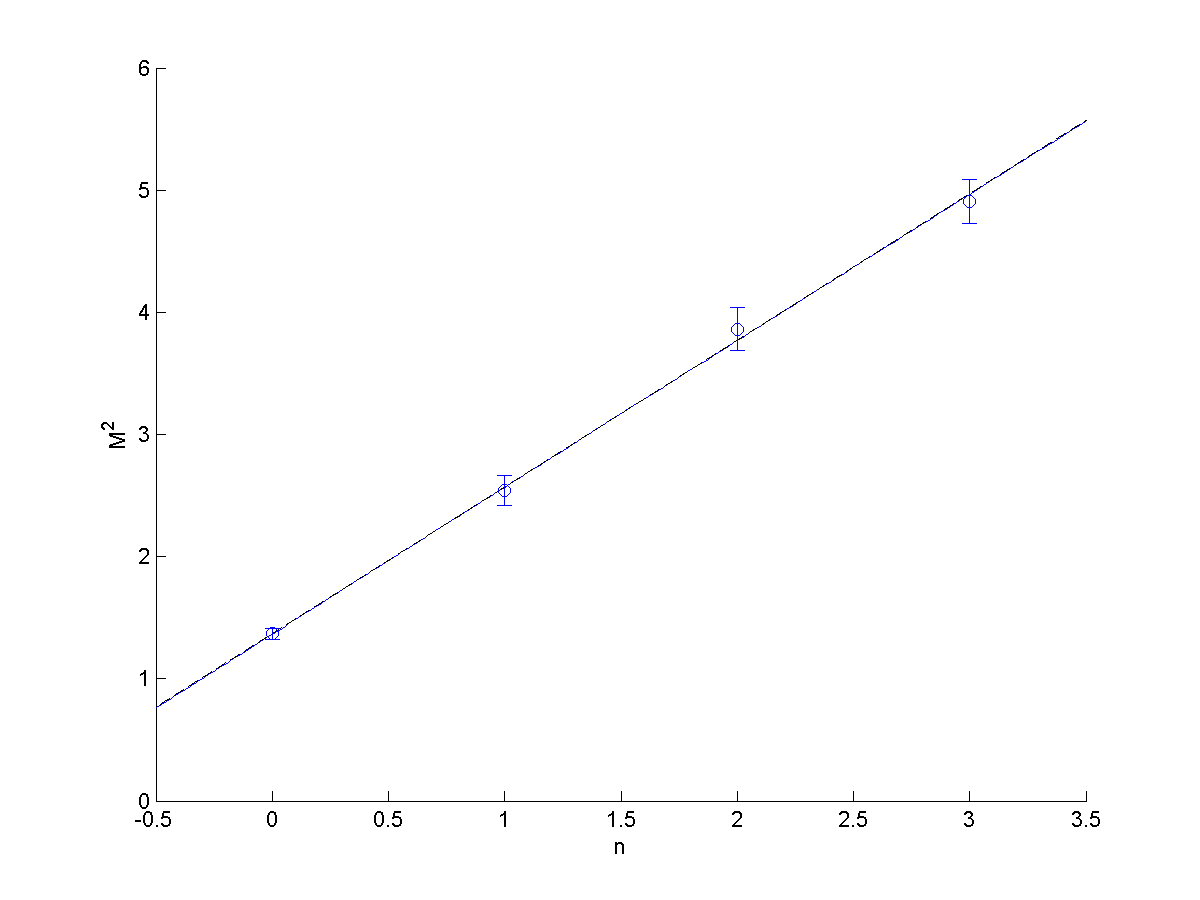}	 \hfill
						\includegraphics[natwidth=1200bp, natheight=900bp, width=.48\textwidth]{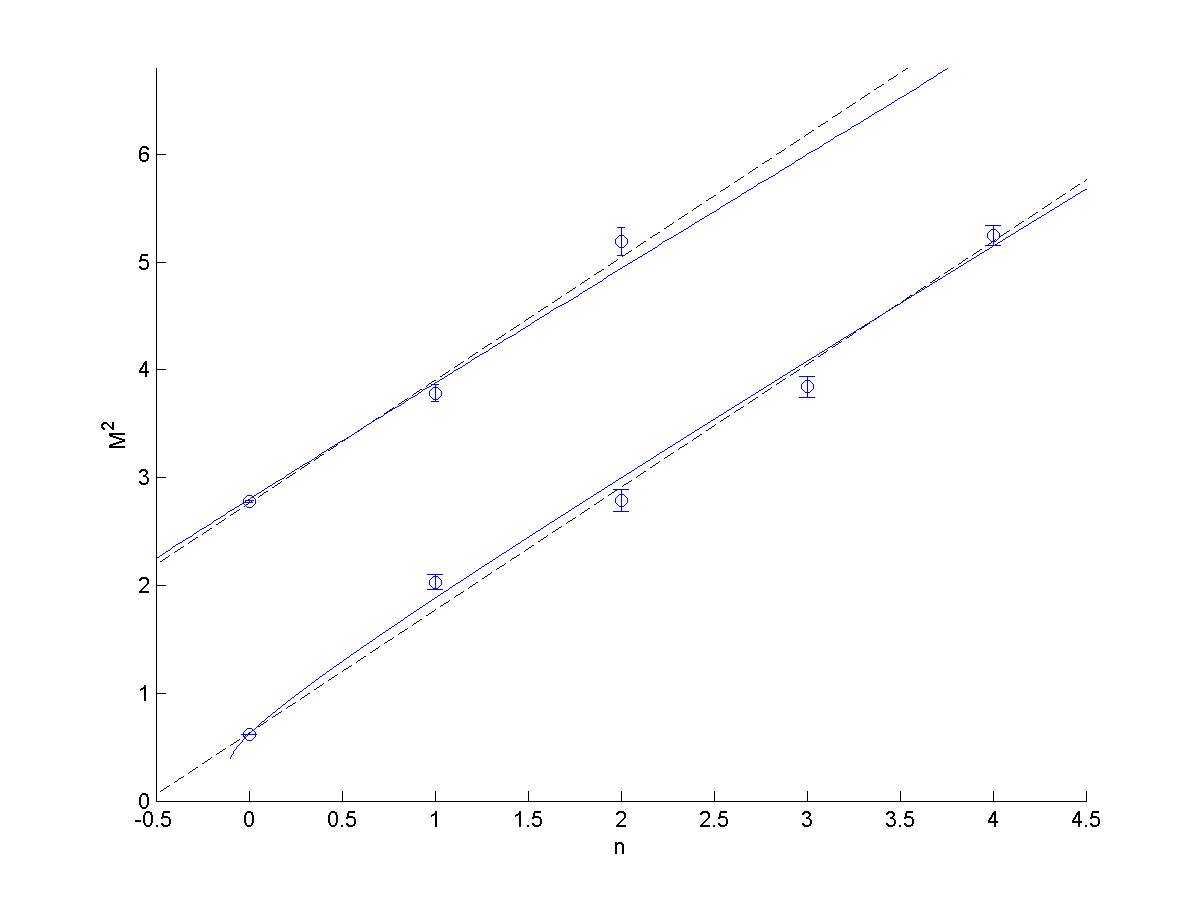} \\
						\caption{\label{fig:wkb_light} The light quark trajectories with their optimal WKB fits. Top left: \(\pi\) (\(0^{-+}\)) and \(\pi_2\) (\(2^{-+}\)). Top right: \(a_1\) (\(1^{++}\)). Bottom left: \(h_1\) (\(1^{+-}\)). Bottom right: \(\omega\) (\(1^{--}\)) and \(\omega_3\) (\(3^{--}\)).}
				\end{figure}

\begin{figure}[tbp] \centering
						\includegraphics[natwidth=1200bp, natheight=900bp, width=.48\textwidth]{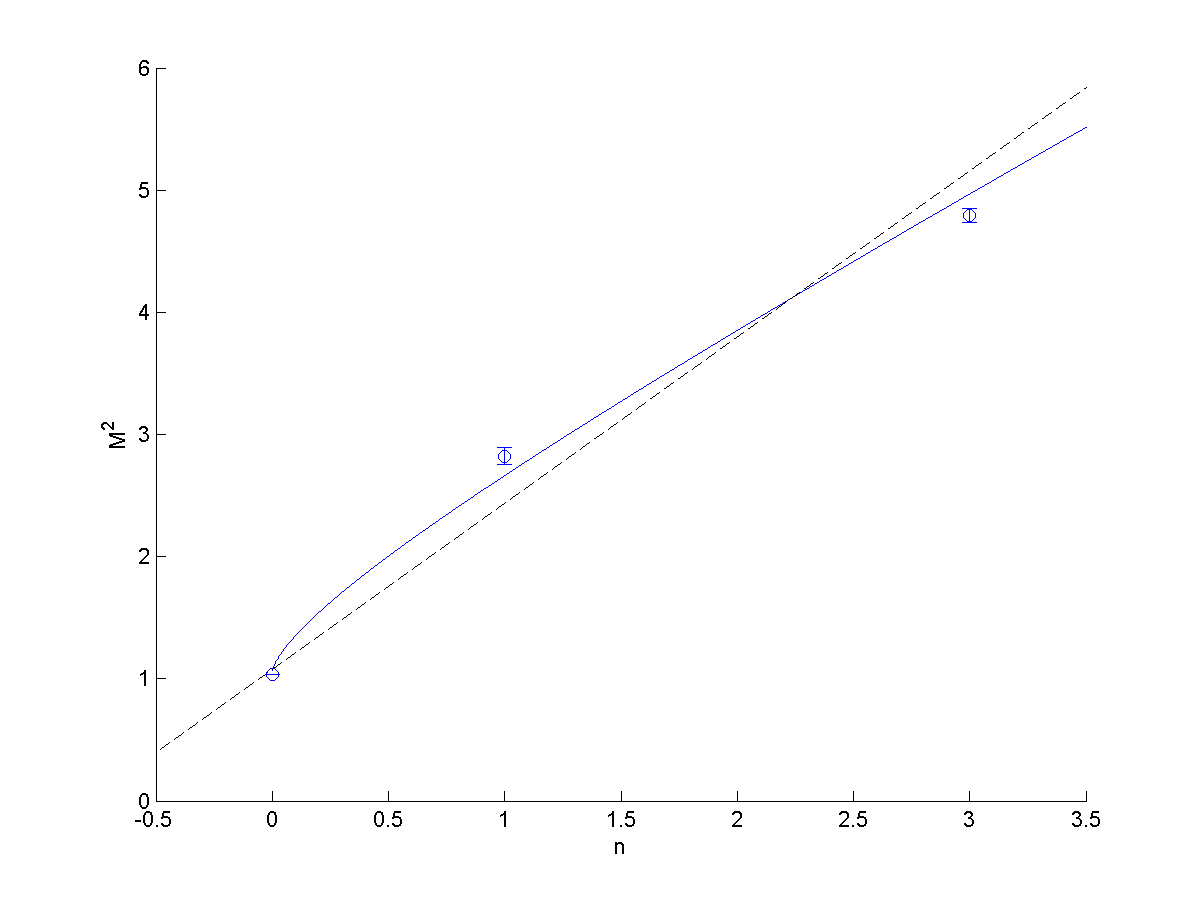}	 \hfill
						\includegraphics[natwidth=1200bp, natheight=900bp, width=.48\textwidth]{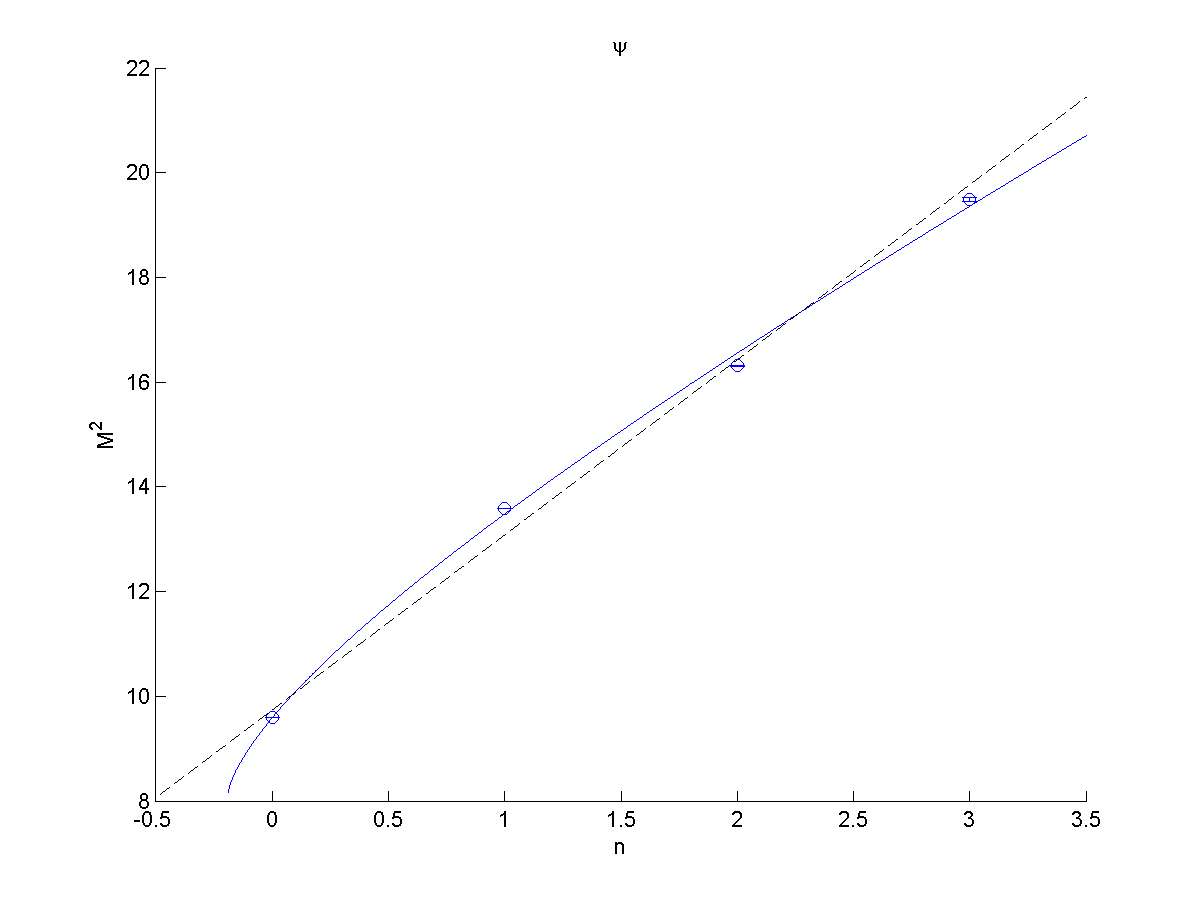} \\
						\includegraphics[natwidth=1200bp, natheight=900bp, width=.48\textwidth]{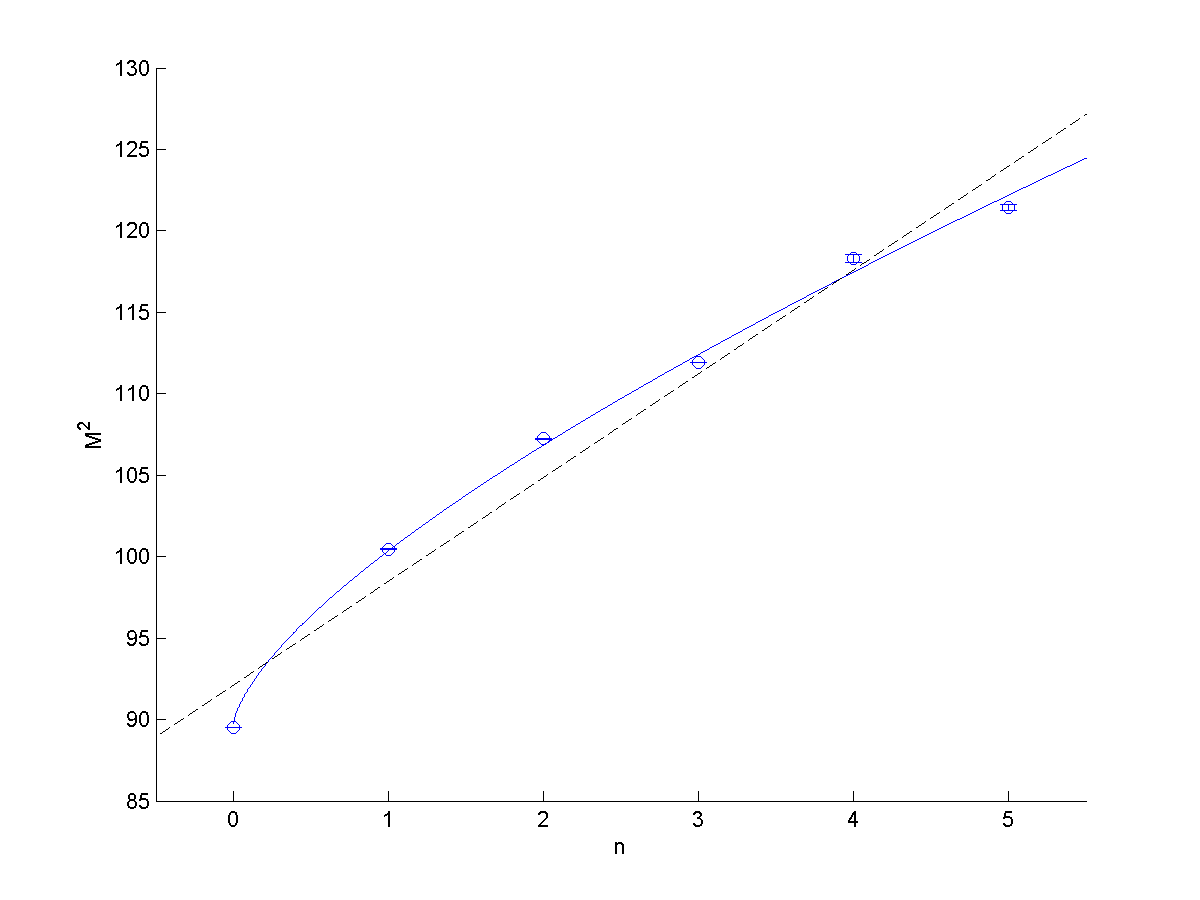}	 \hfill
						\includegraphics[natwidth=1200bp, natheight=900bp, width=.48\textwidth]{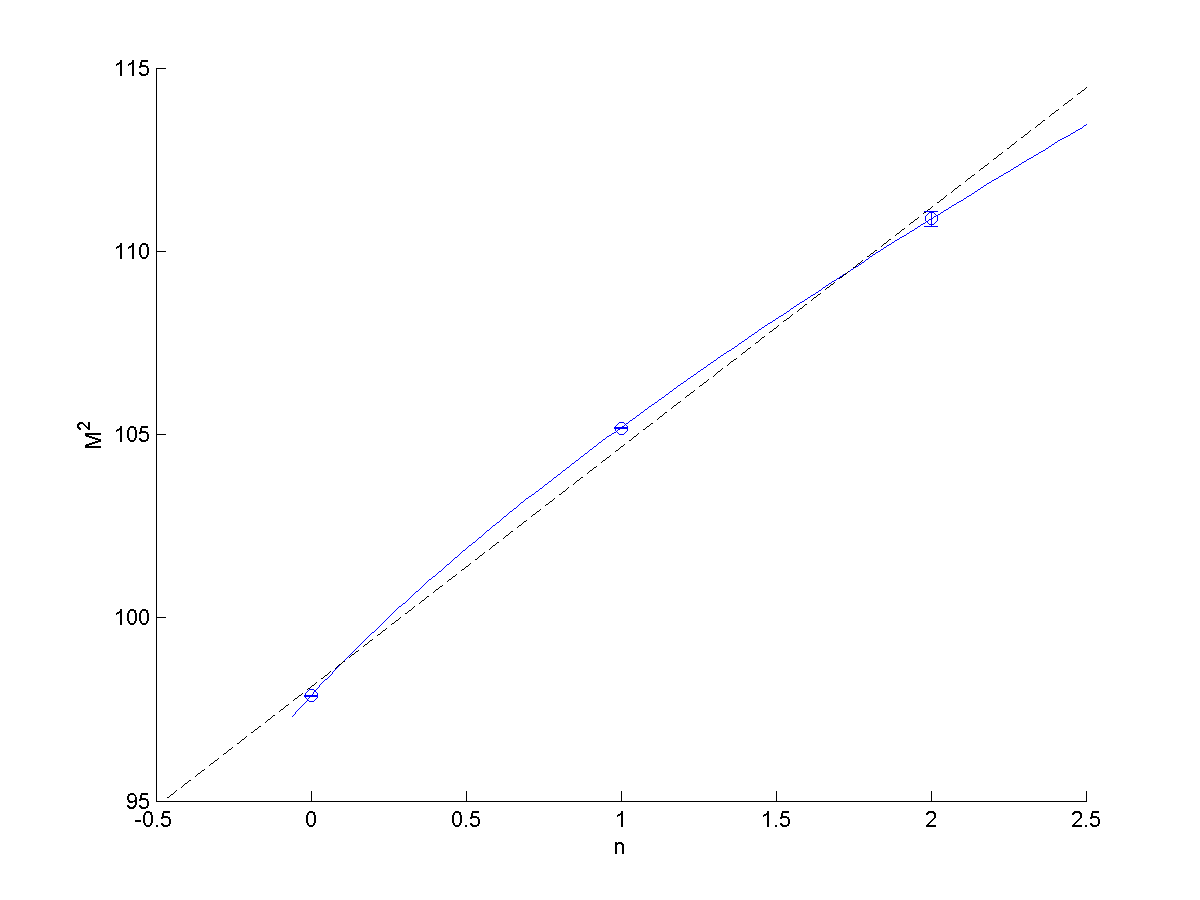} \\
						\caption{\label{fig:wkb_heavy} The heavier quark trajectories with their optimal WKB fits. Top left: \(\phi\) (\(\ssb\), \(1^{--}\)). Top right: \(\psi\) (\(\ccb\), \(1^{--}\)). Bottom left: \(\upsilon\) (\(\bbb\) \(1^{--}\)). Bottom right: \(\chi_b\) (\(\bbb\), \(1^{++}\)).}
				\end{figure}

				\paragraph{WKB fit plots:} Lastly, we include, in figures (\ref{fig:wkb_light}) and (\ref{fig:wkb_heavy}), the plots of all the \((n,M^2)\) trajectories and their respective WKB fits.
				
\clearpage				
\section{Universal fit: Calculated vs. measured masses} \label{app:universal}
We present, in table (\ref{tab:universal_masses}), the values of the masses obtained from the universal slope fit in section \ref{sec:universal} vs. their experimental values. The plots of the 9 trajectories used are in figure (\ref{fig:multiFit_j}).

\begin{table}[tpb] \centering
	\begin{tabular}{|c|c|c|c|c|c|c|c|c|} \hline
		Traj. & \(J^{PC}\) & Exp. & Calc. & \qquad & Traj. & \(J^{PC}\) & Exp. & Calc. \\ \hline\hline
		
		\(\pi/b\) & \(1^{+-}\) & 1229 & 1257 & 			     &	\(K^*\) & \(1^-\)    & 892  & 892  \\
		          & \(2^{-+}\) & 1672 & 1650 &				            &	& \(2^+\)    & 1426 & 1415 \\
		          & \(3^{+-}\) & 2032 & 1965 &				            &	& \(3^-\)    & 1776 & 1783 \\
							& \(4^{-+}\) & 2250 & 2236 &				            &	& \(4^+\)    & 2045 & 2084 \\ \cline{1-4}
		\(\rho/a\)& \(1^{--}\) & 776  & 776  &				            &	& \(5^-\)    & 2382 & 2345 \\ \cline{6-9}
							& \(2^{++}\) & 1318 & 1324 &			   &\(\phi/f'\) & \(1^{--}\) & 1020 & 1019 \\
							& \(3^{--}\) & 1689 & 1701 &				            &	& \(2^{++}\) & 1525 & 1514 \\
							& \(4^{++}\) & 1996 & 2008 &				            &	& \(3^{--}\) & 1854 & 1870 \\ \cline{6-9}
							& \(5^{--}\) & 2330 & 2274 &				    &	\(D\)   & \(0^-\)    & 1865 & 1862 \\
							& \(6^{++}\) & 2450 & 2511 &				            &	& \(1^+\)    & 2421 & 2408 \\ \cline{1-4}
		\(\eta/h\)& \(0^{-+}\) & 548  & 545  &				            &	& \(2^-\)    & 2737 & 2752 \\ \cline{6-9}
							& \(1^{+-}\) & 1170 & 1206 &				  &	\(D^*_s\) & \(1^-\)    & 2112 & 2112 \\
							& \(2^{-+}\) & 1617 & 1612 &				            &	& \(2^+\)    & 2572 & 2563 \\
							& \(3^{+-}\) & 2025 & 1933 &				            &	& \(3^-\)    & 2862 & 2881 \\ \cline{6-9}
							& \(4^{-+}\) & 2328 & 2208 &					 &	\(\Psi\)& \(1^{--}\) & 3097 & 3080 \\ \cline{1-4}
		\(\omega/f\)&\(1^{--}\)& 783  & 768  &				            &	& \(1^{++}\) & 3494 & 3535 \\
							& \(2^{++}\) & 1275 & 1319 &				            &	& \(1^{--}\) & 3778 & 3824 \\ \cline{6-9}
							& \(3^{--}\) & 1667 & 1698 &				            &	& & & \\
							& \(4^{++}\) & 2018 & 2006 &				            &	& & & \\
							& \(5^{--}\) & 2250 & 2271 &				            &	& & & \\
							& \(6^{++}\) & 2469 & 2509 &				            &	& & & \\ \hline
	\end{tabular}
	\caption{\label{tab:universal_masses} Comparison of calculated and measured masses for all the states used in the universal trajectory fit. There are 38 states in total, and 13 fitting parameters (one slope, 3 quark endpoint masses, and 9 intercepts). We use the values \(m_{u/d} = 60\), \(m_s = 220\), and \(m_c = 1500\) MeV for the masses, \(\alp = 0.884\) GeV\(^{-2}\) for the slope. For the \(c\bar{c}\) \(\Psi\) we use states with equal \(J\), but increasing \(L\).}
\end{table}

\section{Predictions for higher states}
	In this section we list our predictions, based on our fits, for the masses of the next higher states in each trajectory. The values used to compute these predictions are the same values that were used in the summary tables ((\ref{tab:mes_j}) and (\ref{tab:mes_n})) in section 4.4.
	Table (\ref{tab:predictions}) has the predictions for the \((J,M^2)\) trajectories, for higher \(J\) states. In table (\ref{tab:predictions_n}) we list the predictions for the \((n,M^2)\) trajectories, for highly excited states with fixed \(J^{PC}\).
	
	\begin{table}[tpb] \centering
					\begin{tabular}{|c|cc|cc|} \hline
						
						Trajectory & \multicolumn{4}{|c|}{Next states} \\ \hline
						
						\(\pi/b\) & \(5^{+-}\): & \(2525-2540\) & \(6^{-+}\): & \(2750-2770\) \\
						
						\(\rho\) & \(7^{--}\): & \(2695-2720\) & \(8^{++}\): & \(2890-2920\) \\
						
						\(\eta/h\) & \(5^{+-}\): & \(2495-2520\) & \(6^{-+}\): & \(2720-2750\) \\
						
						\(\omega/f\) & \(7^{--}\): & \(2680-2685\) & \(8^{++}\): & \(2875-2885\) \\
						
						\(K^*\) & \(6^+\) & \(2580-2590\) & \(7^-\) & \(2790-2810\) \\
						
						\(\phi\) & \(4^{++}\): & \(2120\) & \(5^{--}\): & \(2350\) \\
						
						\(D\) & \(3^-\) & \(2990\) & \(4^+\) & \(3205\) \\
						
						\(\Psi\) & \(4^{++}\): & \(4000\) & \(5^{--}\): & \(4195\) \\
						
						\(\Upsilon\) & \(4^{++}\): & \(10380\) & \(5^{--}\): & \(10570\) \\
						
					\hline \end{tabular}
					\caption{\label{tab:predictions} Predictions for the next states in the \((J,M^2)\) plane based on the optimal massive fits, with their \(J^{PC}\) and mass (in MeV) values. The ranges listed correspond to the ranges in table (\ref{tab:mes_j}).}
					\end{table}
					
						\begin{table}[tpb] \centering
					\begin{tabular}{|c|c|cc|cc|} \hline
						
						Traj. & \(J^{PC}\) & \multicolumn{4}{|c|}{Next states} \\ \hline
						
						\(\pi\) & \(0^{-+}\) & \(n = 5\): & \(2635-2675\) & \(n = 6\): & \(2870-2910\) \\
						
						\(\pi_2\) & \(2^{-+}\) & \(n = 3\): & \(2425-2475\) & \(n = 4\): & \(2680-2725\) \\
						
						\(a_1\) & \(1^{++}\) & \(n = 4\): & \(2535-2575\) & \(n = 5\): & \(2750-2810\) \\
						
						\(h_1\) & \(1^{--}\) & \(n = 4\): & \(2470-2485\) & \(n = 5\): & \(2690-2715\) \\
					
						\(\omega\) & \(1^{--}\) & \(n = 5\): & \(2535-2540\) & \(n = 6\): & \(2740\) \\
						
						\(\omega_3\) & \(3^{--}\) & \(n = 3\): & \(2375\) & \(n = 4\): & \(2600\) \\
						
						\(\phi\) & \(1^{--}\) & \(n = 2\): & \(1965\) & \(n = 4\): & \(2450-2460\) \\
						
						\(\Psi\) & \(1^{--}\) & \(n = 4\): & \(4670-4700\) & \(n = 5\): & \(4925-4975\) \\
						
						\(\Upsilon\) & \(1^{--}\) & \(n = 6\): & \(11245-11260\) & \(n = 7\): & \(11430-11450\) \\
						
						\(\chi_b\) & \(1^{++}\) & \(n = 3\): & \(10765\) & \(n = 4\): & \(10980\) \\
						
					\hline \end{tabular}
					\caption{\label{tab:predictions_n} Predictions for the next states in the \((n,M^2)\) plane based on the optimal massive fits. We use an assignment where the ground state has \(n = 0\), and masses are in MeV. For the \(\phi\), where we have assigned the three known states the values \(n = 0\), \(1\), and \(3\), one of the masses is that of the missing \(n = 2\) state. The ranges listed correspond to the ranges in table (\ref{tab:mes_n}).}
					\end{table}
					\clearpage
\bibliographystyle{JHEP}
\bibliography{Spectrum}
\end{document}